%% file: main.tex
\documentclass[aps,prx,twocolumn,10pt,superscriptaddress,nofootinbib,longbibliography,floatfix]{revtex4-2}
\usepackage{amsmath,amssymb,amsthm,mathtools,bm}
\usepackage{graphicx}
\usepackage{physics}
\usepackage{microtype}
\usepackage[hidelinks]{hyperref}
\makeatletter
\AtBeginDocument{%
\setlength{\@fptop}{0pt}
\setlength{\@dblfptop}{0pt}
\setlength{\@fpbot}{0pt plus 1fil}
\setlength{\@dblfpbot}{0pt plus 1fil}
\setlength{\@fpsep}{14pt plus 2pt minus 2pt}
\setlength{\@dblfpsep}{14pt plus 2pt minus 2pt}
}
\makeatother

\renewcommand{\Tr}{\operatorname{Tr}}
\newcommand{\TV}{\operatorname{TV}}
\renewcommand{\dd}{\mathrm d}
\newtheorem{theorem}{Theorem}
\newtheorem{proposition}{Proposition}

\begin{document}
\title{Multitime Equilibration and Multitime Mpemba Effects}
\author{He Wang}
\email{dr\_hewang19@163.com}
\begin{abstract}
Equilibration of a quantum state leaves open whether its response to later
interventions has also equilibrated. We establish multitime equilibration
for a continuous process on a fixed observation window, retaining the
autonomous system--environment dynamics and all internal time differences.
A finite-time mean-square bound controls the full continuous-process norm
through the effective energy dimension and energy-gap correlations. A
factorial temporal-order bound lifts fixed-sector convergence to the full
process, while a process-visible spectral measure distinguishes
mean-square from pointwise equilibration. This framework defines multitime
Mpemba effects by comparing preparations relative to one stationary
process. In a semicircle-spectrum model, a preparation filter changes the
decay exponent from $3/2$ to $5/2$ in every fixed sector and in the summed
norm, despite a larger initial distance. A fixed unitary control and
terminal measurement exhibit the same exponent improvement in their
output probabilities. Noncommuting baths and a many-body spin environment
separate this all-sector mechanism from selective finite-time acceleration.
\end{abstract}
\maketitle
\raggedbottom

\section{Introduction}
An equilibrated state need not have an equilibrated response. A system can
look stationary under every undisturbed observation and still reveal a
non-equilibrium memory when an operation is inserted later in the
experiment. The intervention exposes correlations with hidden degrees of
freedom. Quantum combs and process tensors retain this response and the
memory carried by an
unobserved environment~\cite{Chiribella2008,Chiribella2009,Pollock2018,MilzModi2021}.
The question is whether the complete response approaches a stationary process.

Finite-resolution process equilibration has been established through
time-averaged statistics and finite-time distinguishability
bounds~\cite{FigueroaRomero2020,DowlingRelaxation2023,DowlingEquilibration2023}.
Their independent interval averages lead to a process with energy
dephasing between interventions. To retain continuous response within an
experiment, we instead translate one observation window along an
autonomous trajectory. The reference uses a stationary preparation and
the same Hamiltonian. This choice preserves finite-delay correlations
and makes all temporal orders marginals of one process. It also brings
the complete response into a single continuous process tensor (cPT).

The continuous limit raises a physical question: does equilibration at
every fixed temporal order control the response to arbitrary bounded
continuous interventions? The operational continuum limit of combs and
the continuum dilation theorem supply the process and instrument
spaces~\cite{Wassner2026,Fuksa2026}. Their continuous matrix product
state (cMPS) representation organizes the entire hierarchy
\cite{Verstraete2010,Haegeman2013}. We combine this structure with
energy-gap estimates to obtain a full-process mean-square bound.
A summable temporal-order estimate then connects fixed-sector
convergence to the complete process.

Once equilibration is established, different preparations can be compared
by the parts of this response that decay slowly. The Mpemba mechanism
allows an initially more distant preparation to approach stationarity
faster~\cite{Mpemba1969,LuRaz2017,Klich2019}. Quantum examples include
Markovian relaxation~\cite{Carollo2021,Nava2024,Moroder2024}, non-Markovian
dynamics~\cite{Strachan2025}, isolated integrable and nonintegrable many-body evolution
\cite{Rylands2024,Bhore2025}, experimental trapped-ion and quantum-simulation realizations
\cite{Joshi2024,AharonyShapira2024,ZhangStrong2025}, and correlated system--reservoir preparations
\cite{LonghiEnt2025}. Recent reviews organize these mechanisms around slow-mode filtering,
symmetry restoration, and entanglement asymmetry~\cite{AresReview2025}.
Here the comparison concerns the full temporal
response. A preparation can suppress a slow component in selected
sectors or throughout the hierarchy. We derive the spectral filtering
criterion for this distinction and solve a model with an exponent
improvement in every sector. A fixed control and measurement make the
same improvement observable. Noncommuting and many-body examples
identify the different information contained in individual sectors.

\section{A stationary reference that retains continuous response}
\label{sec:continuous}
\subsection{Translating an observation window}
Let $S$ be a finite-dimensional system and $E$ an environment, with
autonomous evolution $\mathcal U_t(X)=e^{-iHt}Xe^{iHt}$ and $\hbar=1$.
The preparation evolves freely for a waiting time $\tau$. Operations
then act on $S$ inside a window $[0,T]$. For bounded system maps
$\mathcal A_1,\ldots,\mathcal A_k$ at ordered relative times,
the scalar process evaluates
\begin{equation}
 \begin{split}
 p_{\tau,T}(\mathbf A;\mathbf t)
 =\Tr\big[\mathcal A_k\mathcal U_{t_k-t_{k-1}}\cdots
 \mathcal A_1\mathcal U_{t_1}(\rho_\tau)\big],\\
 \rho_\tau=\mathcal U_\tau(\rho).
 \end{split}
 \label{eq:window-process}
\end{equation}
Each operation is extended by the identity on $E$.
All insertion bounds below use the completely bounded trace norm, so
$\|J_\nu\otimes\operatorname{id}_E\|_{1\to1}\le R$ uniformly in the
environment dimension.
Replacing $\rho_\tau$ by a stationary $\omega$, $[H,\omega]=0$,
defines $\Omega_T$. The reference keeps every propagator inside the
window. It is stationary under a common time translation and can still
have memory and oscillating finite-delay correlations.
Equilibration concerns this stationary temporal response. Thermalization
additionally identifies its local reference with a Gibbs state at the
conserved energy density. The process comparison itself applies to both
thermal and other stationary references.

A simple example fixes this physical meaning. For a single qubit with
$H=\nu Z/2$ and $\omega=I/2$, two $X$ measurements separated by $s$ give
\begin{equation}
 p_*(x,y;s)=\frac14[1+xy\cos(\nu s)].
 \label{eq:stationary-correlation}
\end{equation}
Equilibration preserves this stationary correlation. An additional
independent average over $s$ removes it and recovers the energy-dephased
interval reference. The window formulation therefore extends the
multitime equilibration question while changing the variable that is
averaged relative to Refs.~\cite{FigueroaRomero2020,DowlingRelaxation2023,DowlingEquilibration2023}.

For a finite joint system a natural reference is
$\omega=\mathcal D_H(\rho)$. A comparison of two preparations fixes
one $\omega$. The spectral-filter and local-chain constructions below
also enforce identical energy-dephased states.
The same process construction applies when the complete $SE$ system is
generated by a GKSL semigroup. In that case the Hamiltonian propagator is
replaced by $\Phi_t=e^{t\mathcal L}$ and a stationary state satisfies
$\mathcal L(\omega)=0$. Appendix~\ref{app:gksl} proves the corresponding
sector and full-cPT bounds, and treats fixed-point manifolds, peripheral
oscillations, and dark subspaces. The two-qubit semigroups below are finite
dimensional examples of this extension.
The unitary dilation makes every finite-time marginal completely
positive and causal. Inserting an identity leaves the same process.
The cPT and its common-start-time averages therefore have compatible
marginals; the continuum dilation framework supplies their cMPS
representation~\cite{Wassner2026,Fuksa2026}.

\subsection{Temporal sectors and their physical content}
Choose a fixed Hilbert--Schmidt orthonormal superoperator basis, with
$q=d_S^4-1$ nonidentity species $J_\nu$. The coefficients of the
continuous process generated by an initial operator $X$ are
\begin{equation}
 \begin{split}
 C^{(n)}_{\bm\nu}(\bm t;X)=\Tr\big[
 J_{\nu_n}\mathcal U_{t_n-t_{n-1}}\cdots
 J_{\nu_1}\mathcal U_{t_1}(X)\big],\\
 \bm t\in\Delta_n(T)=\{0<t_1<\cdots<t_n<T\}.
 \end{split}
 \label{eq:coeff}
\end{equation}
Their ordered-time wavefunctions form
\begin{equation}
 \begin{split}
 \mathcal F_T(X)=\bigoplus_{n\ge0}F_{T,n}(X),\\
 \Gamma_T=\bigoplus_{n\ge0}
 L^2(\Delta_n(T);\mathbb C^{q^n}).
 \end{split}
 \label{eq:fock-space}
\end{equation}
Time is continuous before any numerical contraction. The cMPS
self-overlap sums squared coefficients, and its number-resolved terms
give the Fock sectors (Appendix~\ref{app:transfer}). For qubits we fix
$J_{\mu\nu}(X)=\sigma_\mu X\sigma_\nu/2$, omitting $(\mu,\nu)=(0,0)$.
Thus $q=15$, $R=\max_\nu\|J_\nu\|_{1\to1}=1/2$.

Write the process deviation and its distances as
\begin{align}
 \Delta\Upsilon_T^\rho(\tau)&=\mathcal F_T(\rho_\tau-\omega),
 \label{eq:delta}\\
 e_n^\rho(\tau)&=\|P_n\Delta\Upsilon_T^\rho(\tau)\|,\qquad
 E^\rho(\tau)=\|\Delta\Upsilon_T^\rho(\tau)\|.
 \label{eq:distancehierarchy}
\end{align}
The scalar zero sector is a trace and vanishes for a state difference.
For a qubit, the first sector is the integrated freely evolving state
distance:
\begin{equation}
 \begin{aligned}
 e_1^\rho(\tau)^2&=4\int_0^T d_\rho(\tau+t)^2\,\dd t,\\
 d_\rho(t)&=\tfrac12\|\Tr_E(\rho_t-\omega)\|_1.
 \end{aligned}
 \label{eq:selective-window}
\end{equation}
Higher sectors resolve the response after earlier insertions. They can
detect slowly relaxing correlations invisible to $d_\rho$.
These sectors are coordinates of a continuous response, not probabilities
of classical outcomes. A deterministic unitary control has one control
branch, yet contracts a coherent combination of many Fock sectors.

Mean-square equilibration controls the waiting-time average
$L^{-1}\int_0^L E^\rho(\tau)^2\,\dd\tau$; pointwise equilibration means
$E^\rho(\tau)\to0$. A small mean-square deviation implies proximity to the
stationary response for most window starts
\cite{Reimann2008,Linden2009,ShortFarrelly2012,GogolinEisert2016}.
Table~\ref{tab:distances} summarizes the response and readout distances.
Retaining the terminal system defines the output hierarchy, while a fixed
control and POVM contract that hierarchy into measured probabilities
(Sec.~\ref{sec:operational}).

\begin{table}[t]
\caption{Distances for continuous multitime equilibration and readout.}
\label{tab:distances}
\begin{ruledtabular}
\renewcommand{\arraystretch}{1.12}
\begin{tabular}{@{}lp{0.52\columnwidth}@{}}
Quantity & \raggedright Physical content\tabularnewline
$e_n$, $E$ & \raggedright Scalar response at fixed order and across all orders\tabularnewline
$\|\Delta F_T^{\rm out}\|$ & \raggedright Response hierarchy with terminal system retained\tabularnewline
$\TV(p_A^{G,M},p_*^{G,M})$ & \raggedright Output of one fixed control and POVM\tabularnewline
\end{tabular}
\end{ruledtabular}
\end{table}

\input{equilibration_main.tex}

\section{Multitime Mpemba effects from preparation filtering}
\label{sec:mpemba}
\subsection{Comparing preparations relative to one process}
The equilibration results provide a common object toward which different
preparations can relax. Fix $H,T,\{J_\nu\}$ and $\Omega_T$. A finite-time
process Mpemba crossing occurs when
\begin{equation}
 E_A(0)>E_B(0),\qquad E_A(\tau_*)<E_B(\tau_*),
 \label{eq:mpemba-crossing}
\end{equation}
with an analogous definition in a chosen sector. Long-time acceleration
compares rates in addition to this initial ordering. When
$e_{j,n}=\Theta(\tau^{-\alpha_{j,n}})$, an exponent advantage means
$\alpha_{A,n}>\alpha_{B,n}$. Exponential rates can be compared in the
same way when exponential asymptotics apply.

We use \emph{all-sector exponent improvement} when this advantage holds
in every fixed active sector and in the full norm. An active sector has
a nonzero asymptotic deviation for the pair under comparison.
\emph{Sector-selective acceleration} means that the advantage is confined
to a specified subset. These definitions classify temporal-order
coverage. Conventional strong Mpemba effects classify removal of a
slowest discrete relaxation mode~\cite{LuRaz2017,Klich2019,Carollo2021}.
In a continuous spectrum, an edge or low-frequency tail replaces that
mode, and an exponent improvement measures how the preparation changes
its visible weight.

Finite-time contraction has a separate definition,
\begin{equation}
 \kappa_{j,n}^{[a,b]}=
 \frac{\log e_{j,n}(a)-\log e_{j,n}(b)}{b-a}.
 \label{eq:intervalrate}
\end{equation}
It characterizes a specified interval, including finite-size crossings.
Equation~\eqref{eq:full-tail} controls the response omitted by a
finite-order comparison.

\subsection{One preparation filter acts throughout the hierarchy}
For the local preparation operators
$Q_m=(-i\operatorname{ad}_H)^mO$, spectral calculus gives the exact identities
\begin{equation}
 f_{Q_m}=\partial_\tau^m f_O,\qquad
 \mu_{T,Q_m}(\dd\omega)=(-i\omega)^m\mu_{T,O}(\dd\omega).
 \label{eq:filter}
\end{equation}
They hold before sector projection. A polynomial filter can therefore
remove a common slow spectral contribution in every order. A filter
that removes only the projections visible to selected sectors changes
their relative relaxation. Positivity is imposed at preparation:
bounded traceless Hermitian $Q$ defines a positive perturbation of the
tracial reference at sufficiently small amplitude.

A rate follows when the visible Fourier tail and its derivatives are
controlled. For example, suppose in the full Fock norm
\begin{equation}
 \begin{split}
 \partial_\tau^j f_O(\tau)
 =c(-1)^j(\alpha)_j\tau^{-\alpha-j}W_T
 +o(\tau^{-\alpha-j}),\\
 j=0,1,2,\qquad \alpha>0,\quad c\ne0,\quad W_T\ne0.
 \end{split}
 \label{eq:tailhyp}
\end{equation}
Then the $Q_1$ and $Q_2$ preparations have exponents $\alpha+1$ and
$\alpha+2$ in the full norm and every fixed sector with $P_nW_T\ne0$.
Appendix~\ref{app:preps} constructs positive local preparations in the
spin chain with the same dephased reference and a rigorous initial
distance ordering. The following model supplies an
explicit visible spectrum, a summable rate bound, and matching lower
bounds.

\input{semicircle_main.tex}

\begin{figure*}[!t]
\includegraphics[width=\textwidth]{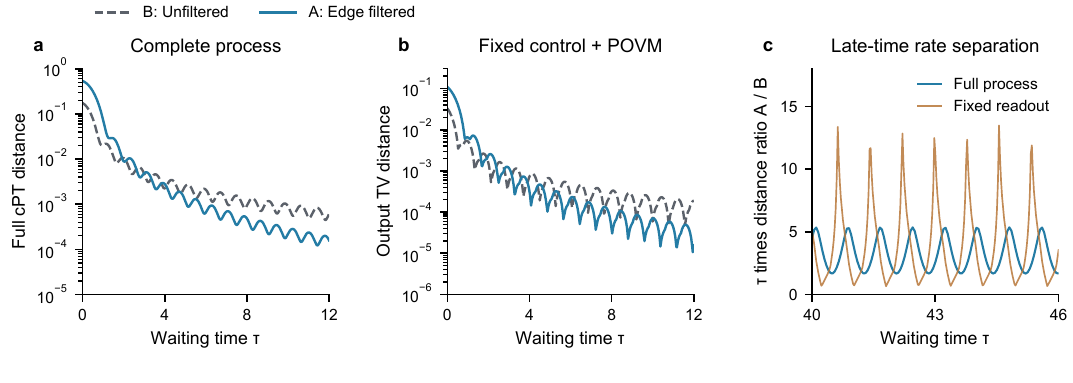}
\caption{One spectral filter accelerates both a complete continuous process
and a fixed measurement output. The preparations are
$\epsilon_A=0.8,g_A=1-b^2/4$ and $\epsilon_B=0.2,g_B=1$, with
$T=0.4$. (a) Full cPT distances from the exact all-order cMPS self-contraction.
(b) Actual output TV distances for constant $0.5Y_S$ control and the fixed
POVM in Eq.~\eqref{eq:fixedpovm}. (c) The late-time ratio multiplied by
$\tau$, for the full process and the same measured output.
Preparation $A$ starts farther in both distances and eventually becomes
closer. Theorems~\ref{thm:edge} and \ref{thm:operational} establish the
respective $3/2$ and $5/2$ exponents independently of the displayed interval.}
\label{fig:edgeoperational}
\end{figure*}

\section{From process equilibration to measured probabilities}
\label{sec:operational}
\subsection{The scalar, output, and fixed-readout descriptions}
An experiment accesses a contraction of the process. Retaining the
system at the end of the window gives operator-valued coefficients
\begin{equation}
 \begin{split}
 F^{\rm out}_{T,n}(X)=\Tr_E\big[
 \mathcal U_{T-t_n}J_{\nu_n}\cdots
 J_{\nu_1}\mathcal U_{t_1}(X)\big].
 \end{split}
\end{equation}
Their output Hilbert--Schmidt norms, integrated over times and summed
over species and order, define $\|\mathcal F_T^{\rm out}(X)\|$.
Its zero sector is the freely evolved state.

For qubit Pauli species, summation over the final scalar insertion
gives an exact norm comparison. Separate the output hierarchy into its
zero-insertion and positive-insertion parts,
\begin{equation}
\begin{aligned}
 \mathcal B_{T,0}(X)^2&=\int_0^T
 \|F^{\rm out}_{t,0}(X)\|_2^2\,\dd t,\\
 \mathcal B_{T,+}(X)^2&=\int_0^T\dd t\sum_{n\ge1}
 \int_{\Delta_n(t)}\|F^{\rm out}_{t,n}(X)\|_2^2\,\dd\mathbf t,\\
 \mathcal B_T(X)^2&=\mathcal B_{T,0}(X)^2+\mathcal B_{T,+}(X)^2.
\end{aligned}
\label{eq:main-output-split}
\end{equation}
Removing the last scalar insertion leaves one fewer insertion in the
retained output. The positive scalar hierarchy therefore corresponds to
the complete output hierarchy, including its zero-insertion part:
\begin{equation}
 \tfrac32\mathcal B_T(X)^2
 \le\|\mathcal F_T^+(X)\|^2
 \le2\mathcal B_T(X)^2.
 \label{eq:mainnormbridge}
\end{equation}
The scalar norm has the separate decomposition
\begin{equation}
 \|\mathcal F_T(X)\|^2=|\Tr X|^2+\|\mathcal F_T^+(X)\|^2.
\label{eq:main-scalar-split}
\end{equation}
Thus a traceless state difference satisfies
$\mathcal F_T^+(X)=\mathcal F_T(X)$, while its retained output may still
have a nonzero zero-insertion component $\mathcal B_{T,0}(X)$.
This component supplies the scalar first sector and is included in
Eq.~\eqref{eq:mainnormbridge}. The bridge compares hierarchies integrated
over termination time. A fixed
termination time is treated by its own output contraction.

\begin{proposition}[Continuous control and output TV]
\label{prop:tvbound}
Expand a bounded control generator as
$-i[h_S(t),\cdot]=a_0(t)I+\sum_\nu a_\nu(t)J_\nu$.
Its coherent instrument cMPS has squared norm
\begin{equation}
 K_G^2=\exp\!\left[\int_0^T
 \left(2\Re a_0(t)+\sum_\nu|a_\nu(t)|^2\right)\dd t\right].
\end{equation}
For any terminal POVM $M$,
\begin{equation}
 \TV(p_\rho^{G,M},p_\omega^{G,M})
 \le\tfrac12\sqrt{d_S}\,K_G
 \|\Delta F_T^{\rm out}\|.
 \label{eq:tv}
\end{equation}
\end{proposition}

The coherent cMPS contraction gives the controlled state.
Cauchy--Schwarz and measurement data processing give
Eq.~\eqref{eq:tv}. These contractions implement the operational
continuum description~\cite{Wassner2026}; controlled process
characterization also probes quantum memory experimentally
\cite{White2020,Pollock2018Markov}. Appendix~\ref{app:readout} proves
the bridge and gives jump and diffusive record contractions
\cite{Gough2009,WisemanMilburn2010,JacobsSteck2006,Dalibard1992}.

The finite-time equilibration theorem also applies directly to the
output hierarchy. Expanding its system output in matrix units, whose
operator norms are one, gives
\begin{equation}
 \frac1L\int_0^L\|\Delta F_T^{\rm out}(\tau)\|^2\,\dd\tau
 \le d_S^2\frac{g_L}{d_{\rm eff}}e^x.
 \label{eq:output-ms}
\end{equation}
Together with Eq.~\eqref{eq:tv}, this bounds the mean-square TV
distance for each bounded deterministic control at a fixed termination
time. The factor $e^x$ includes the unperturbed output zero sector.

For an output hierarchy $r(\tau)W(\tau)+o(r(\tau))$, a fixed control and
POVM inherit the two-sided rate when their contraction of the complete
leading vector $W(\tau)$ stays bounded away from zero. The next theorem
establishes this visibility condition for the spectral-edge filter.

\begin{figure*}[!t]
\includegraphics[width=\textwidth]{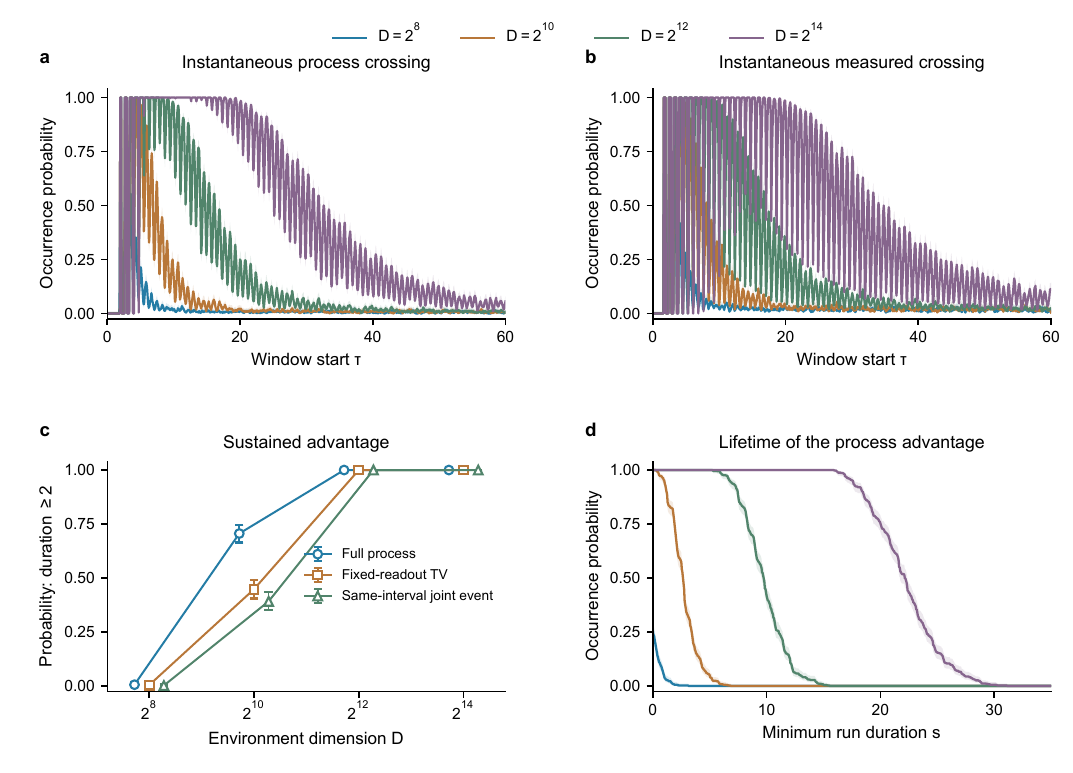}
\caption{Occurrence and duration of multitime acceleration for one fixed
preparation pair across random environments. There are $500$ independent
GUE realizations at each dimension, with the preparations of
Theorem~\ref{thm:edge}, $T=0.4$, and the fixed readout of
Theorem~\ref{thm:operational}. (a,b) Instantaneous crossing probabilities
for the complete cPT norm and measured TV distance, respectively.
(c) Probabilities that the root-mean-square advantage in
Eq.~\eqref{eq:occurrence-envelope} lasts at least two waiting-time units.
The joint event requires a common interval for the process and TV
advantages. (d) Probability of a root-mean-square full-process crossed
run lasting at least $s$. Shading and error bars show pointwise $95\%$ Wilson intervals.
Crossings and durations use a waiting-time spacing of $0.05$ on $[0,60]$;
the complete cPT norm comes from the independent $z=1$ transfer.}
\label{fig:mpembaprobability}
\end{figure*}

\input{semicircle_control_main.tex}

\begin{figure*}[!t]
\includegraphics[width=\textwidth]{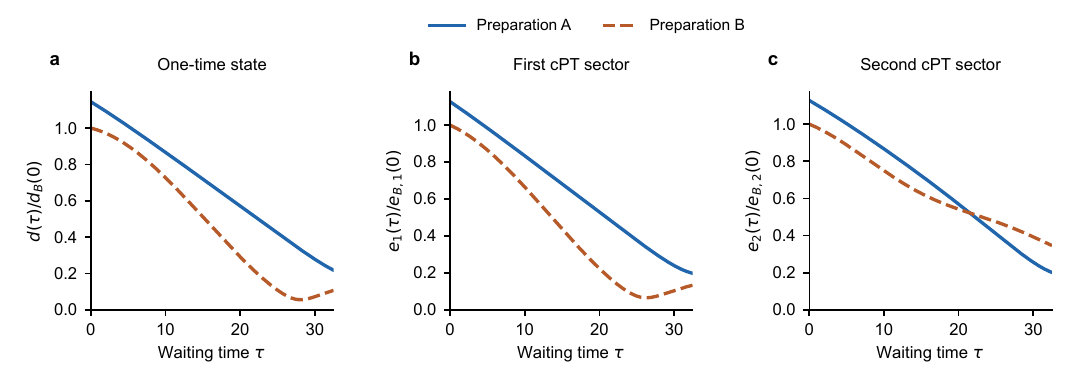}
\caption{A preparation can accelerate the response to interventions
without accelerating its freely evolving state. In the fixed
noncommuting bath, $A$ starts farther from the common stationary
reference. The state and first-sector distances retain their
ordering, while the second-sector distances cross. Each panel uses
preparation $B$'s initial value for normalization. The sector
distances come from number-resolved doubled cMPS contractions;
the rate comparison concerns $[0,32.5]$.}
\label{fig:randomselective}
\end{figure*}

\input{mpemba_probability_main.tex}

\section{Mechanisms beyond the solvable spectral model}
\label{sec:mechanisms}
Noncommuting environments can expose a preparation advantage only
after an intervention. In the qubit--bath model of
Appendix~\ref{app:supporting} (with its detailed construction in
Appendix~\ref{app:random}), take $D=48$, $K=12$ and $T=4$.
The two positive preparations have the same stationary shell state.
Preparation $A$ starts farther in the state distance and both
computed sectors. The state and first-sector orderings persist on
$[0,32.5]$, while the second-sector distances change from
$(0.08852,0.07848)$ to $(0.01582,0.02717)$ for $(A,B)$.
The second-sector contractions are $\kappa_A=0.05299$ and
$\kappa_B=0.03264$. In contrast, $B$ contracts faster in the state
and first sector. Figure~\ref{fig:randomselective} shows this
finite-interval sector selectivity.

Noncommuting couplings thus make the preparation advantage depend on
temporal order, complementing state-based measures of
quantum memory~\cite{Breuer2009,Rivas2014,Breuer2016}.

The same filtering picture survives in a local many-body environment. In
the mixed-field Ising chain, local commutator preparations share the same
energy-dephased stationary reference. Their commutator structure filters
the process-visible low-frequency response, and the first four cPT sectors
show the same delayed-versus-rapid
ordering across bath sizes $N=8,10,12,14,16$ (Appendix~\ref{app:supporting},
Fig.~\ref{fig:ethfour}). This is the ETH realization of the mechanism: the
many-body spectrum supplies a dense set of matrix-element frequencies, while
the preparation determines which part of that spectrum the process sees.

Equal-population shell preparations also give a full-norm crossing at
$N=8$ (Appendix~\ref{app:supporting}). An exact two-qubit memory model separates state and
process relaxation by a correlation visible to system operations.
A constrained detailed-balance model then shows how an entangled
preparation removes a slow process component retained by every
allowed separable preparation. Joint GOE/GUE/GSE examples test
realization and symmetry dependence of finite-time sector coverage.
Four-sector statistics quantify this coverage for the same
laboratory preparation rule (Fig.~\ref{fig:dysonprobability}). A fixed
control and binary system measurement also resolve a TV crossing in one
noncommuting GSE realization (Appendix~\ref{app:gseoperational}).

\input{eth_operational_main.tex}

\section{Discussion}
Continuous multitime equilibration means that the response during
an observation window approaches a stationary response, including
its finite-delay correlations. The full-process bound
\eqref{eq:window-ms} places this statement alongside the effective-
dimension picture of state equilibration. It extends finite-order
multitime estimates to a summable continuous hierarchy. The
stationary reference and the average both use one shared window
start, preserving autonomous dynamics within the experiment.

This construction distinguishes the source of equilibration from
the source of a preparation advantage. Phase cancellation in a
visible continuum spectrum makes the response decay. Preparation
filtering changes the weight of the components responsible for
the slow decay. The semicircle example solves both steps and
connects them to measured TV asymptotics. Its edge zero acts on
every temporal order before their norm is summed. Noncommuting
couplings can instead distribute preparation information
differently across orders, leading to selective acceleration.

Locality extends this construction to infinite many-body environments
\cite{LiebRobinson1972,Bravyi2006,Nachtergaele2006}. In the spin-chain
realization, ETH relates the visible spectrum to smooth matrix-element
envelopes~\cite{Deutsch1991,Srednicki1994,Rigol2008Thermal,Dalessio2016}.
The preparation filter acts on these spectral weights through an exact
commutator identity, linking local preparation to the response hierarchy.

The continuous process framework thus connects equilibration,
preparation filtering and operational readout within one
autonomous description. It identifies when faster state
relaxation extends to a complete temporal response, and when
interventions reveal a different preparation ordering. This
provides a process-level criterion for deciding which part of
quantum relaxation has accelerated.

\appendix
\input{derivations.tex}
\input{random_matrix_appendix.tex}

\input{semicircle_appendix.tex}

\input{mpemba_probability_appendix.tex}
\input{dyson_sector_theory.tex}

\section{Supporting physical realizations}
\label{app:supporting}
\input{auxiliary_results.tex}
\clearpage
\makeatletter
\immediate\write\@auxout{\string\citation{apsrev42Control}}
\makeatother
\bibliographystyle{apsrev4-2}
\bibliography{references}
\end{document}

%% file: equilibration_main.tex
\section{Multitime equilibration of the continuous process}
\label{sec:equilibration}
\subsection{Finite-time mean-square equilibration}
Consider first a finite joint Hilbert space. Write
$H=\sum_a E_aP_a$, where the $E_a$ are distinct and the projectors may be
degenerate. For a preparation $\rho$, its energy-dephased state is
$\omega=\sum_aP_a\rho P_a$. Define
\begin{equation}
 p_a=\Tr(P_a\rho),\qquad
 d_{\rm eff}=\left(\sum_a p_a^2\right)^{-1}.
 \label{eq:effective-dimension}
\end{equation}
The effective dimension counts the energy subspaces populated by the
preparation. The other spectral input is the finite-time overlap of Bohr
frequencies. For $G_{ab}=E_a-E_b$, $a\ne b$, let
\begin{equation}
 M_{ab,cd}^{(L)}=\frac1L\int_0^L
 e^{i(G_{ab}-G_{cd})\tau}\,\dd\tau,\qquad
 g_L=\|M^{(L)}\|_{\rm op}.
 \label{eq:gap-Gram}
\end{equation}
Here $L$ is the range of window starting times; $T$ is the duration of
each window. These two times have different physical roles.

\begin{theorem}[Continuous-window multitime equilibration]
\label{thm:window}
Fix the system operation basis, $T<\infty$, and
$\|J_\nu\|_\diamond\le R$. Set $x=qR^2T$. The stationary reference
$\Omega_T=\mathcal F_T(\omega)$ and every averaged window process are
legal cPTs generated by the same autonomous dynamics. In each sector,
\begin{equation}
 \frac1L\int_0^L e_n(\tau)^2\,\dd\tau
 \le\frac{g_L}{d_{\rm eff}}\frac{x^n}{n!},\qquad n\ge1.
 \label{eq:sector-ms}
\end{equation}
The complete hierarchy satisfies
\begin{equation}
 \frac1L\int_0^L E(\tau)^2\,\dd\tau
 \le\frac{g_L}{d_{\rm eff}}(e^x-1).
 \label{eq:window-ms}
\end{equation}
Moreover, the mean process itself converges in the full norm:
\begin{equation}
 \left\|\frac1L\int_0^L\Upsilon_{\tau,T}\,\dd\tau-\Omega_T\right\|
 \longrightarrow0\quad(L\to\infty).
 \label{eq:mean-process}
\end{equation}
\end{theorem}

Physically, the theorem says that one shared waiting-time average controls
the whole response window. The factor $x^n/n!$ is the volume of the ordered
time simplex, so higher insertion orders become progressively less important
when the window and the intervention strength are fixed.

The proof makes the passage from multitime statistics to a continuous
process explicit. Pull a coefficient back to the initial state to obtain
$C^{(n)}_{\boldsymbol\nu}(\mathbf t;X)=\Tr[A_{\boldsymbol\nu}(\mathbf t)X]$,
where $\|A_{\boldsymbol\nu}(\mathbf t)\|\le R^n$.
The energy-gap Gram argument of Ref.~\cite{ShortFarrelly2012} gives
\begin{equation}
 \frac1L\int_0^L|\Tr[A(\rho_\tau-\omega)]|^2\,\dd\tau
 \le\frac{g_L\|A\|^2}{d_{\rm eff}}.
 \label{eq:observable-ms}
\end{equation}
There are $q^n$ species words, and their ordered time simplex has volume
$T^n/n!$. Integration gives Eq.~\eqref{eq:sector-ms}; nonnegative
summation gives Eq.~\eqref{eq:window-ms}. Thus infinitely many temporal
orders are controlled without choosing a time grid. Averaging the common
starting time also gives
$L^{-1}\int\Upsilon_{\tau,T}\dd\tau=\mathcal F_T(\bar\rho_L)$,
where $\bar\rho_L\to\omega$. This proves Eq.~\eqref{eq:mean-process}.
Appendix~\ref{app:windowproof} supplies the mixed-state proof and the
legality argument.

Equation~\eqref{eq:window-ms} expresses typical equilibration of a
continuous response. The fraction of window starts with $E(\tau)>\eta$
is at most $g_L(e^x-1)/(d_{\rm eff}\eta^2)$. For large $L$, $g_L$
approaches the largest nonzero-gap degeneracy $g$. At finite $L$,
Ref.~\cite{ShortFarrelly2012} gives
\begin{equation}
 g_L\le N(\varepsilon)
 \left(1+\frac{8\log_2 m}{\varepsilon L}\right),
 \label{eq:finite-gap-bound}
\end{equation}
where $m$ counts distinct energies and $N(\varepsilon)$ counts the
largest number of ordered nonzero gaps in an interval of width
$\varepsilon$. Broad energy occupation and weak gap crowding therefore
suppress the full multitime deviation. At fixed finite dimension, the
bound controls fluctuations across window starts.
For a many-body family with fixed $S,T,R$, write
$g_N:=\lim_{L\to\infty}g_{L,N}$, namely the largest multiplicity of a
nonzero Bohr gap (with $g_N=0$ when there are no nonzero gaps). The condition
$g_N/d_{{\rm eff},N}\to0$ therefore implies
\begin{equation}
 \lim_{N\to\infty}\limsup_{L\to\infty}
 \frac1L\int_0^L E_N(\tau)^2\,\dd\tau=0.
 \label{eq:thermodynamic-ms}
\end{equation}
Energy occupation and gap crowding thus control the thermodynamic
mean-square deviation.

\subsection{From fixed sectors to the full process}
The factorial weight also controls the hierarchy away from a time average.
For any trace-zero initial difference $X$, trace-norm preservation by free
evolution gives
\begin{align}
 \|F_{T,n}(X)\|^2&\le\frac{x^n}{n!}\|X\|_1^2,
 \label{eq:factorial}\\
 \|\mathcal F_T(X)\|&\le\sqrt{e^x-1}\,\|X\|_1.
 \label{eq:fullbound}
\end{align}
For two normalized preparations, the omitted squared norm above order $K$
is bounded uniformly in the waiting time and environment dimension by
\begin{equation}
 \mathcal R_K(T)=4\sum_{n>K}\frac{x^n}{n!}.
 \label{eq:full-tail}
\end{equation}
This is a bound on the response left unresolved by a finite-order
description. Its physical parameters are the window length and the fixed
normalization of the operation basis.

\begin{theorem}[Lifting sector equilibration]
\label{thm:lift}
Let a family of process differences have a common bound
$\|X_\alpha\|_1\le C$ and fixed $T,q,R$. If every fixed sector converges
to zero, then the full cPT norm converges to zero. The statement holds
both pointwise and in mean square. More generally, a rate $r(\tau)$ lifts
to the full norm if
\begin{equation}
 e_n(\tau)\le b_n r(\tau),\qquad \sum_n b_n^2<\infty.
 \label{eq:rate-lift}
\end{equation}
\end{theorem}

The factorial tail has a simple interpretation. A finite set of temporal
orders captures the process up to a controlled error, uniformly in the
waiting time. Once those sectors have equilibrated, the remaining
high-order response is too small to change the full norm.

Choose a sector cutoff so that its factorial tail is below a prescribed
error, and then take the limit in the remaining finite sum. The same
argument applies after a nonnegative time average. Under
Eq.~\eqref{eq:rate-lift}, summing the squared envelopes gives the full rate.
Theorem~\ref{thm:edge} below gives
an explicit envelope and matching lower bounds for an autonomous model.

\input{equilibration_numerics_main.tex}

\subsection{Pointwise equilibration and visible spectra}
For an infinite environment, the observed response can decay pointwise
when its discrete frequency lines become a continuous visible spectrum.
At the tracial stationary state $\omega_0$, complete the local observable
algebra in $\langle A,B\rangle_0=\omega_0(A^\dagger B)$ and denote the
result by $\mathcal H_0$. Autonomous evolution is a unitary group
$V_\tau=e^{-iK\tau}$ on this space. With the constant scalar sector
removed, $\mathcal F_T:\mathcal H_0\to\Gamma_T^+$ is Hilbert--Schmidt.

\begin{theorem}[Full-process visible spectral criterion]
\label{thm:main}
For a local traceless Hermitian preparation operator $O$, let
\begin{equation}
 \mu_{T,O}(B)=\mathcal F_T E_K(B)O,\qquad
 f_O(\tau)=\mathcal F_TV_\tau O,
 \label{eq:visiblemeasure}
\end{equation}
where $E_K$ is the spectral resolution of $K$. Then
$\|\mathcal F_T\|_{\rm HS}^2\le e^x-1$, the vector measure $\mu_{T,O}$
has finite variation, and
\begin{align}
 f_O(\tau)&=\int e^{-i\omega\tau}\mu_{T,O}(\dd\omega),\\
 \lim_{L\to\infty}\frac1L\int_0^L\|f_O(\tau)\|^2\,\dd\tau
 &=\sum_\omega\|\mu_{T,O}(\{\omega\})\|^2.
 \label{eq:wiener}
\end{align}
Consequently, no visible atoms is equivalent to mean-square
equilibration. A Bochner-integrable density of $\mu_{T,O}$ is sufficient
for pointwise equilibration. These statements hold in the full Fock space
and after any sector projection.
\end{theorem}

In this language, $\mu_{T,O}$ is the spectrum actually seen by the process,
including the later interventions. Its atoms leave persistent oscillations,
whereas a smooth low-frequency distribution dephases; its edge then sets the
long-time power law. The preparation and the later response enter the same
measure, so two states can populate the same energy levels while assigning
different visible weights to them.
Spectral edges and low-frequency singularities determine decay rates
when their Fourier asymptotics are controlled. Appendix~\ref{app:spectral}
derives the vector form of Wiener's identity~\cite{Wiener1930}.
Locality and the factorial tail construct the local cPT thermodynamic
limit for finite-range many-body Hamiltonians (Appendix~\ref{app:local}).

%% file: equilibration_numerics_main.tex
\subsection{Typical equilibration in a finite dephasing bath}
\label{sec:equilibration-example}
A finite dephasing bath makes the two averaging operations physically
distinct. Fix $H=Z_S\otimes B_D$, with nonzero eigenvalues $b_j$ of
$B_D$, and populate $M$ bath levels equally:
\begin{equation}
 \begin{aligned}
 \eta_M&=\frac1M\sum_{j=1}^M|b_j\rangle\langle b_j|,\\
 \rho_M&=\tfrac12(I+\epsilon X_S)\otimes\eta_M,\qquad
 \omega_M=\tfrac12 I\otimes\eta_M.
 \end{aligned}
 \label{eq:finitebath-preparation}
\end{equation}
Each preparation has $d_{\rm eff}=2M$ and stationary reference
$\mathcal F_T(\omega_M)$. We first populate all bath levels, $M=D$,
and increase the bath dimension while keeping the GUE spectral
normalization, the window $T=0.8$, and $\epsilon=0.8$ fixed.
Figure~\ref{fig:ensemble-equilibration} compares
$D=2^8,2^{10},2^{12},2^{14}$, with $500$ independent fixed Hamiltonians
at each dimension. Each realization retains the same Hamiltonian
throughout the waiting time and observation window. Its stationary
reference retains that same Hamiltonian. We contract each process
separately, then summarize the resulting distances by their
median and central $90\%$ range across realizations.

\input{ensemble_equilibration_figure.tex}

Two averages distinguish the mean response from its fluctuations:
\begin{align}
 \mathsf A_L&=\left\|\frac1L\int_0^L
 \Delta\Upsilon_T(\tau)\,\dd\tau\right\|^2,\nonumber\\
 \mathsf B_L&=\frac1L\int_0^L
 \|\Delta\Upsilon_T(\tau)\|^2\,\dd\tau.
 \label{eq:two-process-averages}
\end{align}
The full distance $E_D(\tau)^2$ in
Fig.~\ref{fig:ensemble-equilibration}(a) initially follows the
continuous semicircle response. At finite $D$, the median reaches a
minimum and turns upward as discrete spectral lines become resolved.
Increasing $D$ delays this turnover and lowers the subsequent
fluctuation scale. The spread across realizations resolves the
remaining sample dependence. The mean process approaches its
stationary reference, $\mathsf A_L\to0$, as panel (b) shows.
The averaging in Eq.~\eqref{eq:two-process-averages} is performed in
time within each fixed environment. The broad bands in panel (b)
reflect the sensitivity of finite-range phase cancellation to small
visible frequencies $\pm2b_j$. Panel (c) shows the surviving
mean-square fluctuations. For distinct $b_j$ with
$b_i\ne-b_j$, their exact mean-square plateau is
\begin{equation}
 \mathsf B_\infty=
 \frac{\epsilon^2}{2d_{\rm eff}}
 \left(e^{15T/4}-e^{-T/4}\right).
 \label{eq:finitebath-plateau}
\end{equation}
With $M=D$, the plateau falls as $D^{-1}$. The dotted lines in
panel (c) mark this exact limit for every realization. At $L=8D$,
the ensemble-mean values are $0.9800$, $0.9798$, $0.9798$, and $0.9799$
times their exact plateaus. The spread of these ratios decreases
with $D$. Increasing the occupied dimension dilutes the weight of
each surviving frequency line and suppresses the complete multitime
deviation, as expressed by Theorem~\ref{thm:window}. The largest bath
has $d_{\rm eff}=32768$.

These distances include every temporal order through the cMPS transfer
at $z=1$. Panel (d) resolves
the omitted response $R_K(D,\tau)=E_D(\tau)^2-\sum_{n=1}^K e_{D,n}(\tau)^2$.
For each realization its maximum over the sampled window starts falls
rapidly as $K$ increases. Both this maximum and its variation across
realizations lie beneath the same factorial bound,
Eq.~\eqref{eq:finitebath-tail}. The four dimension curves nearly
coincide: the cutoff required to control the higher-order response
remains independent of the bath dimension. This illustrates the lifting
mechanism of Theorem~\ref{thm:lift}.
For example, retaining sixteen orders bounds the omitted squared norm
by $2.8\times10^{-7}$ at every waiting time and every bath dimension.

Fixed-realization responses in Fig.~\ref{fig:large-bath-equilibration}
of Appendix~\ref{app:finitebath} resolve the oscillations behind the
statistical bands. In those same four Hamiltonians,
Fig.~\ref{fig:effective-dimension-scans} scans every $M=1,\ldots,D$.
Multiplying $\mathsf B_L$ by $d_{\rm eff}=2M$ exposes the common
inverse-dimension plateau and its finite-range interference. Varying
occupation within a fixed spectrum isolates the effect of energy
weight from changes in the environment itself.

%% file: ensemble_equilibration_figure.tex
\begin{figure*}[t]
 \includegraphics[width=\textwidth]{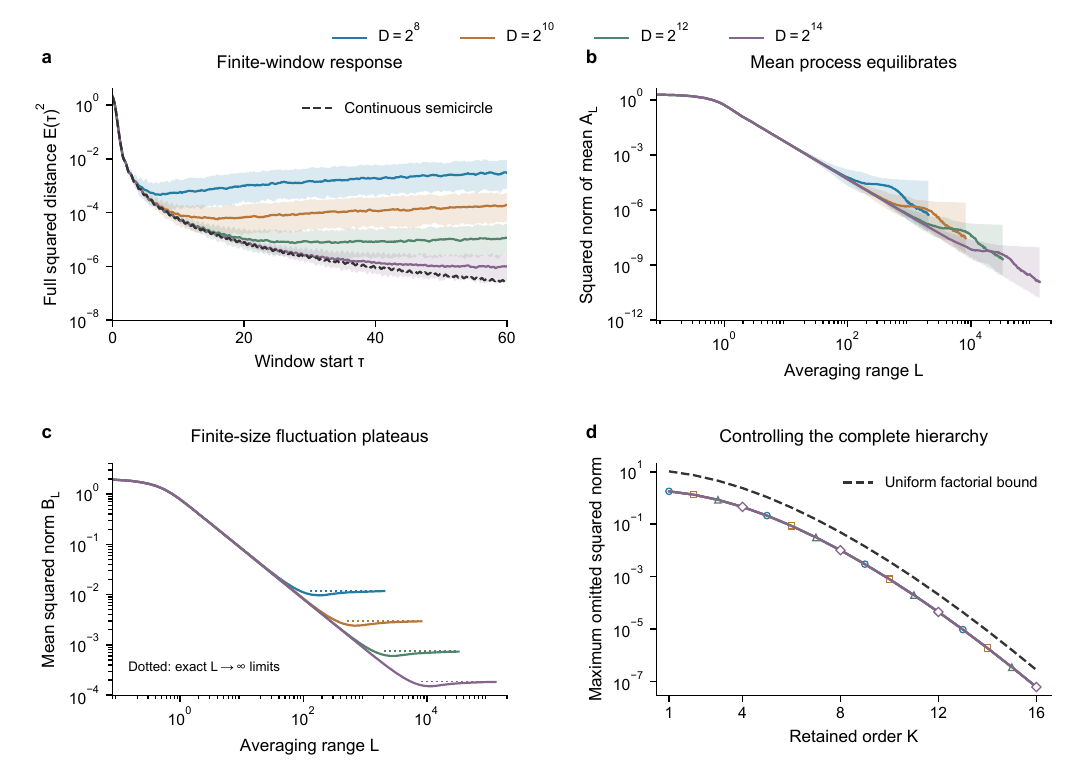}
 \caption{Continuous multitime equilibration over independent GUE environments.
 The model is $H_D=Z_S\otimes B_D$, with full bath occupation,
 $T=0.8$, and $\epsilon=0.8$. Each dimension contains $500$ independent
 fixed Hamiltonians. Solid curves are pointwise medians and shaded
 bands span the $5$th--$95$th percentiles across realizations.
 (a) Full squared cPT distance: the response initially follows the
 continuous semicircle limit (dashed). A finite-size minimum precedes
 the buildup of fluctuations; this turnover moves to later times
 and lower distances as $D$ grows.
 (b) Squared norm of the time-averaged process, $\mathsf A_L$.
 (c) Time average of the squared deviation, $\mathsf B_L$, with its
 exact $D^{-1}$ plateaus (dotted).
 (d) Omitted squared norm above order $K$, first maximized over the
 $1201$ sampled waiting times for each realization. Its median and
 percentile band nearly coincide across dimensions beneath the
 uniform factorial bound (dashed). Staggered markers distinguish
 the overlapping dimension curves.
 Each realization is contracted separately using the all-order
 $z=1$ cMPS transfer; statistical summaries are formed from the
 resulting distances. Panels (b,c) extend to $L=8D$.
 Appendix~\ref{app:finitebath} gives the finite-bath formulas,
 fixed-realization examples, and occupation scans.}
 \label{fig:ensemble-equilibration}
\end{figure*}

%% file: semicircle_main.tex
\section{Semicircle bath: equilibration and all-sector acceleration}
\label{sec:exact}
A spectral filter can accelerate every multitime sector under one autonomous
dynamics and one stationary reference. Consider
\begin{equation}
 \begin{aligned}
 H_D&=Z_S\otimes B_D,\qquad \omega_D=\frac{I}{2D},\\
 \rho_{\ell}&=\frac{I+\epsilon_{\ell}X_S\otimes g_{\ell}(B_D)}{2D},
 \end{aligned}
 \label{eq:wddephasing}
\end{equation}
where $g_0(b)=1$, $g_1(b)=(1-b^2/4)_+$, and $0<\epsilon_{\ell}<1$.
The index $\ell=0,1$ labels the unfiltered and filtered preparations.
Assume that the empirical spectral measure of $B_D$ converges to
\begin{equation}
 d\mu_{\rm sc}(b)=\frac{\sqrt{4-b^2}}{2\pi}
 \mathbf1_{[-2,2]}(b)\,db.
 \label{eq:semicirclemeasure}
\end{equation}
GOE, GUE, and GSE baths with matched second moments supply this limit
\cite{Wigner1958,Dyson1962}. The preparations have the same environment
marginal and the same energy-dephased state, almost surely for these ensembles.
Write $e_{\ell,n}(\tau)$ and $E_{\ell}(\tau)$ for the sector and full scalar
cPT distances in the semicircle limit.

The conserved bath coordinate $b$ marks a waiting-time coherence with
frequency $2b$. Later interventions change its phase path only over the
finite window. A preparation zero at a spectral edge therefore suppresses
the same slowly decaying envelope at every temporal order. This is the
physical origin of the all-sector result.

\begin{theorem}[All-sector exponent improvement]
\label{thm:edge}
The statement is a sequential-limit result: first take the continuum
environment limit $D\to\infty$ on compact waiting-time intervals, and only
then take the long-waiting-time limit $\tau\to\infty$.  For every fixed
$T>0$ and every fixed sector $n\ge1$,
\begin{align}
 e_{\ell,n}(\tau)&=\Theta(\tau^{-\beta_{\ell}}),&
 E_{\ell}(\tau)&=\Theta(\tau^{-\beta_{\ell}}),&
 \beta_{\ell}&=\ell+\tfrac32.
 \label{eq:edgeexponents}
\end{align}
Choose $A=(\ell=1,\epsilon_1=0.8)$ and $B=(\ell=0,\epsilon_0=0.2)$.
For $0<T<1/\sqrt2$,
\begin{align}
 e_{A,n}(0)&\ge(3-4T^2)e_{B,n}(0)>e_{B,n}(0),\nonumber\\
 E_A(0)&\ge(3-4T^2)E_B(0)>E_B(0).
 \label{eq:edgeinitial}
\end{align}
Nevertheless $e_{A,n}/e_{B,n}=O(\tau^{-1})$ for every fixed $n$,
and $E_A/E_B=O(\tau^{-1})$.
The constants and crossover times may depend on $n$.
\end{theorem}

The filter removes weight precisely at the slow spectral edges. Every
finite temporal order samples the same edge envelope, so the extra power of
$\tau^{-1}$ survives when the sectors are summed into the full cPT norm.

The mechanism follows from two exact Fourier integrals:
\begin{equation}
 \phi_0(s)=\frac{J_1(2s)}s,\qquad
 \phi_1(s)=\frac{3J_2(2s)}{2s^2},
 \label{eq:wdphi}
\end{equation}
with continuous values $\phi_0(0)=1$ and $\phi_1(0)=3/4$.
The square-root edges of the unfiltered spectrum give exponent $3/2$.
The extra zero of $g_1$ at each edge raises the exponent to $5/2$.
Every cPT coefficient is a sum of at most two such amplitudes with arguments
$\pm2\tau+s$, $|s|\le2T$. The resulting bound
\begin{equation}
 e_{\ell,n}(\tau)^2\le
 \epsilon_{\ell}^2\frac{(15T/4)^n}{n!}
 \sup_{|s|\le2T}|\phi_{\ell}(2\tau+s)|^2
 \label{eq:edgeupper}
\end{equation}
controls all omitted sectors. A nonzero Pauli word in each sector supplies the matching lower bound after integration over the final insertion time. The integration removes the isolated Bessel zeros from the norm asymptotics. Appendix~\ref{app:edgeproof} proves both bounds and the initial ordering.

The preparation is a spectral-edge filter. On the limiting support, its
preparation operator is
$[1-(\operatorname{ad}_{H})^2/16](X_S\otimes I)$.
The limit has an explicit autonomous dilation with a normal stationary
reference state, given in Appendix~\ref{app:edgeproof}.

%% file: semicircle_control_main.tex
\subsection{The same acceleration in a fixed measurement output}
The exponent improvement is visible with one bounded control and one fixed
terminal measurement. During the window apply
$h_S(t)=hY_S$ and use the six-outcome informationally complete POVM
\begin{equation}
 M_{a,\pm}=\frac{I\pm\sigma_a}{6},\qquad a=x,y,z.
 \label{eq:fixedpovm}
\end{equation}
The uncontrolled dynamics remains $Z_S\otimes B$.
For every waiting time the control and POVM are identical. Let
$D_{\ell}(\tau)=\TV(p_{\ell}(\tau),p_*)$ denote their actual outcome distance from the
common stationary output.

\begin{theorem}[Fixed-control operational Mpemba effect]
\label{thm:operational}
If $h\ne0$ and $\cos(2T\sqrt{4+h^2})\ne1$, then
\begin{equation}
 D_{\ell}(\tau)=\Theta(\tau^{-\beta_{\ell}}),\qquad
 \frac{D_A(\tau)}{D_B(\tau)}=O(\tau^{-1}).
 \label{eq:TVexponents}
\end{equation}
For the preparations of Theorem~\ref{thm:edge}, $T=0.4$ and $h=0.5$,
\begin{equation}
 D_A(0)\ge\frac{0.424}{6}>
 \frac{0.28}{6}\ge D_B(0).
 \label{eq:TVinitial}
\end{equation}
Thus the same model and preparation pair exhibit both all-sector
acceleration and a reversal of their measured output distances.
\end{theorem}

The controlled output Bloch vector has only $x,z$ components and
$D_{\ell}=(|r_{\ell,x}|+|r_{\ell,z}|)/6$. Put
$r_*=\sqrt{4+h^2}$, $c_*=\cos(2r_*T)$ and $s_*=\sin(2r_*T)$.
The two spectral endpoints give
\begin{align}
 \binom{r_{\ell,x}}{r_{\ell,z}}
 ={}&\frac{2\epsilon_{\ell}\Gamma(\beta_{\ell})}
 {\pi(2\tau)^{\beta_{\ell}}}
 [u\cos\theta_{\ell}+v\sin\theta_{\ell}]
 +O(\tau^{-\beta_{\ell}-1}),\nonumber\\
 u={}&\binom{c_*}{-hs_*/r_*},\qquad
 v=\binom{-2s_*/r_*}{2h(1-c_*)/r_*^2},\nonumber\\
 \theta_{\ell}={}&4\tau-\frac{\pi\beta_{\ell}}{2}.
 \label{eq:TVellipse}
\end{align}
Since $\det[u,v]=2h(c_*-1)/r_*^2$, the leading output traces an ellipse
with a strictly positive distance from the origin.
This supplies a lower bound uniform in the oscillation phase.
The control has coherent cMPS coefficients
$a_{y0}=-2ih$, $a_{0y}=2ih$ and norm $K_G=e^{4h^2T}$,
independent of $\tau$.
Appendix~\ref{app:edgecontrol} derives the exact contraction and
Eq.~\eqref{eq:TVinitial}. Figure~\ref{fig:edgeoperational} shows the
full-process and measured-output crossovers.
In panels (a,b), preparation $A$ starts farther and becomes closer in
both distances. Panel (c) removes the leading $1/\tau$ ratio: the
rescaled curves remain of order one with spectral oscillations.
The fixed readout therefore resolves the same one-power exponent
improvement as the complete process norm.

%% file: mpemba_probability_main.tex
\subsection{Occurrence across fixed random environments}
The continuum exponent improvement predicts a preparation advantage before
finite spectral fluctuations dominate. To resolve this competition, we
compare the same filtered and unfiltered preparations in $500$ independent
GUE environments at each $D=2^8,2^{10},2^{12},2^{14}$. Each environment remains
fixed throughout the experiment. We keep $T=0.4$, the control $0.5Y_S$, and
the POVM~\eqref{eq:fixedpovm}. The occurrence probability for a distance $d$
is

\begin{equation}
 P_D^d(\tau)=\Pr_{B_D}\big[d_A(0)>d_B(0),\ d_A(\tau)<d_B(\tau)\big].
 \label{eq:occurrence-probability}
\end{equation}
The initial ordering holds for every sampled pair, both in the full norm
and in the measured TV distance.

Figure~\ref{fig:mpembaprobability}(a,b) shows an interval of frequent
crossings that broadens with $D$. Spectral oscillations modulate this
probability. At later sampled times, finite-size fluctuations make the
opposite ordering more frequent. A brief crossing and a sustained advantage
therefore carry different information. We quantify the latter through the
local root mean square
\begin{equation}
 \overline d_j(\tau)=
 \left[\frac1\Delta\int_\tau^{\tau+\Delta}d_j(s)^2\dd s\right]^{1/2},
 \qquad \Delta=2.
 \label{eq:occurrence-envelope}
\end{equation}
This averages the distance over waiting times while keeping the physical
observation window $T$ fixed.

Requiring $\overline d_A<\overline d_B$ for at least two waiting-time units
gives full-process occurrence rates of $0.6\%$, $70.6\%$, $100\%$, and
$100\%$ in increasing dimension. The corresponding fixed-readout rates
are $0.4\%$, $44.8\%$, $100\%$, and $100\%$. At $D=2^{10}$ the two
advantages persist on a common interval in $39.2\%$ of realizations
[Fig.~\ref{fig:mpembaprobability}(c)]. The median longest full-process
advantage grows from $2.65$ at $D=2^{10}$ to $9.73$ at $D=2^{12}$ and
$22.15$ at $D=2^{14}$ [panel (d)]. These durations describe the finite-size
window in which the spectral filter is effective. Appendix~\ref{app:probability}
gives sector coverage, uncertainty intervals, and the compact-time
continuity result that connects these probabilities to the continuum
theorems.

%% file: eth_operational_main.tex
\subsection{An operational readout in a chaotic many-body environment}
\begin{figure*}[t]
\includegraphics[width=\textwidth]{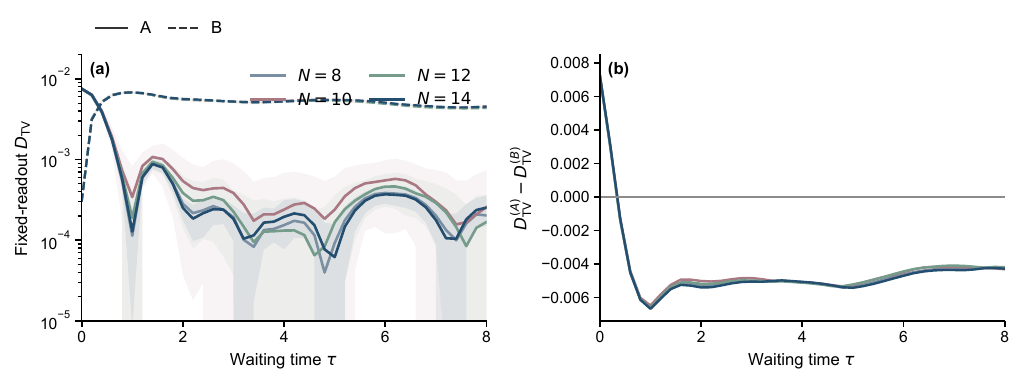}
\caption{Operational process relaxation in the mixed-field Ising chain.
The fixed local control is \(h_{\rm c}=0.3Z_S+0.2X_S\), the observation
window is \(T=0.02\), and the terminal measurement is the Pauli POVM in
Eq.~\eqref{eq:ethoperationalpovm}. Solid and dashed curves denote the two
preparations \(A\) and \(B\); colors denote bath sizes \(N=8,10,12,14\).
(a) TV distances to the common reference under the same readout. Shading
shows two standard errors from random-trace sampling; \(N=8\) uses exact
reduced states. (b) The difference of these distances changes sign and
remains negative over the subsequent sampled interval. Dashed curves appear
only in (a), where they denote preparation \(B\); panel (b) contains one
solid difference curve for each bath size. The waiting
dynamics includes the full chain; the short readout window uses the local
contraction described in Appendix~\ref{app:ethfour}.}
\label{fig:ethoperational}
\end{figure*}

The many-body response can be read out with one fixed control and a
terminal system measurement. We use the same mixed-field Ising Hamiltonian
and commutator preparations as in Eq.~\eqref{eq:localpreps}, apply
\(h_{\rm c}=0.3Z_S+0.2X_S\) for \(T=0.02\), and measure the system with
the six-outcome Pauli POVM
\begin{equation}
 E_{a,\pm}=\frac{I\pm\sigma_a}{6},\qquad a=x,y,z .
 \label{eq:ethoperationalpovm}
\end{equation}
Contracting the process cMPS with the coherent control cMPS and each
terminal effect gives the probabilities. For each preparation \(j\), we
compare them with those of the common stationary reference
\(\omega_N=I/2^{N+1}\), subjected to the same control and POVM:
\begin{equation}
 D_{\rm TV}^{(j)}(\tau)=
 \frac12\sum_{a,\pm}
 \left|p_{j,a,\pm}(\tau)-p_{*,a,\pm}\right| .
 \label{eq:ethoperationaltv}
\end{equation}
Preparation \(A\) starts farther from this reference, with
\(D_{\rm TV}^{(A)}(0)=7.55\times10^{-3}\) and
\(D_{\rm TV}^{(B)}(0)=3.02\times10^{-4}\). Their ordering is reversed
at every sampled waiting time from \(\tau=0.4\) to \(8\), for all four
bath sizes \(N=8,10,12,14\) [Fig.~\ref{fig:ethoperational}].
The initially visible response of \(A\) falls rapidly. In \(B\), hidden
system--bath correlations first develop into a larger system response.
This delayed response makes the process-ordering reversal accessible
through a fixed laboratory readout in a chaotic many-body environment.

%% file: derivations.tex
\section{Fock normalization and exact cMPS contractions}
\label{app:transfer}
We use the ordered simplex $\Delta_n(T)=\{0<t_1<\cdots<t_n<T\}$ with Lebesgue measure. The process coefficient of an initial operator $X$ is
\begin{equation}
 C^{(n)}_{\boldsymbol\nu}(\mathbf t;X)
 =\langle\!\langle I|e^{Q(T-t_n)}J_{\nu_n}e^{Q(t_n-t_{n-1})}
 \cdots J_{\nu_1}e^{Qt_1}|X\rangle\!\rangle,
 \label{eq:appcoefficient}
\end{equation}
where $Q=-i[H,\cdot]$ for the autonomous model. A trace-preserving joint semigroup can replace $e^{Qt}$ in the auxiliary benchmarks. The final propagator disappears against the trace boundary. With $B_X=|X\rangle\!\rangle\langle\!\langle I|$, the cMPS is specified by $(Q,\{J_\nu\},B_X)$. For two cMPS with parameters $(Q_l,R_{l,\nu},B_l)$ and $(Q_r,R_{r,\nu},B_r)$, their overlap is
\begin{align}
 \langle l|r\rangle&=\Tr[(B_l^*\otimes B_r)e^{T\mathbb T_{lr}}],\\
 \mathbb T_{lr}&=Q_l^*\otimes I+I\otimes Q_r
 +\sum_\nu R_{l,\nu}^*\otimes R_{r,\nu}.
 \label{eq:apptransfer}
\end{align}
The star denotes entrywise conjugation; this convention follows column vectorization. Dyson expansion orders all insertion times and sums all species. Therefore
\begin{equation}
 \|F_{T,n}(X)\|^2=[z^n]\Tr[(B_X^*\otimes B_X)e^{T\mathbb T(z)}],
 \label{eq:appsectors}
\end{equation}
where $\mathbb T(z)=Q^*\otimes I+I\otimes Q+z\sum_\nu J_\nu^*\otimes J_\nu$. No additional $n!$ appears in the ordered convention. Setting $z=1$ gives the exact full norm.

For a qubit, the fixed local species are $J_{\mu\nu}(X)=\sigma_\mu X\sigma_\nu/2$, excluding $(\mu,\nu)=(0,0)$. They are Hilbert--Schmidt orthonormal on local Liouville space, with $q=15$ and complete induced trace norm $R=1/2$. The omitted basis element is $I_{\rm Liouville}/2$; it is distinct from a physical absence of an insertion. All comparisons retain this normalization.

Each coefficient in Eq.~\eqref{eq:appcoefficient} obeys $|C^{(n)}|\le R^n\|X\|_1$. Multiplication by $q^n$ species and simplex volume $T^n/n!$ gives
\begin{equation}
 \|F_{T,n}(X)\|^2\le \frac{x^n}{n!}\|X\|_1^2,
 \qquad x=qR^2T.
 \label{eq:appfactorial}
\end{equation}
For traceless $X$, the zeroth scalar coefficient vanishes, and the full bound is $(e^x-1)\|X\|_1^2$. For any cutoff $k$,
\begin{equation}
 \sum_{n>k}\|F_{T,n}(X)\|^2
 \le\|X\|_1^2\sum_{n>k}\frac{x^n}{n!}.
 \label{eq:apptail}
\end{equation}
If $k+2>x$, the remaining series is at most $x^{k+1}/[(k+1)!(1-x/(k+2))]$.

Sector-to-full convergence follows from the same bound. Suppose $\|P_n\Delta_\alpha\|\to0$ for every fixed $n$ and $\|P_n\Delta_\alpha\|^2\le B_n$ with $\sum_nB_n<\infty$. Given $\varepsilon>0$, choose $k$ with $\sum_{n>k}B_n<\varepsilon^2/2$. For this fixed $k$, choose $\alpha$ so that $\sum_{n\le k}\|P_n\Delta_\alpha\|^2<\varepsilon^2/2$. The full norm is below $\varepsilon$. For normalized physical state differences, Eq.~\eqref{eq:appfactorial} supplies $B_n=4x^n/n!$ uniformly in waiting time and environment size.

\input{window_equilibration_proof.tex}

\section{The full-process spectral criterion}
\label{app:spectral}
Let $\omega_0$ be the stationary trace state on the quasi-local algebra, and complete the observables in $\|O\|_{2,0}^2=\omega_0(O^\dagger O)$. The autonomous group $V_\tau O=\alpha_{-\tau}(O)$ is unitary on this Hilbert space. Denote its self-adjoint generator by $K$, so $V_\tau=e^{-iK\tau}$. A local insertion with a finite multiplication expansion $J_\nu(X)=\sum_a A_{\nu a}XB_{\nu a}$ is assumed to satisfy
\begin{equation}
 \sum_a\|A_{\nu a}\|\|B_{\nu a}\|\le R.
 \label{eq:appbudget}
\end{equation}
The Pauli species meet this condition with one term. Moving the free propagators to the Heisenberg picture and using cyclicity gives
\begin{align}
 (\mathcal F_TO)_{n,\boldsymbol\nu}(\mathbf t)
 &=\omega_0(A_{\boldsymbol\nu}(\mathbf t)O),\\
 A_{\boldsymbol\nu}(\mathbf t)
 &=\sum_{\mathbf a}\alpha_{t_1}(B_1)\cdots\alpha_{t_n}(B_n)\nonumber\\
 &\hspace{12mm}\times\alpha_{t_n}(A_n)\cdots\alpha_{t_1}(A_1).
 \label{eq:appkeldysh}
\end{align}
Thus $\|A_{\boldsymbol\nu}\|\le R^n$. We use positive sectors in $\mathcal F_T$; for a traceless perturbation the omitted scalar sector is zero.

For an orthonormal basis $\{e_j\}$ of $\mathcal H_0$, Parseval and Tonelli give
\begin{align}
 \|\mathcal F_T\|_{\rm HS}^2
 &=\sum_{n\ge1}\sum_{\boldsymbol\nu}\int_{\Delta_n(T)}
 \sum_j|\langle A_{\boldsymbol\nu}^\dagger,e_j\rangle_0|^2\dd\mathbf t\nonumber\\
 &=\sum_{n\ge1}\sum_{\boldsymbol\nu}\int_{\Delta_n(T)}
 \|A_{\boldsymbol\nu}\|_{2,0}^2\dd\mathbf t
 \le e^x-1.
 \label{eq:appHS}
\end{align}
For a local $O$, define $\mu(B)=\mathcal F_TE_K(B)O$. If $\{B_j\}$ is a disjoint partition, Cauchy--Schwarz gives
\begin{align}
 \sum_j\|\mu(B_j)\|
 &\le\sum_j\|\mathcal F_TE_K(B_j)\|_{\rm HS}\|E_K(B_j)O\|_{2,0}\nonumber\\
 &\le\|\mathcal F_T\|_{\rm HS}\|O\|_{2,0}.
 \label{eq:appvariation}
\end{align}
Hence the process-valued spectral measure has finite total variation, and its Fourier integral is well defined in $\Gamma_T$.

Write $\mu(\dd\omega)=h(\omega)|\mu|(\dd\omega)$ using its vector Radon--Nikodym derivative. The mean squared norm is the double integral of $\langle h(\omega'),h(\omega)\rangle$ times
\begin{equation}
 K_L(\omega-\omega')=L^{-1}\int_0^L e^{-i(\omega-\omega')\tau}\dd\tau.
\end{equation}
This kernel has modulus at most one and tends pointwise to the indicator of $\omega=\omega'$. Dominated convergence applies to the finite product variation measure. The diagonal mass is precisely
\begin{equation}
 \lim_{L\to\infty}L^{-1}\int_0^L\|f_O(\tau)\|^2\dd\tau
 =\sum_\omega\|\mu(\{\omega\})\|^2.
 \label{eq:appWiener}
\end{equation}
The statement is in the full process space before any projection. If $\mu$ has a Bochner $L^1$ density, the vector Riemann--Lebesgue lemma gives pointwise norm convergence. A second sufficient condition is $V_\tau O\rightharpoonup0$ in $\mathcal H_0$: Eq.~\eqref{eq:appHS} makes $\mathcal F_T$ compact, so it maps this bounded weak convergence to norm convergence.

For a finite-range Hamiltonian, every fixed nested commutator $Q_m=(-i\operatorname{ad}_H)^mO$ is local and lies in the domain of $K^m$. The spectral calculus gives
\begin{equation}
 \mu_{T,Q_m}(\dd\omega)=(-i\omega)^m\mu_{T,O}(\dd\omega),
 \quad f_{Q_m}(\tau)=\partial_\tau^mf_O(\tau).
 \label{eq:appfilter}
\end{equation}
In particular $m\ge1$ eliminates the zero-frequency atom, and
\begin{equation}
 |\mu_{T,Q_m}|([-\omega_c,\omega_c])
 \le\omega_c^m\sqrt{e^x-1}\|O\|_{2,0}.
\end{equation}
This is a spectral variation bound. A pointwise time-domain bound between the two process norms does not follow by replacing the integral with a single frequency.

\section{Joint GKSL dynamics: sector and full-process equilibration}
\label{app:gksl}
The continuous-process construction also applies when the complete system and
environment evolve under a finite-dimensional GKSL generator,
\begin{equation}
 \begin{aligned}
 \mathcal L(X)&=-i[H,X]+\sum_a\left(L_aXL_a^\dagger-
 \frac12\{L_a^\dagger L_a,X\}\right),\\
 \Phi_t&=e^{t\mathcal L}.
 \end{aligned}
 \label{eq:gkslgenerator}
\end{equation}
This is the standard finite-dimensional quantum Markov semigroup form
\cite{Davies1974,AlickiLendi2007}.
Interventions act on the system factor, while the environment remains an
internal memory. The cPT coefficient is obtained from
Eq.~\eqref{eq:appcoefficient} by replacing each $e^{Qt}$ by $\Phi_t$.
The same ordered cMPS contraction then represents the complete response.

\subsection{Relaxing semigroups}
Assume that $\Phi_t$ has a unique stationary state $\omega$ and that its
nonstationary spectrum lies in $\operatorname{Re}z\le-\gamma$ with
$\gamma>0$. If $r$ is the largest Jordan-block size in this part of the
spectrum, finite-dimensional spectral calculus gives constants
$C_{\mathcal L}$ and $0<\gamma_0\le\gamma$ such that, for every Hermitian
traceless $X$,
\begin{equation}
 \|\Phi_\tau(X)\|_1\le C_{\mathcal L}(1+\tau^{r-1})e^{-\gamma_0\tau}\|X\|_1.
 \label{eq:gkslsemigroupdecay}
\end{equation}
The polynomial factor is absent when the decaying part of $\mathcal L$ is
diagonalizable.

Let $X_\tau=\Phi_\tau(\rho-\omega)$ and let $F_{T,n}$ denote the $n$th cPT
sector map. With $q$ insertion species, budget $R$, and $x=qR^2T$, the
factorial estimate gives
\begin{equation}
 \begin{aligned}
 \|F_{T,n}(X_\tau)\|^2
 &\le \frac{x^n}{n!}\|X_\tau\|_1^2\\
 &\le \frac{x^n}{n!}C_{\mathcal L}^2
 (1+\tau^{r-1})^2e^{-2\gamma_0\tau}\|\rho-\omega\|_1^2 .
 \end{aligned}
 \label{eq:gkslsectorbound}
\end{equation}
Summing all positive sectors yields
\begin{equation}
 \begin{aligned}
 \|\Upsilon_{\rho,\tau,T}-\Omega_{\omega,T}\|_{\rm Fock}
 &\le C_{\mathcal L}\sqrt{e^x-1}\,(1+\tau^{r-1})e^{-\gamma_0\tau}\\
 &\quad\times\|\rho-\omega\|_1 .
 \end{aligned}
 \label{eq:gkslfullbound}
\end{equation}
Thus fixed-sector convergence lifts to the full cPT whenever the finite-window
insertion budget is bounded. The pointwise exponential estimate is stronger
than the Hamiltonian time-average statement; integrating the square of
Eq.~\eqref{eq:gkslsectorbound} gives the corresponding mean-square result.

The physical interpretation is direct. A relaxing joint Lindblad evolution
removes the preparation through its visible Liouvillian modes. Bounded
interventions propagate these modes through every temporal order, while the
factorial sector factor controls the infinite hierarchy.

\subsection{Multiple stationary states and conserved sectors}
If the zero eigenspace of $\mathcal L$ is multidimensional, let $\Pi_0$ be
the spectral projection onto the fixed-point space. When the remaining
spectrum has real part at most $-\gamma_0<0$,
\begin{equation}
 \|\Phi_\tau(X)-\Pi_0X\|_1\le
 C_{\mathcal L}(1+\tau^{r-1})e^{-\gamma_0\tau}\|X\|_1
 \label{eq:gkslmultistable}
\end{equation}
for Hermitian $X$. Each preparation has the stationary reference
$\Omega_{\Pi_0\rho,T}$. A common reference for a pair exists when
$\Pi_0\rho_A=\Pi_0\rho_B$; this is the open-system counterpart of fixing
the conserved energy populations. If the projections differ, one must either
compare each preparation with its own stationary process or regard the change
of stationary target as part of the comparison.

\subsection{Peripheral modes, decoherence-free subspaces, and dark states}
If $\mathcal L$ has peripheral eigenvalues $i\Omega_\ell$ with
$\Omega_\ell\ne0$, write
\begin{equation}
 \begin{aligned}
 \Phi_\tau&=\Phi^{\rm per}_\tau+\mathcal R_\tau,\\
 \|\mathcal R_\tau\|_{1\to1}&\le
 C(1+\tau^{r-1})e^{-\gamma_0\tau}.
 \end{aligned}
 \label{eq:gkslperipheral}
\end{equation}
The persistent peripheral contribution defines a legal, generally
quasiperiodic reference process. Indeed, boundedness of $\Phi_t$ excludes
Jordan blocks at peripheral eigenvalues. Simultaneous recurrence of the
finitely many phases supplies $s_k\to\infty$ with
$e^{i\Omega_\ell s_k}\to1$ for every $\ell$. Thus, for each $t\ge0$,
$\Phi_{t+s_k}\to\Phi_t^{\rm per}$ in operator norm. Complete positivity
and trace preservation pass to this finite-dimensional limit, so
$\Phi_t^{\rm per}(\rho)$ is a state generating compatible cPT marginals.
Relative to the static reference $\mathcal F_T(\Pi_0\rho)$, pointwise
process equilibration holds precisely when
$\mathcal F_T\Pi_{i\Omega_\ell}\rho=0$ at every nonzero peripheral
frequency. Here $\Pi_{i\Omega_\ell}$ denotes the full spectral projection
at that frequency. Necessity follows by averaging the squared norm of
the finite peripheral Fourier sum; sufficiency follows from the
decaying remainder. A common-time average instead removes those
nonzero frequencies. A
decoherence-free subspace and a dark state are special cases: a component
there remains visible whenever a system intervention overlaps with it. The
cPT estimate applies to $\mathcal R_\tau$ and cannot suppress a peripheral
component that the process can detect.

\subsection{Continuous GKSL Mpemba effect}
For two preparations with a common stationary reference, define
\begin{equation}
 \begin{aligned}
 d_{j,n}(\tau)&=\|F_{T,n}[\Phi_\tau(\rho_j-\omega)]\|,\\
 d_j(\tau)&=\|\Upsilon_{\rho_j,\tau,T}-\Omega_{\omega,T}\|_{\rm Fock}.
 \end{aligned}
 \label{eq:gkslmpembadist}
\end{equation}
The visible decay modes include generalized eigenvectors when the generator
has Jordan blocks. Let $X_j=\rho_j-\omega$,
$N_\lambda=(\mathcal L-\lambda I)\Pi_\lambda$, and let $r_\lambda$
be the size of its largest Jordan block. Spectral calculus gives
\begin{equation}
 \begin{aligned}
 F_{T,n}\Phi_\tau X_j
 =\sum_\lambda e^{\lambda\tau}
 \sum_{k=0}^{r_\lambda-1}\frac{\tau^k}{k!}
 F_{T,n}N_\lambda^k\Pi_\lambda X_j .
 \end{aligned}
 \label{eq:gksl-visible-jordan}
\end{equation}
The largest real part among eigenvalues with a nonzero coefficient sets the
sector's exponential envelope; Jordan blocks also supply polynomial factors.
An improvement of the exponential rate requires cancellation of every
coefficient at the slow real part, including all peripheral frequencies
and all powers $k$ present there. Removing only the highest Jordan power
can improve a polynomial factor without improving that rate.

If $A$ cancels these coefficients in selected sectors and the next visible
real part is strictly smaller than for $B$, this gives a sector-selective
rate advantage. In finite dimension there are finitely many spectral
coefficients. If each slow coefficient vanishes in every scalar sector,
its full-Fock vector vanishes as well; applying the full map to the same
finite expansion therefore proves the improved full-norm exponential
envelope. All-sector improvement additionally requires the rate advantage
in every fixed active sector. These statements compare exponential
envelopes. An eventual pointwise advantage also requires a matching lower
bound for $B$ uniform in its oscillation phase, while a Mpemba crossing
requires the initial ordering in the norm being compared. For
infinite-dimensional extensions, a summable envelope at the improved rate
is an additional requirement. Cancellation in the reduced state alone
does not imply cancellation after later interventions.

The two-qubit models that follow realize these alternatives. The hidden
correlation model contains a slow mode invisible in the undisturbed system
state but visible after a system rotation and delayed readout. The
detailed-balance model has a stationary Gibbs state and an explicit
constrained preparation that removes the slow process mode. Both are finite
dimensional joint GKSL semigroups, so their cPT contractions fall under the
bounds above.

\section{Local positive preparations, initial ordering, and conditional rates}
\label{app:preps}
\subsection{Physical preparations with the same stationary reference}
For the spin chain in the main text, local commutators give
\begin{align}
 Q_1={}&-0.3Y_SX_1,\\
 Q_2={}&0.4575X_SX_1-0.09Z_S\nonumber\\
 &-0.33Y_SY_1-0.6Y_SY_1Z_2.
 \label{eq:localpreps}
\end{align}
The preparations $\rho_{B,N}=(I+Q_1)/d_N$ and $\rho_{A,N}=(I+Q_2/2)/d_N$ are strictly positive, have the same bath marginal, and satisfy
\begin{equation}
 \mathcal D_{H_N}(\rho_{A,N})=
 \mathcal D_{H_N}(\rho_{B,N})=I/d_N.
\end{equation}
They therefore share every energy population and the stationary cPT reference. Their contrast remains local as $N\to\infty$. At $T=0.02$, an analytic first-sector lower bound and a full-sector upper bound give, uniformly for $N\ge2$,
\begin{equation}
 \|\Delta\Upsilon_A(0)\|>0.0057371,
 \quad \|\Delta\Upsilon_B(0)\|<0.0005024.
 \label{eq:initialorder}
\end{equation}
Under Eq.~\eqref{eq:tailhyp}, the initial ordering in Eq.~\eqref{eq:initialorder} gives a full-process Mpemba crossing, with an exponent advantage in every fixed sector satisfying $P_nW_T\ne0$. Sectorwise crossings follow for sectors with the corresponding initial ordering. The estimates below also give volume-independent $L^{-1}$ and $L^{-2}$ bounds for a common smooth random waiting protocol.

The norm bounds $\|Q_1\|=0.3$ and $\|Q_2\|\le1.4775$ imply
\begin{equation}
 \rho_{B,N}\succeq0.7I/d_N,\qquad
 \rho_{A,N}\succeq0.26125I/d_N.
\end{equation}
All terms are traceless and contain a nonidentity system Pauli, so both preparations have normalized trace and the same bath marginal. Every commutator vanishes within an equal-energy block: $P_E[H,X]P_E=0$. Therefore their energy-dephased states are exactly $I/d_N$, even with spectral degeneracies. Since $Q_1,Q_2$ are supported on $S,1,2$, the local prepared states stabilize for all larger volumes and have nonzero thermodynamic contrast.

To bound the initial process distance of $B$, remove only the coupling $gX_SX_1$. Without it, every system-only insertion leaves a traceless environment factor, and all trace-boundary process coefficients of $Q_1/d_N$ vanish. Duhamel telescoping over the free intervals gives
\begin{equation}
 |C^{(n)}_B|\le2gTR^n\|Q_1/d_N\|_1=0.09TR^n.
\end{equation}
The interval lengths sum to at most $T$, irrespective of the insertion number. Summing all sectors yields
\begin{equation}
 E_B(0)\le0.09T\sqrt{e^{15T/4}-1}.
 \label{eq:appBinitial}
\end{equation}
For $A$, set $r_A(t)=\Tr[Z_S\mathcal U_t(Q_2/(2d_N))]$. Its initial value is $-0.045$, and
\begin{equation}
 |r_A'(t)|\le\|i[H,Z_S]\|\|Q_2/(2d_N)\|_1
 \le0.221625.
\end{equation}
Pauli completeness implies $e_{A,1}^2\ge\int_0^T|r_A(t)|^2\dd t$. Hence
\begin{equation}
 E_A(0)\ge\tfrac12(0.09-0.44325T)\sqrt T
\end{equation}
whenever the parenthesis is positive. At $T=0.02$ this lower bound exceeds $0.00573710$, while Eq.~\eqref{eq:appBinitial} is below $0.00050235$.

The common-tail condition is stated directly in the full process norm.
Suppose $\alpha>0$, $c\ne0$, and $W_T\ne0$, and for $j=0,1,2$,
\begin{align}
 \partial_\tau^j f_{Z_S}(\tau)
 &=c(-1)^j(\alpha)_j\tau^{-\alpha-j}W_T+r_j(\tau),\nonumber\\
 \|r_j(\tau)\|&=o(\tau^{-\alpha-j}).
 \label{eq:appmixing}
\end{align}
For every fixed $n$ with $P_nW_T\ne0$, the exact derivative identity gives
\begin{align}
 e_{B,n}(\tau)&\sim\alpha|c|\|P_nW_T\|\tau^{-\alpha-1},\\
 e_{A,n}(\tau)&\sim\tfrac12\alpha(\alpha+1)|c|
 \|P_nW_T\|\tau^{-\alpha-2}.
\end{align}
The same asymptotic formulas hold for the full norm with $\|W_T\|$, giving $E_A/E_B\sim(\alpha+1)/(2\tau)$.

To specify a sufficient local mixing input, let $\mathbb E_{K_\ell}$ be
the tracial conditional expectation onto the buffer $K_\ell$ around the
accessible region, and define
\begin{equation}
 \|A\|_{{\rm loc},\mu}=\|A\|+
 \sup_{\ell\ge0}e^{\mu\ell}
 \|A-\mathbb E_{K_\ell}A\| .
 \label{eq:quasilocal-mixing-norm}
\end{equation}
Choose $\mu>0$ within the locality exponent of
Appendix~\ref{app:local}. Its buffer approximation and contractivity of
$\mathbb E_{K_\ell}$ imply
$\|A_{\boldsymbol\nu}\|_{{\rm loc},\mu}
\le R^n(1+c_{T,\mu}n)$, with a constant independent of $n$ and volume.
Suppose there is a linear functional $\ell$ bounded in this norm such
that, for $Q_j=(-i\operatorname{ad}_H)^jZ_S$, $j=0,1,2$, and
$\delta>0$,
\begin{equation}
 \begin{aligned}
 \mathfrak r_j(A,\tau)={}&\omega_0(A V_\tau Q_j)\\
 &-c(-1)^j(\alpha)_j\tau^{-\alpha-j}\ell(A),\\
 |\mathfrak r_j(A,\tau)|\le{}&
 C_j\|A\|_{{\rm loc},\mu}\tau^{-\alpha-j-\delta} .
 \end{aligned}
 \label{eq:local-mixing-sufficient}
\end{equation}
This estimate holds for every $A$ of finite local norm and every
$\tau\ge\tau_0$, with $\tau_0$ and $C_j$ independent of $A$.
The coefficients $\ell(A_{\boldsymbol\nu})$ define $W_T$. If this
vector is nonzero, integration and species summation give
\begin{equation}
 \|r_j(\tau)\|\le C_j\tau^{-\alpha-j-\delta}
 \left[\sum_{n\ge1}(1+c_{T,\mu}n)^2\frac{x^n}{n!}\right]^{1/2}.
\end{equation}
This sum is finite. It supplies the required passage from a quantitative local mixing law to a full-process rate.

A distinct exact result concerns a common random waiting time $\tau=Lu$, with density $w(u)=30u^2(1-u)^2$ on $[0,1]$. The mixed process remains physical. Both $w$ and $w'$ vanish at the endpoints, so integration by parts gives
\begin{align}
 \left\|\int_0^1w(u)\Delta\Upsilon_B(Lu)\dd u\right\|
 &\le\frac{15}{4L}\sqrt{e^x-1},\\
 \left\|\int_0^1w(u)\Delta\Upsilon_A(Lu)\dd u\right\|
 &\le\frac{20}{\sqrt3L^2}\sqrt{e^x-1}.
\end{align}
Here $\|w'\|_1=15/4$, $\|w''\|_1=40/\sqrt3$, and the second preparation carries the factor $1/2$. These bounds hold in finite and infinite volume. They describe the norm of the averaged process; they are distinct from averaging the squared process distance.

\section{Continuous control, terminal measurements, and records}
\label{app:readout}
Retain the system output at time $T$ and define
\begin{equation}
 F^{\rm out}_{T,n}(X;\mathbf t,\boldsymbol\nu)
 =\Tr_E[\mathcal U_{T-t_n}J_{\nu_n}\cdots J_{\nu_1}\mathcal U_{t_1}(X)].
\end{equation}
The norm is the $L^2$ Fock sum with Hilbert--Schmidt norm on the output. Partial trace is contractive in trace norm, and $\|A\|_2\le\|A\|_1$, so the same factorial bound holds, now including the possibly nonzero $n=0$ output sector.

\subsection{An exact scalar--output norm bridge}
For a qubit and any complex system matrix $Y$, Pauli completeness gives
\begin{equation}
 \sum_{(\mu,\nu)\ne(0,0)}
 |\Tr R_{\mu\nu}(Y)|^2
 =2\|Y\|_2^2-\frac14|\Tr Y|^2.
 \label{eq:Paulioutputidentity}
\end{equation}
The normalization follows by summing all sixteen species, which gives
$2\|Y\|_2^2$, and then removing $R_{00}(Y)=Y/2$.
Define the termination-time-integrated output norm
\begin{equation}
 \mathcal B_T(X)^2=
 \int_0^Tdt\sum_{n\ge0}\int_{\Delta_n(t)}
 \|F^{\rm out}_{t,n}(X)\|_2^2\,d\mathbf t .
 \label{eq:integratedoutput}
\end{equation}
The system output at time $t$ includes all operations before $t$.
Separate the output zero sector from the positive insertion sectors:
\begin{equation}
 \begin{aligned}
 \mathcal B_{T,0}(X)^2&=
 \int_0^T\|F^{\rm out}_{t,0}(X)\|_2^2\,\dd t,\\
 \mathcal B_{T,+}(X)^2&=
 \int_0^T\!\dd t\sum_{n\ge1}\int_{\Delta_n(t)}
 \|F^{\rm out}_{t,n}(X)\|_2^2\,\dd\mathbf t,\\
 \mathcal B_T(X)^2&=\mathcal B_{T,0}(X)^2\\
 &\quad+\mathcal B_{T,+}(X)^2 .
 \end{aligned}
 \label{eq:outputsector-split}
\end{equation}
Decompose each scalar coefficient of order $m\ge1$ according to its final
insertion time. Before that insertion the retained output has order
$n=m-1\ge0$. Applying Eq.~\eqref{eq:Paulioutputidentity} gives
\begin{align}
 \|\mathcal F_T^+(X)\|^2
 ={}&\sum_{n\ge0}\int_0^Tdt\int_{\Delta_n(t)}d\mathbf t\nonumber\\
 &\times\left[2\|F^{\rm out}_{t,n}(X)\|_2^2
 -\frac14|\Tr F^{\rm out}_{t,n}(X)|^2\right].
 \label{eq:exactnormbridge}
\end{align}
Since $|\Tr Y|^2\le2\|Y\|_2^2$,
\begin{equation}
 \frac32\mathcal B_T(X)^2\le
 \|\mathcal F_T^+(X)\|^2\le2\mathcal B_T(X)^2.
 \label{eq:normbridge}
\end{equation}
The scalar Fock norm has the separate zero-sector decomposition
\begin{equation}
 \|\mathcal F_T(X)\|^2=|\Tr X|^2+\|\mathcal F_T^+(X)\|^2 .
 \label{eq:scalar-zero-split}
\end{equation}
The output zero sector is already included in $\mathcal B_T$. Its
contribution to the scalar hierarchy is exactly
\begin{equation}
 \|F_{T,1}(X)\|^2
 =2\mathcal B_{T,0}(X)^2-\frac{T}{4}|\Tr X|^2,
 \label{eq:firstsector-outputzero}
\end{equation}
because the free evolution preserves the trace. If only positive output
orders are retained, the matching scalar hierarchy starts at order two:
\begin{equation}
 \frac32\mathcal B_{T,+}(X)^2
 \le\sum_{m\ge2}\|F_{T,m}(X)\|^2
 \le2\mathcal B_{T,+}(X)^2.
 \label{eq:positiveoutputbridge}
\end{equation}
Equations~\eqref{eq:normbridge}--\eqref{eq:positiveoutputbridge} apply to
general inputs. The scalar zero sector is handled separately through
Eq.~\eqref{eq:scalar-zero-split}. For a state difference, $\Tr X=0$ and
$\mathcal F_T^+=\mathcal F_T$, while
$\|F_{T,1}(X)\|^2=2\mathcal B_{T,0}(X)^2$.
A fixed-time output is instead retained
explicitly in the operational norm; a point evaluation is not inferred from
an $L^2$ time norm.

\subsection{Deterministic coherent control}

Expand a bounded local Hamiltonian control as $\mathcal G_t=a_0(t)I+\sum_\nu a_\nu(t)J_\nu$. Dyson expansion of the controlled joint propagator pairs its coefficients with $F^{\rm out}_{T,n}$. The dual coherent cMPS has scalar virtual dimension one, drift $a_0^*$, and species amplitudes $a_\nu^*$. Its squared norm is
\begin{equation}
 \|J_G\|^2=\exp\int_0^T\left[2\operatorname{Re}a_0(t)+\sum_\nu|a_\nu(t)|^2\right]\dd t.
\end{equation}
Vector-valued Cauchy--Schwarz in the direct sum gives $\|\Delta\sigma^G\|_2\le\|J_G\|\|\Delta F^{\rm out}\|$. For any terminal POVM $\{M_y\}$,
\begin{align}
 \TV(p,q)&=\tfrac12\sum_y|\Tr(M_y\Delta\sigma^G)|\\
 &\le\tfrac12\|\Delta\sigma^G\|_1
 \le\tfrac12\sqrt{d_S}\|J_G\|\|\Delta F^{\rm out}\|.
\end{align}
This proves a bound uniform over any control class with a fixed $\|J_G\|$ budget. The retained output is essential: a trace-boundary contraction with a deterministic trace-preserving control alone returns normalization.

The bridge also gives a direct operational estimate from the scalar norm.
Choose a terminal time $t\in[0,T]$ with a fixed density $w(t)$, record it,
and at that time perform a POVM after bounded continuous control.
Writing $K_t$ for its coherent-control norm, Cauchy--Schwarz gives
\begin{equation}
\TV(P_\rho,P_\sigma)\le
 \frac12\sqrt{d_S}\,
 \left(\int_0^T w(t)^2K_t^2\,dt\right)^{1/2}
 \|\mathcal F_T(X)\|.
 \label{eq:randomterminationTV}
\end{equation}
Here $X=\rho-\sigma$, and the estimate holds for square
integrable $wK$. The prefactor is the general terminal-POVM bound; for the
six-outcome Pauli POVM used in the main text it can be sharpened by replacing
$\frac12\sqrt{d_S}$ with $1/\sqrt6$. It relates a scalar process norm to
this finite-budget family of observations. The fixed terminal time in
Theorem~\ref{thm:operational} uses the separate exact model calculation.

\subsection{Jump and diffusive records}

For jump monitoring, use the quantum-jump expansion~\cite{Dalibard1992}. Let $\mathcal J_j(X)=L_jXL_j^\dagger$ and $Q_0=Q-\frac12\{\sum_jL_j^\dagger L_j,\cdot\}$. A record $(\mathbf j,\mathbf t)$ on $\Delta_n(T)$ has density
\begin{equation}
 p_\rho(\mathbf j,\mathbf t)=\langle\!\langle I|e^{Q_0(T-t_n)}\mathcal J_{j_n}\cdots\mathcal J_{j_1}e^{Q_0t_1}|\rho\rangle\!\rangle.
\end{equation}
The omitted factors between consecutive jumps are $e^{Q_0(t_{k+1}-t_k)}$. This is a cMPS contraction with the no-click transfer and jump insertions. Summing all records gives the trace of $e^{(Q_0+\sum_j\mathcal J_j)T}\rho$, equal to one. If $\|\sum_jL_j^\dagger L_j\|\le\lambda$, the event-count tail is bounded by $\Pr[\operatorname{Pois}(\lambda T)>k]$.

For a diffusive measurement, choose a Wiener reference measure and the standard unnormalized filtering equation~\cite{JacobsSteck2006},
\begin{equation}
 \dd\sigma_t=(Q+\mathcal D[L])\sigma_t\dd t
 +(L\sigma_t+\sigma_tL^\dagger)\dd Y_t.
\end{equation}
Its likelihood is $Z_\rho(Y)=\Tr\sigma_T(Y)$, and the exact record distance is
\begin{equation}
 \TV(P_\rho,P_\sigma)=\tfrac12\mathbb E_W|Z_\rho(Y)-Z_\sigma(Y)|.
\end{equation}
For Hermitian $L$, the finite-step Kraus operator $K_h(y)=\exp(Ly-L^2h)$ obeys $\mathbb E_{y\sim N(0,h)}K_h^\dagger K_h=I$; its expansion gives the preceding stochastic equation. Thus diffusive and jump monitoring define legitimate continuous record instruments. A full-record Fock-to-$L^1$ bound additionally requires boundedness of that record map; ordinary white-noise paths are not finite-norm deterministic coherent controls.

\section{Exact hidden-mode benchmark and finite-width tester}
\label{app:twoqubit}
The swap, twirl, and reset maps commute, and the latter two are idempotent channels. Their generator is $-ig[\mathsf S,\cdot]+\gamma(\mathcal T-I)+\epsilon(\mathcal R-I)$, with $\mathcal R(X)=\Tr(X)I_4/4$. Exponentiating the commuting maps yields the convex formula in the main text.

For a product preparation with system Bloch vector $\mathbf r$ and memory Bloch vector $\mathbf s$, partial swap gives
\begin{equation}
 \mathbf r(t)=e^{-(\gamma+\epsilon)t}
 [c^2\mathbf r+s^2\mathbf s+cs(\mathbf s\mathbin{\times}\mathbf r)].
\end{equation}
Its squared norm for $\mathbf r=\hat x$ and $\mathbf s=m\hat z$ is $e^{-2(\gamma+\epsilon)t}(c^4+m^2c^2s^2+m^2s^4)$. For $\mathbf r=b\hat z$ it is $e^{-2(\gamma+\epsilon)t}(bc^2+ms^2)^2$. More precisely,
\begin{equation}
 4e^{2(\gamma+\epsilon)t}(d_A^2-d_B^2)
 =(1-b^2)c^4+m(m-2b)c^2s^2.
 \label{eq:appnocross}
\end{equation}
For the actual parameters $m=0.6$, $b=0.2$, both coefficients are nonnegative, so the reduced-state ordering cannot reverse. Equality is possible at isolated exchange times where $c=0$.

The singlet weight of a product state is $(1-\mathbf r\cdot\mathbf s)/4$. Thus $\mathcal T\rho_A=I/4$ and $\mathcal T\rho_B=I/4+bmR/12$, where $R=XX+YY+ZZ$. This slow correlation has zero partial trace on both subsystems. A local $Z$ rotation followed by a delay changes its readout to
\begin{equation}
 \Tr[Z_S\mathcal E_s(V_z(R/4)V_z^\dagger)]
 =e^{-(\gamma+\epsilon)s}\sin(2gs).
\end{equation}
The right side is nonzero on an interval around $s=0.35$.

For a width $w=0.02$, define averaged observables in the Heisenberg picture,
\begin{align}
 O_x&=w^{-1}\int_0^w\mathcal E_u^*(X_S)\dd u,\\
 O_z&=w^{-2}\int_0^w\dd u\int_d^{d+w}\dd s\,
 \mathcal E_u^*[V_z^\dagger\mathcal E_s^*(Z_S)V_z].
\end{align}
Recording the branch and the binary outcome, but not the times, gives
\begin{equation}
 D_J(\tau)=\tfrac12\{0.8|\Tr[O_x\Delta\rho_\tau]|
 +0.2|\Tr[O_z\Delta\rho_\tau]|\}.
\end{equation}
The stationary outcomes are unbiased. The $B$ preparation has a late-time coefficient
\begin{align}
 D_{J,B}(\tau)&\sim\frac{bm}{30}e^{-\epsilon\tau}A_w,\\
 A_w&=\frac1{w^2}\int_0^w\!\dd u\int_d^{d+w}\!\dd s\,
 e^{-\epsilon u-(\gamma+\epsilon)s}\sin(2gs)>0.
\end{align}
The factor $bm/30$ equals the branch weight $0.2$ times $bm/3$ times $1/2$. The $A$ preparation has no $\mathcal T$ component and is bounded by $C_Je^{-(\gamma+\epsilon)\tau}$. Initial readout distances are $0.4020079$ and $0.0182465$, so continuity and the different exponents prove an eventual readout reversal. The integrable timing kernels define bounded finite-sector functionals of the continuous process. Smoothing their endpoints or replacing a rotation by a finite pulse preserves strict endpoint inequalities by continuity of the finite-dimensional propagator.

\subsection{Complete cMPS sectors and a deterministic terminal readout}
The same hidden-mode construction can be evaluated without selecting a
single scalar tester. We use the local Pauli species
$J_{\mu\nu}(X)=\sigma_\mu X\sigma_\nu/2$ and the cMPS self-contraction on
$[0,T]$ with $T=0.4$. The resulting Fock components are denoted by
$E_{j,n}(\tau)=\|P_n\mathcal F_T(\mathcal E_\tau(\rho_j)-\rho_*)\|$.
The full norm is evaluated directly from the transfer exponential at
insertion fugacity $z=1$; only the displayed sector-resolved curves are
truncated at $n=4$. The factorial estimate in Eq.~\eqref{eq:apptail}
controls the omitted sectors.
For an operational readout we apply the fixed system rotation
$V_z=\exp(-i\pi Z_S/4)$, wait for $d=0.35$, and measure the six-outcome
informationally complete POVM
\begin{equation}
 M_{a,\pm}=\frac{I\pm\sigma_a}{6},\qquad a=x,y,z .
 \label{eq:hiddenicpovm}
\end{equation}
Let $r_{j,a}(\tau)=\Tr[\sigma_a\mathcal E_d(V_z
(\mathcal E_\tau(\rho_j)-\rho_*)V_z^\dagger)]$ denote the difference of
the output Bloch components from the equally controlled reference.
The output total
variation distance is
\begin{equation}
 D_{\rm TV}^{(j)}(\tau)=\frac16\sum_{a=x,y,z}|r_{j,a}(\tau)| .
 \label{eq:hiddentv}
\end{equation}
This is a deterministic control followed by one terminal measurement; no
random timing mixture is needed.

The complete readout follows directly from partial-swap propagation.
Write $a=\gamma+\epsilon$, $c=\cos(g\tau)$, $s=\sin(g\tau)$,
$C=\cos(gd)$, and $S=\sin(gd)$. Relative to the equally controlled
stationary reference, the three output components for $A$ are
\begin{align}
 r_{A,x}&=e^{-a(\tau+d)}\big[-mC^2cs+S^2s^2
                  -CS(cs+mc^2)\big],\nonumber\\
 r_{A,y}&=e^{-a(\tau+d)}\big[C^2c^2-mS^2cs
                  -CS(cs+ms^2)\big],\nonumber\\
 r_{A,z}&=m e^{-a(\tau+d)}(C^2s^2+S^2c^2).
 \label{eq:hiddenclosedA}
\end{align}
For $B$, $r_{B,x}=r_{B,y}=0$ and
\begin{align}
 r_{B,z}=e^{-ad}\Big\{&e^{-a\tau}
 [C^2(bc^2+ms^2)+S^2(mc^2+bs^2)]\nonumber\\
 &+\frac{bm}{3}(e^{-\epsilon\tau}-e^{-a\tau})\sin(2gd)\Big\}.
 \label{eq:hiddenclosedB}
\end{align}
For example, with $U_\tau=cI-is\mathsf S$,
\begin{align*}
 U_\tau(XI)U_\tau^\dagger&=c^2XI+s^2IX+cs(YZ-ZY),\\
 U_\tau(XZ)U_\tau^\dagger&=c^2XZ+s^2ZX+cs(YI-IY).
\end{align*}
Applying $V_z$, propagating for $d$, and tracing the memory gives
Eqs.~\eqref{eq:hiddenclosedA} and \eqref{eq:hiddenclosedB}.
At $d=0.35$ the slow coefficient in $r_{B,z}$ is positive, whereas
$r_{A,z}\ge m e^{-a(\tau+d)}\min(C^2,S^2)>0$.
Consequently the fixed readout has the two-sided asymptotic orders
\begin{equation}
 D_{\rm TV}^{(A)}=\Theta(e^{-a\tau}),\qquad
 D_{\rm TV}^{(B)}=\Theta(e^{-\epsilon\tau}).
 \label{eq:hiddenreadoutrates}
\end{equation}
Its initial ordering is also analytic:
\begin{align}
 D_{\rm TV}^{(A)}(0)-D_{\rm TV}^{(B)}(0)
 =\frac{e^{-ad}}6\big[(1-b)C^2+m|CS|\big]>0.
 \label{eq:hiddenreadoutinitial}
\end{align}
Here $m>b>0$ and $C,S>0$ for the stated parameters.
Thus the same fixed control resolves a strict Mpemba reversal.

The numerical cMPS contractions give
\begin{align*}
 &(E_A,E_B)_{\tau=0}=(0.76353,0.17021),\\
 &(E_A,E_B)_{\tau=20}=(1.91\times10^{-8},1.77\times10^{-3}),\\
 &(D_{\rm TV}^{(A)},D_{\rm TV}^{(B)})_{\tau=0}
   =(0.14188,0.03058),\\
 &(D_{\rm TV}^{(A)},D_{\rm TV}^{(B)})_{\tau=20}
   =(4.43\times10^{-9},1.17\times10^{-3}).
\end{align*}
The state-distance ordering remains fixed. On the displayed waiting-time
grid, the full process first crosses in the interval $4.2<\tau<4.3$ and the
fixed-readout distance crosses in $3.8<\tau<3.9$. The same grid shows that
$n=1$ retains the state ordering, whereas the $n=2,3,4$ sectors cross at
approximately $4.1$--$4.3$. Thus the observed reversal is a genuinely
multitime effect carried by correlations invisible in the unperturbed
marginal. A strict full-norm initial-order bound is given in
Eq.~\eqref{eq:qubitcertifiedorder}.

\begin{figure*}[t]
\includegraphics[width=\textwidth]{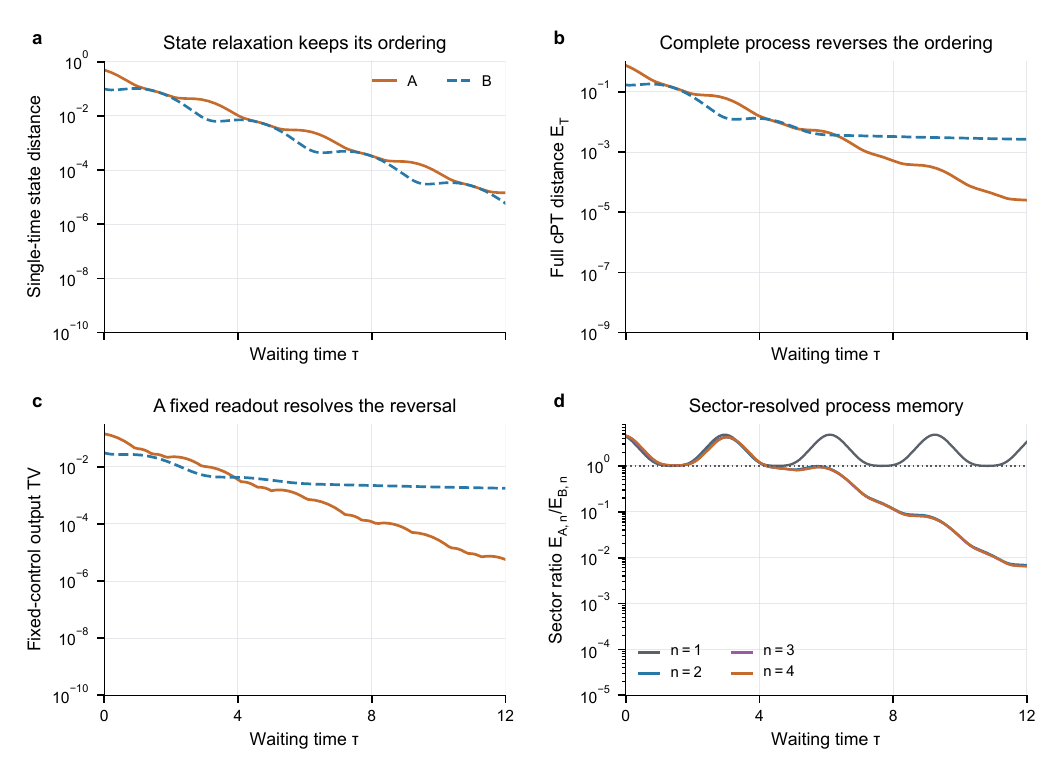}
\caption{Complete hidden-correlation benchmark. (a) The single-time system
distance keeps its ordering. (b) The cMPS Fock sum reverses the ordering.
(c) A fixed unitary control and informationally complete terminal POVM
resolve the same reversal. (d) Ratios of the $n=1,2,3,4$ sector norms show
that the higher sectors carry the slow correlation responsible for the
process-level crossing.}
\label{fig:hidden_complete}
\end{figure*}

\section{Sharp constrained separable optimization and slow-mode visibility}
\label{app:separable}
Use $H=gZZ$, $\sigma=(I-mZZ)/4$, $p_*=(1+m)/4$, and $\sigma_t=(\sigma-p_*P_s)/(1-p_*)$. The projection $\mathcal P(X)=\Tr(P_sX)P_s+\Tr(P_tX)\sigma_t$ and reset $\mathcal R_\sigma(X)=\Tr(X)\sigma$ commute with each other and with the Hamiltonian channel. Consequently
\begin{align}
 \mathcal E_t(X)={}&e^{-(\gamma+\epsilon)t}U_tXU_t^\dagger
 +(e^{-\epsilon t}-e^{-(\gamma+\epsilon)t})\mathcal P(X)\nonumber\\
 &+(1-e^{-\epsilon t})\Tr(X)\sigma.
\end{align}
The invariant slow vector is $R_s=P_s-\sigma_t$, and for a normalized preparation
\begin{align}
 \mathcal E_t(\rho)-\sigma
 ={}&e^{-\epsilon t}c_\rho R_s\nonumber\\
 &+e^{-(\gamma+\epsilon)t}U_t(\rho-\mathcal P\rho)U_t^\dagger,\nonumber\\
 c_\rho={}&\Tr(P_s\rho)-p_*.
 \label{eq:appsplit}
\end{align}

To exhibit detailed balance, choose the Bell energy eigenbasis with index zero the singlet and stationary probabilities $p_a$. Rank-one jumps $\sqrt{r_{ab}}|a\rangle\langle b|$ with
\begin{equation}
 r_{ab}=\epsilon p_a+
 \begin{cases}
 \gamma,&a=b=0,\\
 \gamma p_a/(1-p_*),&a,b>0,\\
 0,&\text{otherwise}
 \end{cases}
\end{equation}
realize the dissipator exactly. Indeed each column sums to $\gamma+\epsilon$, its recycling map is $\gamma\mathcal P+\epsilon\mathcal R_\sigma$, and $r_{ab}p_b=r_{ba}p_a$. The energy probabilities have ratio $e^{-\beta(E_a-E_b)}$ with $\beta g=\operatorname{atanh}m$.

For the optimization, write the computational-basis diagonal as $(a,b,c,d)$. The marginal constraints imply $a=d=(1-v)/4$ and $b=c=(1+v)/4$. Set $z=\rho_{01,10}$. Positivity of the $00,11$ principal minor of $\rho^{T_M}$ gives $|z|\le(1-v)/4$. Therefore
\begin{equation}
 \Tr(P_s\rho)=\frac{1+v}{4}-\operatorname{Re}z\ge\frac v2.
\end{equation}
The state
\begin{equation}
 \rho_s=\tfrac14[I+\tfrac{1-v}{2}(XX+YY)-vZZ]
\end{equation}
attains equality. Its Bell-diagonal correlation magnitudes sum to one, or equivalently its partial transpose is positive, so it is separable. This proves the complete constrained minimum, including states with other coherences.

For $m=1/50$, $v=23/25$, the entangled state in the main text is Bell diagonal. Its eigenvalues are
\begin{equation}
 \frac{1-v}{4},\quad \frac{1-v}{4},\quad
 \frac{1+m}{4},\quad \frac{1+2v-m}{4},
\end{equation}
which are strictly positive. The eigenvalues of its partial transpose are
\begin{equation}
 \frac{1-2v+m}{4},\quad \frac{1-m}{4},\quad
 \frac{1+v}{4},\quad \frac{1+v}{4}.
\end{equation}
The first is negative because $2v>1+m$, so $\rho_e$ is entangled. Its singlet weight is exactly $p_*$, hence $c_{\rho_e}=0$. Every constrained separable state has $c_\rho\ge41/200$.

Slow-mode visibility is a system-only statement. Let $V_x=(I-iX_S)/\sqrt2$. Direct Pauli propagation gives
\begin{equation}
 \Tr[X_S\mathcal E_d(V_xR_sV_x^\dagger)]
 =-\frac{4(1-m)}{3-m}e^{-(\gamma+\epsilon)d}\sin(2gd),
 \label{eq:appslowvisible}
\end{equation}
nonzero at $d=0.35$. Finite timing windows preserve this nonzero signal, so
$\mathcal F_T(R_s)\ne0$ for a window containing the protocol. Its first
scalar sector vanishes because all unperturbed system marginals vanish;
higher sectors resolve it. The rotation in Eq.~\eqref{eq:appslowvisible}
expands as $R_{00}+R_{xx}+iR_{0x}-iR_{x0}$. Its $R_{00}$ contribution
vanishes on the freely evolving marginal of $R_s$, so the nonzero
readout proves that some second-order scalar coefficient is nonzero.
Moving the rotation slightly inside the window preserves this signal.
Moreover, every species $R_{\mu\nu}$ is a nonzero multiple of a product
$R_{r0}R_{s\nu}$ of two nonidentity species: choose $r\ne0,\mu$ and
$s\ne0$ such that $\sigma_r\sigma_s=e^{i\theta}\sigma_\mu$. Then
$R_{r0}R_{s\nu}=e^{i\theta}R_{\mu\nu}/2$.
Splitting one insertion into two nearby ordered times, strong continuity
preserves a nonzero coefficient. This raises the order from $n$ to
$n+1$ and proves $\|F_{T,n}(R_s)\|>0$ for every $n\ge2$.
The first sector remains zero; each later order has a nonzero constant.

Finally set $C_T=\sqrt{e^{15T/4}-1}$ and $c_T=\|\mathcal F_T(R_s)\|>0$. Since $\mathcal P$ is CPTP, $\|\rho-\mathcal P\rho\|_1\le2$. The reverse triangle inequality in Eq.~\eqref{eq:appsplit} gives, uniformly over all constrained separable preparations,
\begin{align}
 E_{\rm sep}(\tau)&\ge\tfrac{41}{200}c_Te^{-\epsilon\tau}
 -2C_Te^{-(\gamma+\epsilon)\tau},\\
 E_e(\tau)&\le C_T\|\rho_e-\sigma\|_1e^{-(\gamma+\epsilon)\tau}.
\end{align}
A finite common waiting threshold therefore makes $E_e$ smaller than every constrained separable distance, proving uniform eventual preparation advantage.

\subsection{Full sectors, fixed readout, and the uniform separable bound}
For the operational comparison we retain the same detailed-balance generator
and use the constrained separable optimizer $\rho_s^\star$ together with the
entangled preparation
\begin{equation}
 \rho_e^{(\kappa)}=\rho_e+\frac{\kappa}{4}Y\otimes Z,
 \qquad \kappa=\frac18 .
 \label{eq:entprepoperational}
\end{equation}
The perturbation has zero one-body marginals and zero $ZZ$ expectation, so
it preserves both initial constraints and leaves $\mathcal P(\rho_e^{(\kappa)})=\sigma$.
In the Bell basis, positivity reduces to two $2\times2$ minors with
off-diagonal magnitude $|\kappa|/4$, on the pairs
$(\Psi^-,\Phi^-)$ and $(\Psi^+,\Phi^+)$. Their determinants are
\begin{equation}
 \frac{(1-v)(1+m)-\kappa^2}{16},\qquad
 \frac{(1-v)(1+2v-m)-\kappa^2}{16}.
 \label{eq:entprepminors}
\end{equation}
For $0\le m\le v<1$ both are positive when
$\kappa^2<(1-v)(1+m)$, as for the stated parameters. Its unchanged
singlet weight $p_*<v/2$ also proves that the perturbed preparation
remains entangled by Eq.~\eqref{eq:sepbound}.
The initial marginal equality does not persist under the $ZZ$ interaction:
\begin{equation}
 \begin{aligned}
 d_e(\tau)&=\frac{|\kappa|}{2}e^{-(\gamma+\epsilon)\tau}
                   |\sin(2g\tau)|,\\
 d_s^\star(\tau)&=0 .
 \end{aligned}
 \label{eq:perturbedlocaldistance}
\end{equation}
Indeed, $U_\tau(YZ)U_\tau^\dagger
=\cos(2g\tau)YZ-\sin(2g\tau)XI$; the second term gives the local signal.
We apply $V_y=\exp(-i\pi Y_S/4)$, wait for $d=0.35<T=0.4$, and use the POVM
in Eq.~\eqref{eq:hiddenicpovm}. Its slow vector has a nonzero readout
overlap, with
\begin{equation}
 \begin{aligned}
 q_d&=\frac16\sum_{a=x,y,z}\left|\Tr\!\left[
 \sigma_a\,\mathcal E_d\!\left(V_yR_sV_y^\dagger\right)\right]\right|
 \\&=\frac{2(1-m)}{3(3-m)}e^{-(\gamma+\epsilon)d}|\sin(2gd)|
 \\&=0.104893\ldots .
 \end{aligned}
 \label{eq:sepslowreadout}
\end{equation}

\subsection{Analytic output curves and strict TV reversal}
Let $a=\gamma+\epsilon$, $h=v-m$, $s_d=\sin(2gd)$,
$c_{\min}=41/200$, and $G=4(1-m)c_{\min}/(3-m)$.
Both preparations are compared with the reference after the same
$V_y$ rotation and delay. Its output Bloch vector is
\begin{equation}
 \mathbf r_\sigma^{\rm out}=(0,-m e^{-ad}s_d,0).
 \label{eq:dbcontrolledreference}
\end{equation}
The Bloch differences from this reference are
\begin{align}
 r_{e,x}&=-\kappa e^{-a(\tau+d)}s_d\cos(2g\tau),\nonumber\\
 r_{e,y}&=-h e^{-a(\tau+d)}s_d,\nonumber\\
 r_{e,z}&=\kappa e^{-a(\tau+d)}\sin(2g\tau),\nonumber\\
 r_{s,y}^\star&=-e^{-ad}s_d
       [Ge^{-\epsilon\tau}+(h-G)e^{-a\tau}],\nonumber\\
 r_{s,x}^\star&=r_{s,z}^\star=0.
 \label{eq:dbclosedbloch}
\end{align}
To obtain these expressions, the Bell-diagonal part commutes with $ZZ$,
while the perturbation propagates as
$U_\tau YZU_\tau^\dagger=\cos(2g\tau)YZ-\sin(2g\tau)XI$.
The rotation sends $X$ to $-Z$, and the subsequent $ZZ$ evolution
converts the rotated correlations into local components.
Since $h>G>0$, the exact TV distances are
\begin{align}
 D_{\rm TV}^{e}(\tau)
 &=\frac{e^{-a(\tau+d)}}6\big[h|s_d|
      +|\kappa s_d\cos(2g\tau)|\nonumber\\
 &\hspace{38mm}+|\kappa\sin(2g\tau)|\big],\nonumber\\
 D_{\rm TV}^{s^\star}(\tau)
 &=\frac{e^{-ad}|s_d|}{6}
       [Ge^{-\epsilon\tau}+(h-G)e^{-a\tau}].
 \label{eq:dbclosedtv}
\end{align}
They have asymptotic orders $\Theta(e^{-a\tau})$ and
$\Theta(e^{-\epsilon\tau})$, respectively, and
\begin{equation}
 D_{\rm TV}^{e}(0)-D_{\rm TV}^{s^\star}(0)
       =\frac{|\kappa|}{6}e^{-ad}|s_d|>0.
 \label{eq:dbreadoutinitial}
\end{equation}
Moreover, $|\cos\theta|+|\sin\theta|/|s_d|
\le\sqrt{1+s_d^{-2}}$ gives the sufficient final-order threshold
\begin{equation}
 \tau>\frac1\gamma\log\!\left[
      1+\frac{|\kappa|}{G}\sqrt{1+s_d^{-2}}\right]
      \simeq0.772977.
 \label{eq:dbpairfinalthreshold}
\end{equation}
Beyond this time the entangled preparation stays closer in this fixed
readout. The threshold bounds the final reversal; the first crossing
occurs at $\tau\simeq0.717646$.

The full cMPS contractions yield
the full norm from the $z=1$ transfer exponential and the displayed sectors
from the corresponding number-resolved blocks:
\begin{align*}
 &(E_e,E_s^\star)_{\tau=0}=(0.12157,0.11828),\\
 &(E_e,E_s^\star)_{\tau=20}=(5.33\times10^{-9},1.36\times10^{-2}),\\
 &(D_{\rm TV}^{(e)},D_{\rm TV}^{(s^\star)})_{\tau=0}
   =(0.08173,0.07177),\\
 &(D_{\rm TV}^{(e)},D_{\rm TV}^{(s^\star)})_{\tau=20}
   =(3.72\times10^{-9},7.91\times10^{-3}).
\end{align*}
On the displayed waiting-time grid, the cPT crossing lies in
$0.7<\tau<0.8$, as does the fixed-readout crossing.
For the separable benchmark $E_{s,1}^\star\equiv0$, while
$E_{e,1}^{(\kappa)}$ contains only the fast local signal in
Eq.~\eqref{eq:perturbedlocaldistance}. The ratio of these first-sector
norms is undefined. We therefore display the sector norms themselves.
Sectors $n=2,3,4$ see the slow correlation and separate the preparations
at long times.

The uniform statement is stronger than the plotted comparison. With
$c_T=\|\mathcal F_T(R_s)\|$ and $C_T=\sqrt{e^{15T/4}-1}$, every constrained
separable preparation obeys
\begin{equation}
 E_{\rm sep}(\tau)\ge
 \frac{41}{200}c_Te^{-\epsilon\tau}
 -2C_Te^{-(\gamma+\epsilon)\tau},
 \label{eq:uniformseploweroperational}
\end{equation}
whereas the entangled preparation obeys
$E_e(\tau)\le C_T\|\rho_e^{(\kappa)}-\sigma\|_1
e^{-(\gamma+\epsilon)\tau}$. The lower bound is a statement about the
complete cPT norm; Eq.~\eqref{eq:sepslowreadout} supplies the nonzero overlap
that makes the same slow component visible in the fixed terminal record.
More explicitly, write $q_d$ for Eq.~\eqref{eq:sepslowreadout} and
$D_e=\|\rho_e^{(\kappa)}-\sigma\|_1$. Applying the same rotation,
propagator, and POVM to the decomposition in Eq.~\eqref{eq:appsplit}
and using trace-norm contractivity gives
\begin{equation}
 \begin{aligned}
 D_{\rm TV}^{\rm sep}(\tau)&\ge
 \left[\frac{41}{200}q_d e^{-\epsilon\tau}
                  -e^{-(\gamma+\epsilon)\tau}\right]_+,\\
 D_{\rm TV}^{e}(\tau)&\le\frac{D_e}{2}
                                e^{-(\gamma+\epsilon)\tau}.
 \end{aligned}
 \label{eq:uniformseptvlower}
\end{equation}
Here $[x]_+=\max(0,x)$. The remainder bound uses
$\|\rho-\mathcal P\rho\|_1\le2$. The first inequality holds for the
entire constrained separable set, with the same fixed readout and the
reference $\sigma$ subjected to that readout. Consequently the entangled
output is eventually closer than every admissible separable output.
The plotted $\rho_s^\star$ attains the constrained minimum of the slow
amplitude.

The bounds give explicit waiting times for the entire constrained set:
\begin{align}
 \tau_E^{\rm unif}
 &=\frac1\gamma\log\frac{C_T(2+D_e)}{c_{\min}c_T}
       \simeq6.23863,\nonumber\\
 \tau_{\rm TV}^{\rm unif}
 &=\frac1\gamma\log\frac{1+D_e/2}{c_{\min}q_d}
       \simeq5.26881.
 \label{eq:dbuniformthresholds}
\end{align}
Here $T=0.4$, $c_T=0.18018595\ldots$, and
$D_e=0.91140661\ldots$. For every $\tau>\tau_E^{\rm unif}$ the
entangled full-process distance is smaller than every constrained
separable distance; for every $\tau>\tau_{\rm TV}^{\rm unif}$ the same
statement holds for the single fixed readout. These follow by comparing
the fast upper bound with the slow lower bound, including its fast
remainder. For each $n\ge2$, writing
$A_n=\sqrt{(15T/4)^n/n!}$ and $c_n=\|F_{T,n}(R_s)\|>0$ similarly gives
\begin{align}
 E_{{\rm sep},n}(\tau)&\ge
   [c_{\min}c_ne^{-\epsilon\tau}-2A_ne^{-a\tau}]_+,\nonumber\\
 E_{e,n}(\tau)&\le A_nD_e e^{-a\tau}.
 \label{eq:dbuniformsectorbounds}
\end{align}
Each fixed higher sector therefore resolves the same separation of
slow and fast decay rates.

\subsection{Strict initial ordering of the full process norms}
The transfer formula also gives a finite-matrix proof of the initial
ordering in both two-qubit comparisons. In column-vectorized coordinates
write $x=\operatorname{vec}(\rho-\omega)$, with $\omega=I_4/4$ or
$\sigma$ as appropriate, and define
\begin{align}
 M&=T\left(\overline{\mathbb L}\otimes I_{16}
       +I_{16}\otimes\mathbb L
       +\sum_\nu\overline{\mathbb J_\nu}\otimes\mathbb J_\nu\right),
       \nonumber\\
 u&=\bar x\otimes x,\qquad
 h=\overline{\operatorname{vec}I_4}\otimes\operatorname{vec}I_4.
 \label{eq:qubitinitialtransfer}
\end{align}
Then $E(0)^2=h^\dagger e^M u$ by Eq.~\eqref{eq:apptransfer}.
All entries are complex rational numbers at the stated parameters and
$T=2/5$. Let $M_{\mathbb R}$ and $u_{\mathbb R}$ denote their real
representations, with real and imaginary parts in separate blocks.
The induced column-sum norm satisfies
$\|M_{\mathbb R}\|_1\le4$ in both models, and the real trace row has
dual norm one. Thus, with $P_N(M)=\sum_{k=0}^N M^k/k!$,
\begin{equation}
 \left|h^\dagger[e^M-P_N(M)]u\right|
 \le 3^4\frac{4^{N+1}}{(N+1)!}\|u_{\mathbb R}\|_1.
 \label{eq:qubitrationalremainder}
\end{equation}
Indeed, the exponential-series tail is at most
$e^4 4^{N+1}/(N+1)!$, and $e^4<3^4$.
At $N=80$ each of the four scalar remainders is below
$2.1\times10^{-70}$. Evaluating the finite polynomial in rational
arithmetic and applying this bound yields
\begin{align}
 E_A(0)^2-E_B(0)^2&>0.5540,\nonumber\\
 E_e(0)^2-E_s^\star(0)^2&>0.0007894.
 \label{eq:qubitcertifiedorder}
\end{align}
These strict initial inequalities, together with the slow-mode lower
bounds and fast-mode upper bounds, establish the full-process reversals.
Equations~\eqref{eq:hiddenreadoutinitial} and
\eqref{eq:dbreadoutinitial} independently establish the initial ordering
for the two fixed terminal readouts.

\begin{figure*}[t]
\includegraphics[width=\textwidth]{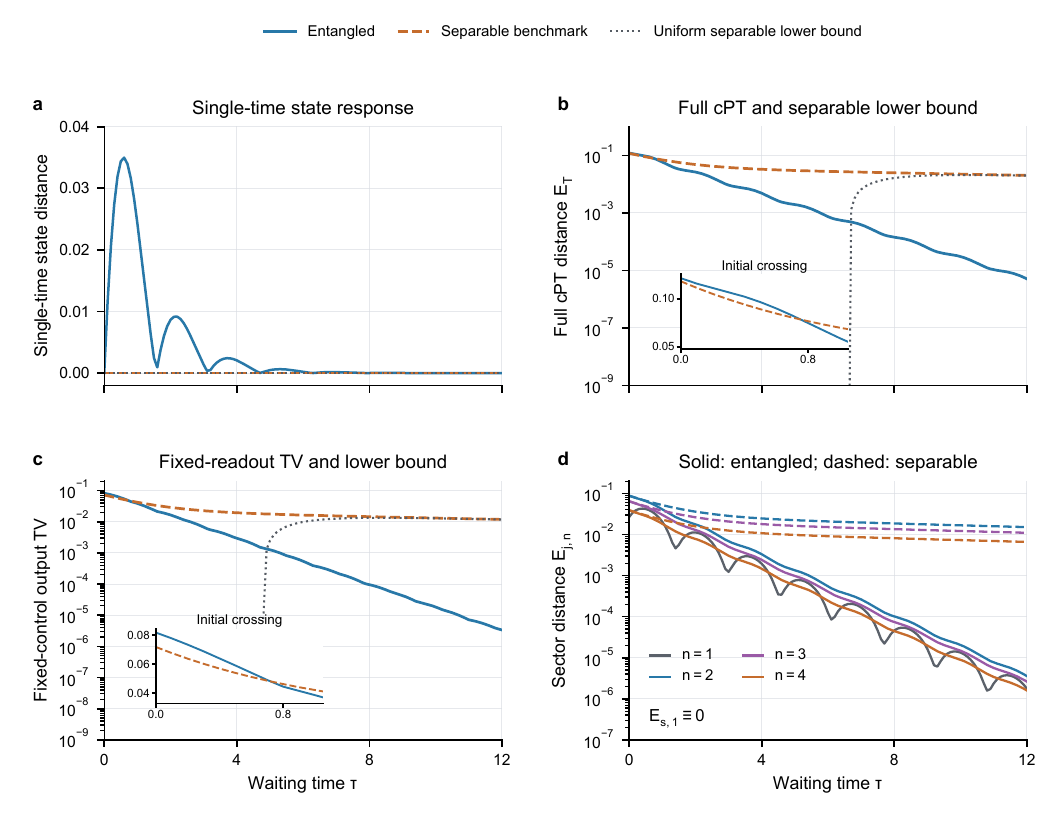}
\caption{Detailed-balance benchmark with constrained preparations.
(a) The fast perturbation generates a transient system signal, while
the separable benchmark has a stationary system marginal.
(b) The full cPT norm crosses despite the initially larger entangled
distance. The dotted curve is a uniform lower bound over all constrained
separable preparations, using the conservative remainder $4C_Te^{-(\gamma+\epsilon)\tau}$.
(c) A fixed $V_y$ rotation, delay $d=0.35$, and
terminal POVM resolve a TV crossing, together with the uniform bound in
Eq.~\eqref{eq:uniformseptvlower}. (d) Solid and dashed curves show the
entangled and separable norms for $n=1,2,3,4$. The separable first sector
is identically zero; the higher sectors retain its slow correlation.
The separable benchmark minimizes the asymptotic slow amplitude.}
\label{fig:separable_complete}
\end{figure*}

\begin{figure*}[t]
\includegraphics[width=\textwidth]{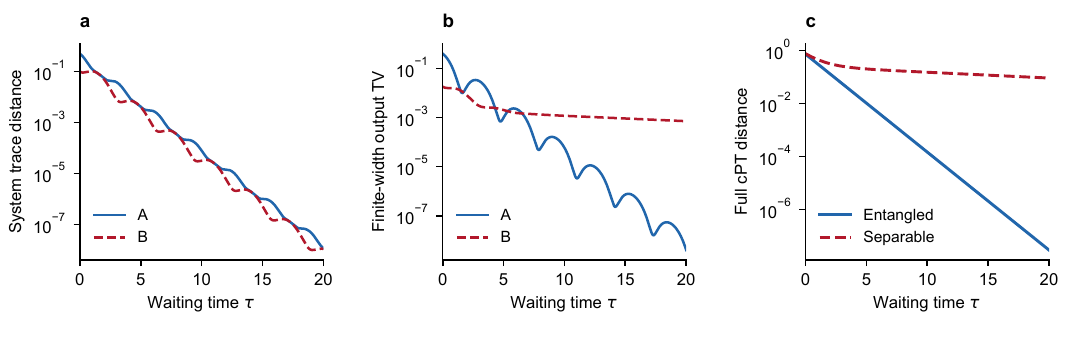}
\caption{Two exact process mechanisms. (a) System trace distances remain ordered for all waiting times. (b) A finite-width multitime tester reverses the ordering because it resolves a correlation with decay rate $\epsilon$ that is absent from the unperturbed system marginal. Panels (a,b) use the same two preparations and dynamics. (c) In the detailed-balance model, an entangled preparation removes the slow component while a constrained separable preparation with the smallest singlet weight retains it.}
\label{fig:twoqubit}
\end{figure*}

\section{ETH-chain sector contractions and finite-time comparisons}
\label{app:eth}
Let $\sigma_N=K^{-1}\sum_{a\in I}|E_a\rangle\langle E_a|$ and $|\psi_\theta\rangle=K^{-1/2}\sum_{a\in I}e^{i\theta_a}|E_a\rangle$. They satisfy $\mathcal D_{H_N}(|\psi_\theta\rangle\langle\psi_\theta|)=\sigma_N$ exactly. The shell-state difference has trace norm $D=2(1-1/K)$, independent of its phases and waiting time. These joint shell preparations fix all energy populations; they do not impose identical system or environment marginals.

The transfer formula can be factorized before constructing a large doubled Liouville matrix. Pauli completeness gives, for $X_t=\mathcal U_t(X)$,
\begin{equation}
 e_1^2=2\int_0^T\|\Tr_E X_t\|_2^2\dd t.
 \label{eq:appe1}
\end{equation}
For the second sector define
\begin{equation}
 Z_{\mu\nu}(t,s)=\Tr_E\mathcal U_s[(\sigma_\mu\otimes I)X_t(\sigma_\nu\otimes I)].
\end{equation}
Then
\begin{align}
 e_2^2=\int_0^T\dd t\int_0^{T-t}\dd s
 \sum_{(\mu,\nu)\ne(0,0)}\left[
 \tfrac12\|Z_{\mu\nu}\|_2^2-\tfrac1{16}|\Tr Z_{\mu\nu}|^2\right].
 \label{eq:appe2}
\end{align}
Summing the final 16 Pauli directions gives $\sum_{\mu,\nu}|\Tr(\sigma_\mu Y\sigma_\nu)|^2=8\|Y\|_2^2$. Removing the identity direction and retaining the factor $1/2$ from each insertion yields Eqs.~\eqref{eq:appe1}--\eqref{eq:appe2}, the exact insertion-number projections of the cMPS self-contraction.

For a quadrature remainder, let $W$ bound the full spectral width and $\delta$ the initial shell width. Iterated commutator bounds are $\|\operatorname{ad}_H^rX\|_1\le\delta^rD$ before an insertion and at most $W^r\|Y\|_1$ thereafter. With Gauss order $m$, $r=2m$, and $c_m=(m!)^4/[(2m+1)((2m)!)^3]$, the substitution $t=Tu$, $s=T(1-u)v$ gives
\begin{align}
 \varepsilon_1&\le c_m\tfrac{15TD^2}{4}(2T\delta)^r,\\
 \varepsilon_2&\le c_m\tfrac{225T^2D^2}{16}\left\{
 [2T(\delta+W)]^r\right.\nonumber\\
 &\hspace{12mm}\left.+r[2T(\delta+W)]^{r-1}+(2TW)^r\right\}.
\end{align}
The middle term comes from the Jacobian $1-u$; the sum is the two successive one-dimensional Gauss remainders. The full-norm interval is
\begin{equation}
 \sqrt{\max(0,S-\varepsilon_1-\varepsilon_2)}
 \ \le E\le\sqrt{S+\varepsilon_1+\varepsilon_2+B_{>2}},
\end{equation}
where $S=e_1^2+e_2^2$ and $B_{>2}=D^2(e^x-1-x-x^2/2)$. These bounds control the omitted sectors and the quadrature remainder.

At $T=2$, the finite-interval comparison on $[0,2]$ is resolved by two pairs. For pair $(33,20)$ the sector contractions of $A$ and $B$ are respectively $(0.49818,0.38418)$ and $(0.09661,0.01234)$. For pair $(1,38)$ they are $(0.36101,0.28171)$ and $(0.21130,0.31296)$. The latter pair reverses in sector one but not sector two. These quantities are logarithmic contractions across the stated interval, not fitted asymptotic exponents.

\begin{figure}[t]
\includegraphics[width=\linewidth]{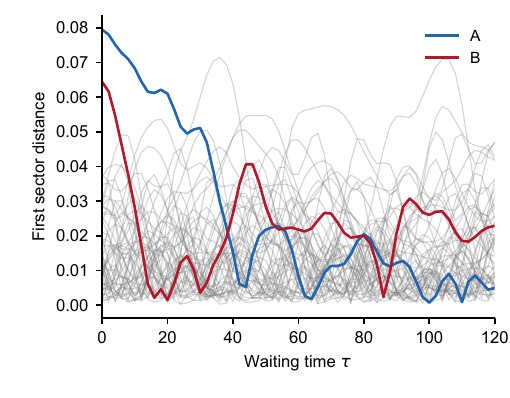}
\caption{Phase-sensitive process relaxation in a common energy shell at $N=8$ and $T=0.02$. The 50 gray curves are independent random-phase preparations; colored curves are the pair in Fig.~\ref{fig:n8}. Every curve has the same shell populations and reference process. Their separation resolves preparation information carried by energy coherences.}
\label{fig:ensemble}
\end{figure}

\input{eth_four_sector_methods.tex}

\section{Finite spectra and recurrence of the whole process}
\label{app:recurrence}
For a finite closed system, group the initial operator by distinct Bohr frequencies: $X_g=\sum_{E_a-E_b=g}P_aXP_b$, $v_g=\mathcal F_T(X_g)$. Then
\begin{equation}
 \Delta\Upsilon(\tau)=\sum_g e^{-ig\tau}v_g,\qquad
 \overline{\|\Delta\Upsilon\|^2}=\sum_g\|v_g\|^2.
\end{equation}
Grouping precedes the norm, so degeneracies are treated exactly. A single
visible nonzero $v_g$ prevents pointwise convergence to zero. Simultaneous
recurrence of the finitely many phases gives arbitrarily late times with
$\mathcal U_s\rho$ close to $\rho$ in trace norm. Quantitatively, put
$\delta(s)=\inf_\phi\|U_s-e^{i\phi}I\|$. Then
\begin{equation}
 \|\Upsilon_{s,T}^\rho-\Upsilon_{0,T}^\rho\|
 \le2\sqrt{e^x-1}\,\delta(s).
 \label{eq:process-recurrence}
\end{equation}
For an initial preparation margin $\Delta=E_A(0)-E_B(0)>0$, the initial
ordering is restored whenever
$\delta(s)<\Delta/[4\sqrt{e^x-1}]$. Recurrence therefore acts on the
full hierarchy and can undo a previous finite-time ordering inversion.

The number of distinct energies also gives a coarse recurrence estimate.
For any $t_0>0$ and integer $Q>1$, simultaneous Dirichlet approximation
of $(E_a-E_1)t_0/(2\pi)$ supplies an integer $1\le q\le Q^{m-1}$ with
$\delta(qt_0)\le2\pi/Q$. Thus a recurrence with the corresponding
full-process error occurs in $[t_0,t_0Q^{m-1}]$. The estimate depends on
the entire finite spectrum; a local equilibration envelope can persist
over much shorter times.

\begin{figure}[t]
\includegraphics[width=\linewidth]{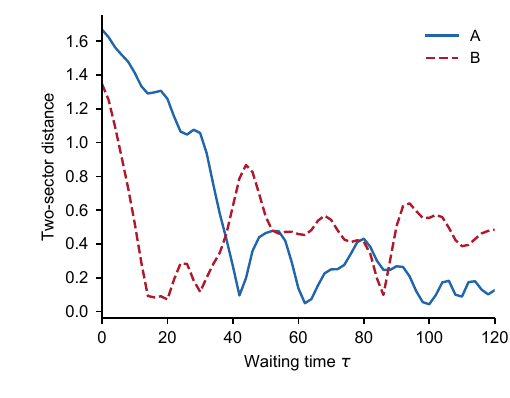}
\caption{Finite-size cPT envelopes over a long waiting interval. The distance is the norm of the first two sectors at $T=2$. Nonmonotone revivals resolve finite-size temporal structure; the exact late-time recurrence follows from the finite Bohr spectrum.}
\label{fig:recurrence}
\end{figure}

If the nonzero directed gaps within a phase-preparation shell are nondegenerate, their mean-square sector distances reduce to
\begin{equation}
 \overline{e_{\theta,n}^2}=\frac1{K^2}\sum_{a\ne b\in I}
 \|F_{T,n}(|E_a\rangle\langle E_b|)\|^2.
\end{equation}
The phases cancel from this expression. Thus phase-controlled finite-time ordering crossings need not imply different infinite-time plateaus. Figure~\ref{fig:recurrence} shows the finite-size revival structure; a thermodynamic decay law concerns a different limiting spectrum.

\section{Local thermodynamic limit with multiple insertions}
\label{app:local}
Assume uniformly bounded finite-range interactions on a spin lattice and a fixed accessible region $\Lambda$. Finite propagation speed and quasi-locality~\cite{LiebRobinson1972,Bravyi2006,Nachtergaele2006} give the Lieb--Robinson finite-volume comparison
\begin{equation}
 \|\alpha_t^{(N)}(A)-\alpha_t(A)\|
 \le C_\Lambda\|A\|e^{-\mu(d_N-v|t|)},
 \qquad A\in\mathcal A_\Lambda,
\end{equation}
where $d_N$ is the distance to the far boundary. Uniformly bounded terms at that boundary can be included in the same estimate. A buffer region $K_\ell$ gives the analogous error $C_\Lambda\|A\|e^{-\mu(\ell-v|t|)}$.

Each branch of Eq.~\eqref{eq:appkeldysh} contains $2n$ time-evolved local factors. For bounded factors, the product difference telescopes one factor at a time. With the multiplication budget in Eq.~\eqref{eq:appbudget}, it follows that
\begin{equation}
 \|A_{N,\boldsymbol\nu}-A_{\infty,\boldsymbol\nu}\|
 \le2nC_\Lambda R^ne^{-\mu(d_N-vT)}.
\end{equation}
All times are at most $T$; adding insertions increases the number of factors, not the total light-cone time. The buffer approximation has the same form with $d_N$ replaced by $\ell$.

Let $\omega_N$ converge locally to $\omega_\infty$ and set
\begin{equation}
 \varepsilon_N(K_\ell)=\sup_{\|A\|\le1,\ A\in\mathcal A_{K_\ell}}
 |(\omega_N-\omega_\infty)(A)|.
\end{equation}
Insert the buffer product between the finite- and infinite-volume products. The resulting three errors are the dynamics difference, the buffer truncation in the state difference, and the local initial-state difference. Since both state functionals have norm one,
\begin{align}
 |C_N^{(n)}-C_\infty^{(n)}|\le R^n[\varepsilon_N(K_\ell)
 +2nC_\Lambda e^{-\mu(d_N-vT)}\nonumber\\
 +4nC_\Lambda e^{-\mu(\ell-vT)}].
\end{align}
Define $A_T=\sqrt{e^x-1}$ and $D_T=e^{x/2}\sqrt{x^2+x}$. Squaring by sector, then using the direct-sum triangle inequality and the exact identity $\sum_{n\ge1}n^2x^n/n!=e^x(x^2+x)$, gives
\begin{align}
 \|\Upsilon_N-\Upsilon_\infty\|
 \le{}&A_T\varepsilon_N(K_\ell)
 +2C_\Lambda D_Te^{-\mu(d_N-vT)}\nonumber\\
 &+4C_\Lambda D_Te^{-\mu(\ell-vT)}.
 \label{eq:applocalbound}
\end{align}
First choose a fixed large $\ell$, then send $N\to\infty$, and finally increase $\ell$. This proves local full-cPT convergence from local convergence of the initial states. Its factorial domination is uniform in $N$. Positivity and causality of bounded finite-order tests pass to the limit from the finite physical dilations.

Retaining a terminal output adds one local factor $\alpha_T(O_a)$ for each member of a fixed system operator basis. The same proof uses $2n+1$ factors and the convergent sum $\sum_n(2n+1)^2x^n/n!$. A waiting time $\tau$ is included by replacing the propagation horizon $T$ with $T+\tau$, giving the spatial limit at every fixed $\tau,T$. Under Eq.~\eqref{eq:appmixing}, this limiting process obeys the long-time filter law derived in Appendix~\ref{app:preps}.

%% file: window_equilibration_proof.tex
\section{Continuous-window equilibration and summation over temporal orders}
\label{app:windowproof}
\subsection{A common autonomous reference}
For $H=\sum_aE_aP_a$, let $\omega=\sum_aP_a\rho P_a$. The finite-dimensional
unitary dilation of $\mathcal F_T(\omega)$ supplies completely positive causal
finite-time marginals, compatible under identity insertions. Purifying
$\omega$ with an inert ancilla gives the corresponding pure dilation.
Strong continuity and bounded insertions give the finite-window cMPS
representation~\cite{Wassner2026,Fuksa2026}. A mixture over window starts
has initial state
\begin{equation}
 \bar\rho_L=\omega+\sum_{a\ne b}
 \frac{1-e^{-i(E_a-E_b)L}}{iL(E_a-E_b)}P_a\rho P_b.
\end{equation}
Thus the averaged process is itself physical, with the same $H$ and
initial state $\bar\rho_L$. At fixed finite dimension $\bar\rho_L\to\omega$
in trace norm. Equation~\eqref{eq:fullbound} proves full-Fock convergence
of this mean process.

\subsection{The mixed-state variance bound}
Pulling the final trace backward through the insertions gives an operator
$A$ with norm at most $R^n$. Set
$v_{ab}=\Tr(AP_a\rho P_b)$ for $a\ne b$. Direct expansion of the squared
Fourier sum gives
\begin{equation}
 \frac1L\int_0^L|\Tr[A(\rho_\tau-\omega)]|^2\,\dd\tau
 \le g_L\sum_{a\ne b}|v_{ab}|^2.
\end{equation}
To bound the last sum for arbitrary mixed $\rho$, purify it to
$|\Psi\rangle$ and define orthonormal vectors
$|\phi_a\rangle=(P_a\otimes I)|\Psi\rangle/\sqrt{p_a}$ for $p_a>0$.
Writing $A_{ba}=\langle\phi_b|A\otimes I|\phi_a\rangle$, one has
$|v_{ab}|^2=p_ap_b|A_{ba}|^2$. The inequality
$p_ap_b\le(p_a^2+p_b^2)/2$ gives
\begin{align}
 \sum_{a\ne b}|v_{ab}|^2
 &\le\frac12\sum_{a,b}(p_a^2+p_b^2)|A_{ba}|^2\nonumber\\
 &\le\|A\|^2\sum_a p_a^2.
\end{align}
The second step uses the row and column squared-norm bounds for a
compression of $A\otimes I$. Hermiticity of $A$ is unnecessary. This is
the observable equilibration argument of Ref.~\cite{ShortFarrelly2012},
applied to each continuous process coefficient.

Integrating over $\Delta_n(T)$ and summing its $q^n$ species gives
Eq.~\eqref{eq:sector-ms}. Tonelli's theorem permits the remaining
nonnegative sum over all $n$, proving Eq.~\eqref{eq:window-ms}.
Chebyshev's inequality gives the bad-start fraction stated in the main text.
As $L\to\infty$, the finite Gram matrix becomes a direct sum of all-one
blocks, one for each distinct nonzero gap. Hence $g_L\to g$ including
gap degeneracies. The case of a single energy has zero deviation.

\subsection{Pointwise and mean-square lifting}
For a family $f_\alpha=\bigoplus_{n\ge1}f_{\alpha,n}$ with
$\|f_{\alpha,n}\|^2\le C^2x^n/n!$, fix $\epsilon>0$ and choose $K$ so that
$C^2\sum_{n>K}x^n/n!<\epsilon/2$. Fixed-sector convergence makes the sum
through $K$ smaller than $\epsilon/2$ for sufficiently large $\alpha$.
This proves Theorem~\ref{thm:lift}. Replacing each squared sector norm by
its time average proves the mean-square version with the same tail.
If $\|f_n(\tau)\|\le b_nr(\tau)$ and $\sum_nb_n^2<\infty$, summation gives
$\|f(\tau)\|\le\|b\|_{\ell^2}r(\tau)$. A nonzero projection with a
matching lower bound supplies a two-sided full-norm rate.
For a quantitative sector cutoff, the ratio of successive omitted
terms gives, whenever $K+2>x$,
\begin{equation}
 \sum_{n>K}\frac{x^n}{n!}
 \le\frac{x^{K+1}}{(K+1)!}\frac1{1-x/(K+2)}.
\end{equation}
This determines a finite temporal order for any prescribed full-norm
truncation tolerance.

For the output-valued hierarchy, expand the final system operator in
matrix units $E_{ab}$. Each scalar coefficient then has a terminal
observable of norm one. Pullback and the same variance estimate bound
it by $g_LR^{2n}/d_{\rm eff}$. Summation over the $d_S^2$ matrix units
and all $n\ge0$ gives Eq.~\eqref{eq:output-ms}, including the
unperturbed terminal state.

An instrument contraction bounded from $\Gamma_T$ into its outcome
$L^1$ space, with norm $K_{\mathcal I}$, obeys
\begin{equation}
 \frac1L\int_0^L\TV(p_\tau^{\mathcal I},p_*^{\mathcal I})^2\,\dd\tau
 \le\frac{K_{\mathcal I}^2g_L}{4d_{\rm eff}}(e^x-1).
\end{equation}
The same bound holds uniformly for a measurable instrument family with
a common contraction norm. A fixed terminal output uses the output-valued
hierarchy described in Appendix~\ref{app:readout}; the scalar--output
bridge integrates over termination time.

\input{finite_bath_equilibration_proof.tex}

%% file: finite_bath_equilibration_proof.tex
\subsection{Exact finite-bath averages and their dimension dependence}
\label{app:finitebath}
Consider Eq.~\eqref{eq:finitebath-preparation}, with $0<\epsilon\le1$,
nonzero $b_j$, and no equalities $b_i=\pm b_j$ for distinct occupied
levels. The initial difference is
\begin{equation}
 X_M=\rho_M-\omega_M=\frac{\epsilon}{2M}
 X_S\otimes\sum_{j=1}^M|b_j\rangle\langle b_j|,
 \qquad \|X_M\|_1=\epsilon.
 \label{eq:finitebath-X}
\end{equation}
Its populations in the joint energy basis are $1/(2M)$, so
$d_{\rm eff}=2M$. Its only off-diagonal energy blocks connect the two
system states at each fixed bath value and have frequencies $\pm2b_j$.

Pauli completeness reduces the double cMPS transfer to four dimensions.
For a bath pair $(b_i,b_j)$ and $\delta_{ij}=b_i-b_j$, define
\begin{align}
 A_{\delta}(z)&=\operatorname{diag}(0,-2i\delta,2i\delta,0)
       +z\left(\mathbf1\mathbf1^{\mathsf T}-I_4/4\right),\nonumber\\
 r_{ij}(z)&=(1,0,0,1)e^{TA_{\delta_{ij}}(z)}.
 \label{eq:finitebath-transfer}
\end{align}
The indices of $r$ below run from zero to three. The same reduction is
used in the continuous-spectrum model of Appendix~\ref{app:edgeproof}.
For complex amplitudes $a=(a_1,\ldots,a_M)$, let
\begin{equation}
 \mathcal Q_z(a)=\frac{\epsilon^2}{4M^2}\sum_{i,j=1}^M
 \left[r_{ij,1}(z)a_i a_j^*
       +r_{ij,2}(z)a_i^*a_j\right].
 \label{eq:finitebath-quadratic}
\end{equation}
The full squared norm is $\mathcal Q_1(a(\tau))$ for
$a_j(\tau)=e^{-2ib_j\tau}$; $[z^n]\mathcal Q_z$ gives its $n$th sector.
This is the exact self-overlap, including all species and insertion times.

Both averaging operations can now be evaluated without a waiting-time
quadrature. Set
\begin{equation}
 k_L(u)=e^{-iuL}\operatorname{sinc}(uL),\qquad
 \operatorname{sinc}v=\frac{\sin v}{v},\quad k_L(0)=1.
 \label{eq:finitebath-sinc}
\end{equation}
Linearity gives
\begin{align}
 \mathsf A_L&=\mathcal Q_1\big(k_L(b_1),\ldots,k_L(b_M)\big),\nonumber\\
 \mathsf B_L&=\frac{\epsilon^2}{4M^2}\sum_{i,j}
 \left[r_{ij,1}(1)k_L(\delta_{ij})
       +r_{ij,2}(1)k_L(\delta_{ij})^*\right].
 \label{eq:finitebath-averages}
\end{align}
For each finite set of nonzero bath values, $k_L(b_j)=O(L^{-1})$.
Consequently $\mathsf A_L=O(L^{-2})$. In the second expression, every
off-diagonal pair vanishes as $L\to\infty$, leaving only $i=j$.

At $\delta=0$, the transfer has a rank-one perturbation of a scalar
matrix. Its two coherence components are equal:
\begin{equation}
 r_{ii,1}(z)=r_{ii,2}(z)
 =\frac12\left(e^{15zT/4}-e^{-zT/4}\right).
 \label{eq:finitebath-diagonal}
\end{equation}
Substitution proves Eq.~\eqref{eq:finitebath-plateau}. Coefficient
extraction also gives the sector-resolved plateau,
\begin{equation}
 \overline{e_n^2}=\frac{\epsilon^2}{4M}
 \left[\left(\frac{15}{4}\right)^n
       -\left(-\frac14\right)^n\right]\frac{T^n}{n!},
 \qquad n\ge1.
 \label{eq:finitebath-sector-plateau}
\end{equation}
Thus the same inverse effective-dimension law holds at each fixed order
and after the complete summation. The surviving diagonal pairs are
the visible frequency atoms of Theorem~\ref{thm:main}.

For a generic bath spectrum, the joint energies $\{b_j,-b_j\}$ have
nonzero-gap degeneracy two. Reflection pairs the gaps
$b_i-b_j=(-b_j)-(-b_i)$, and there are no additional coincidences with
probability one for a GUE draw. Theorem~\ref{thm:window} therefore gives
the infinite-range bound $2(e^{15T/4}-1)/d_{\rm eff}$.
Only the lines $\pm2b_j$ are visible for this preparation, which is why
the exact plateau is smaller than the bound based on every energy gap.

Equation~\eqref{eq:finitebath-X} sharpens the general factorial bound to
\begin{equation}
 0\le E(\tau)^2-\sum_{n=1}^K e_n(\tau)^2
 \le\epsilon^2\sum_{n>K}\frac{(15T/4)^n}{n!},
 \label{eq:finitebath-tail}
\end{equation}
uniformly in the bath dimension, the occupation number, and the window
start. Figure~\ref{fig:ensemble-equilibration}(d) displays this common
bound together with statistics of the sampled maximum tails.
Figure~\ref{fig:large-bath-equilibration}(d) resolves four individual
spectra under the same bound.

\input{large_bath_equilibration.tex}

\input{occupation_scan_methods.tex}

%% file: large_bath_equilibration.tex
\subsubsection{Dimension scaling in fully occupied GUE baths}
The comparisons in Figs.~\ref{fig:ensemble-equilibration}
and~\ref{fig:large-bath-equilibration}
keeps $H_D=Z_S\otimes B_D$, $T=0.8$, and $\epsilon=0.8$.
The GUE normalization is $B_D=(Z+Z^\dagger)/\sqrt{4D}$, with
independent standard real and imaginary Gaussian entries of $Z$.
Here every bath level is occupied: $\eta_D=I_D/D$.
The four bath dimensions are $2^8$, $2^{10}$, $2^{12}$, and $2^{14}$;
including the system gives joint dimensions $512$, $2048$, $8192$,
and $32768$. Figure~\ref{fig:ensemble-equilibration} contains $500$
independent realizations per dimension. Figure~\ref{fig:large-bath-equilibration}
shows one additional fixed spectrum at each dimension, with the same
preparation and spectral normalization.

The initial state and all allowed interventions are invariant under
simultaneous conjugation of the bath Hamiltonian and its basis. Thus
these cMPS contractions depend on the eigenvalues of $B_D$ alone.
They can be sampled from the exact GUE eigenvalue law using the
Hermite tridiagonal representation~\cite{DumitriuEdelman2002}.
In our normalization, its independent entries are
\begin{equation}
 d_i\sim\mathcal N(0,D^{-1}),\qquad
 e_i\sim\frac{\chi_{2(D-i)}}{\sqrt{2D}},\quad 1\le i<D.
 \label{eq:large-bath-sampling}
\end{equation}
This samples the same joint eigenvalue distribution as the dense GUE
matrix with the normalization above. Its correlations,
including level repulsion, are retained.

The large-bath contractions use the same transfer
\eqref{eq:finitebath-transfer}. Let $r(\delta)=r_{ij,1}(1)$ for
$\delta=b_i-b_j$. Equations~\eqref{eq:finitebath-quadratic} and
\eqref{eq:finitebath-averages} reduce to
\begin{equation}
 E_D(\tau)^2=\frac{\epsilon^2}{2D^2}
 \sum_{i,j=1}^D r(b_i-b_j)e^{-2i(b_i-b_j)\tau}.
 \label{eq:large-bath-kernel}
\end{equation}
The sum is real and nonnegative. Only the short window $T$ enters
$r$; the potentially long waiting time occurs in the Fourier phases.
This separation permits a dimension-independent compression of the
transfer kernel.

To see why this compression is controlled, write
\begin{equation}
 \begin{aligned}
 A_\delta(1)&=Q+\frac{15}{4}I_4-2i\delta V,\\
 Q&=\mathbf1\mathbf1^{\mathsf T}-4I_4,\\
 V&=\operatorname{diag}(0,1,-1,0).
 \end{aligned}
 \label{eq:large-bath-markov-kernel}
\end{equation}
Here $Q$ generates a four-state Markov chain, with rate one between
each pair of distinct states. The Feynman--Kac representation of the
transfer, starting in state $1$ and ending in either $0$ or $3$, gives
a positive measure $\xi_T$ such that
\begin{equation}
 \begin{aligned}
 r(\delta)&=\int_{-T}^{T}e^{-2i\delta s}\,\xi_T(\dd s),\\
 \xi_T([-T,T])&=r(0)=\frac{e^{15T/4}-e^{-T/4}}{2}.
 \end{aligned}
 \label{eq:large-bath-positive-kernel}
\end{equation}
Thus $r(b-b')$ is a positive kernel and
$|r^{(p)}(\delta)|\le r(0)(2T)^p$.
For $|b|,|b'|\le B$, tensor Chebyshev interpolation with $p$ nodes
in each variable has the uniform error bound
\begin{equation}
 \|r-r_p\|_\infty
 \le (1+\Lambda_p)\,
       \frac{2r(0)(BT)^p}{p!},
 \label{eq:large-bath-compression}
\end{equation}
where $\Lambda_p$ is the one-variable interpolation Lebesgue constant.
Indeed, the product of the node differences is bounded by
$2^{1-p}B^p$. Applying the derivative bound first in one variable
and then in the other proves Eq.~\eqref{eq:large-bath-compression}.
In the calculations $p=28$ and $B=2.25$. The interval contains every
sampled energy, and $\Lambda_p\le2p$ is sufficient here.

Let $\mathsf T_j$ denote the Chebyshev polynomial of order $j$, and
write the interpolant as
$r_p(b,b')=\sum_{j,k=0}^{p-1}G_{jk}\mathsf T_j(b/B)\mathsf T_k(b'/B)$.
Its coefficient matrix $G$ is positive semidefinite: it is a
congruence transform of the positive kernel matrix on the interpolation
nodes. Consequently
\begin{equation}
 \begin{aligned}
 m_j(\tau)&=\frac1D\sum_{a=1}^D
       \mathsf T_j(b_a/B)e^{-2ib_a\tau},\\
 E_D(\tau)^2&\simeq\frac{\epsilon^2}{2}
             m(\tau)^{\mathsf T}Gm(\tau)^*.
 \end{aligned}
 \label{eq:large-bath-moments}
\end{equation}
The interpolation contribution to the absolute error is at most
$\epsilon^2\|r-r_p\|_\infty/2$, uniformly in $D$ and $\tau$.
Factoring $G$ turns the contraction into a sum of squared Fourier
features. An omitted eigenspace contributes at most
$\epsilon^2p\sum_{\rm omitted}|\lambda_j(G)|/2$.
The all-order norm uses $z=1$ throughout; number resolution through
$n=16$ is computed separately from the coefficients of the same
transfer.

The Fourier features are evaluated on a uniform waiting-time mesh
$h=0.04$ up to $L_{\max}=8D$. Forty-node local interpolation on an
eightfold oversampled Fourier grid evaluates the phase sums. If its
uniform single-phase error is $\eta$, and a feature has spectral
charges $c_{aj}$, the squared-norm error is bounded by
\begin{equation}
 \frac{\epsilon^2}{2}(2\eta+\eta^2)
       \sum_j\left(\sum_a|c_{aj}|\right)^2.
 \label{eq:large-bath-Fourier-error}
\end{equation}
The norm of the averaged process is evaluated directly by replacing
each Fourier phase in $m_j$ with $k_L(b_a)$ from
Eq.~\eqref{eq:finitebath-sinc}.

For the mean squared norm, the time integral is evaluated by composite
Simpson quadrature at even mesh endpoints. The spectral representation
\eqref{eq:large-bath-kernel} has frequencies
$|\omega|\le\Omega=4\max_a|b_a|$ and total variation bounded by
$C=\epsilon^2r(0)/2$. For an individual frequency, the Simpson integral
equals its exact integral times
\begin{equation}
 R_S(\theta)=\frac{\theta(2+\cos\theta)}{3\sin\theta},
 \qquad \theta=\omega h.
 \label{eq:large-bath-Simpson-factor}
\end{equation}
The continuous value is used at $\theta=0$. Since
$|R_S(\theta)-1|\le\theta^4/150$ for $|\theta|\le1/2$ and the
exact frequency integral is bounded by $2/(L|\omega|)$, integration
over the spectral measure gives
\begin{equation}
 \big|\widehat{\mathsf B}_L-\mathsf B_L\big|
 \le \frac{2C\Omega^3h^4}{150L},
 \label{eq:large-bath-quadrature}
\end{equation}
in addition to the kernel and Fourier errors above. The time-average
error therefore decreases with the averaging range.

Figure~\ref{fig:large-bath-equilibration}(a) compares the full cPT
distance with the continuous semicircle limit. Panel (b) resolves the
convergence of the mean process; panel (c) resolves its remaining
finite-size fluctuations. The four mean-square values at $L=8D$ are
$0.992$, $0.982$, $0.980$, and $0.980$ times their respective exact
infinite-time plateaus. The exact plateau in
Eq.~\eqref{eq:finitebath-plateau} falls
from $1.20\times10^{-2}$ at $D=2^8$ to
$1.88\times10^{-4}$ at $D=2^{14}$. In this model, growing the bath
reduces the surviving atomic spectral weight.

The single-realization curves retain oscillations that are smoothed
in the ensemble median. In Fig.~\ref{fig:large-bath-equilibration}(d),
orders above four carry up to $27.3\%$ of the full squared distance
on the waiting grids. Resolving more orders rapidly reduces this
remainder beneath the same dimension-independent bound used for
the ensemble. Figure~\ref{fig:ensemble-equilibration} reports
pointwise medians and $5$th--$95$th percentiles of the distances
computed for each fixed Hamiltonian. In panel (d), the waiting-grid
maximum is taken before the percentile calculation.

\input{large_bath_equilibration_figure.tex}

%% file: large_bath_equilibration_figure.tex
\begin{figure*}[t]
 \includegraphics[width=\textwidth]{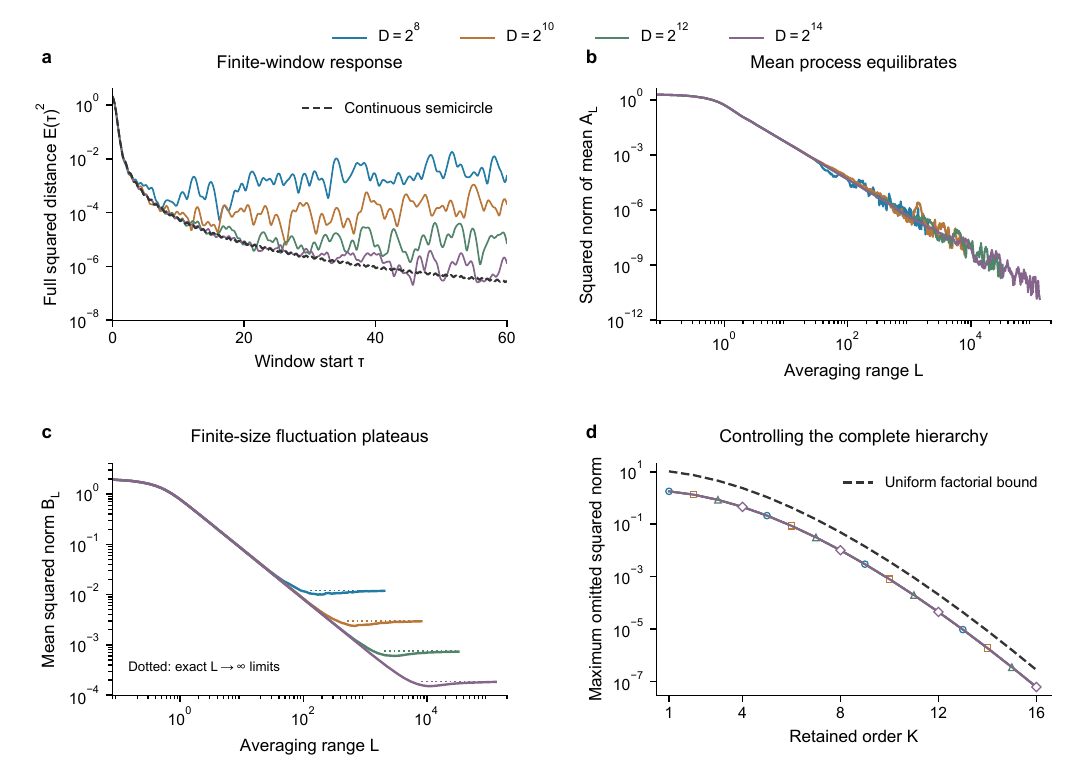}
 \caption{Continuous multitime equilibration for one fixed GUE realization
 at each environment dimension.
 The Hamiltonian is $H_D=Z_S\otimes B_D$, with $B_D$ drawn from the
 GUE at fixed spectral normalization, $T=0.8$, and $\epsilon=0.8$.
 Every bath level is occupied, $\eta_D=I_D/D$, and each dimension
 uses one fixed realization.
 (a) The full squared cPT distance initially follows the continuous
 semicircle response (dashed). Finite-size fluctuations become smaller
 as $D$ increases.
 (b) The squared norm of the mean process, $\mathsf A_L$, approaches
 zero.
 (c) The mean squared norm, $\mathsf B_L$, retains a fluctuation
 plateau proportional to $D^{-1}$. Dotted lines mark the exact
 infinite-time values in Eq.~\eqref{eq:finitebath-plateau}.
 (d) The maximum omitted squared norm above order $K$ over the same
 waiting-time grid. The nearly coincident dimension curves lie below
 the common factorial bound (dashed), which holds for every waiting
 time by Eq.~\eqref{eq:finitebath-tail}. The cutoff needed for a fixed
 absolute accuracy therefore has no bath-dimension dependence.
 All curves use the complete $z=1$ cMPS self-contraction.
 Panel (a) contains 1201 window starts. Panels (b,c) extend to
 $L=8D$ with $960$--$1184$ averaging ranges per dimension.
 Panel (d) resolves cutoffs $K=1,\ldots,16$; markers distinguish the
 four dimensions along their overlapping curves.
 These fixed spectra resolve individual oscillations underlying the
 ensemble statistics of Fig.~\ref{fig:ensemble-equilibration} and are
 also used in the occupation scans of
 Fig.~\ref{fig:effective-dimension-scans}.}
 \label{fig:large-bath-equilibration}
\end{figure*}

%% file: occupation_scan_methods.tex
\subsubsection{Energy occupation within each fixed bath}
\begin{figure*}[t]
 \includegraphics[width=\textwidth]{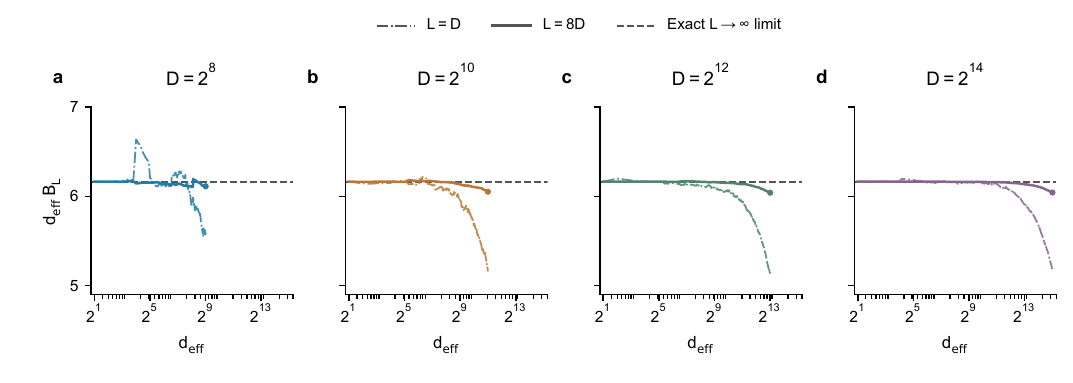}
 \caption{Effective energy dimension within four fixed GUE environments.
 Panels (a)--(d) use $D=2^8,2^{10},2^{12},2^{14}$, respectively,
 with the same Hamiltonians, $T=0.8$, and $\epsilon=0.8$ as
 Fig.~\ref{fig:large-bath-equilibration}.
 Each panel includes every $M=1,\ldots,D$, giving $256$, $1024$,
 $4096$, and $16384$ occupation values, with $d_{\rm eff}=2M$.
 The ordinate $d_{\rm eff}\mathsf B_L$ exposes departures from the
 inverse-dimension law: its exact infinite-range value is the common
 constant $C_T$ in Eq.~\eqref{eq:occupation-rescaling} (dashed).
 Dash-dotted curves use $L=D$; solid curves use $L=8D$.
 The solid endpoint marks full occupation.
 The finite-range excursions arise from off-diagonal gap terms in
 Eq.~\eqref{eq:occupation-pair-sum}. Their reduction at the longer
 averaging range shows the approach to the occupation-controlled
 fluctuation plateau. All values are full $z=1$ cMPS contractions.}
 \label{fig:effective-dimension-scans}
\end{figure*}

The occupation scans in Fig.~\ref{fig:effective-dimension-scans} use
the same four Hamiltonians as Fig.~\ref{fig:large-bath-equilibration}.
A fixed random permutation of the bath levels defines nested occupied
sets $\eta_M=M^{-1}\sum_{j=1}^M|b_{\pi(j)}\rangle\langle b_{\pi(j)}|$.
Every occupation number $M=1,\ldots,D$ is included. The reference for
each preparation is its energy-dephased state in
Eq.~\eqref{eq:finitebath-preparation}.
Keeping the spectrum fixed makes occupation the sole varying
preparation parameter. These scans therefore complement the
independent-Hamiltonian statistics of
Fig.~\ref{fig:ensemble-equilibration}.

The finite-range mean square is obtained from the analytic sinc
average of the same double-cMPS kernel. Write
\begin{equation}
 \begin{aligned}
 W_{ij}^{(L)}&=r(b_{\pi(i)}-b_{\pi(j)})
                 k_L(b_{\pi(i)}-b_{\pi(j)}),\\
 S_M(L)&=\sum_{i,j\le M}W_{ij}^{(L)}.
 \end{aligned}
 \label{eq:occupation-pair-sum}
\end{equation}
Hermiticity gives the recursion
\begin{equation}
 \begin{aligned}
 S_M(L)&=S_{M-1}(L)+r(0)
                  +2\operatorname{Re}\sum_{j<M}W_{Mj}^{(L)},\\
 \mathsf B_L(M)&=\frac{\epsilon^2}{2M^2}S_M(L),\qquad S_0(L)=0.
 \end{aligned}
 \label{eq:occupation-prefix}
\end{equation}
Thus one set of pair contractions resolves the full occupation scan.
The sinc factor performs the time integral directly. Since
$|k_L|\le1$, the kernel-compression bound
\eqref{eq:large-bath-compression} also controls these averages
uniformly in $M$ and $L$.

Multiplication by $d_{\rm eff}=2M$ removes the exact inverse-dimension
trend:
\begin{equation}
 d_{\rm eff}\mathsf B_\infty(M)
 =C_T,\qquad
 C_T=\frac{\epsilon^2}{2}
          \left(e^{15T/4}-e^{-T/4}\right).
 \label{eq:occupation-rescaling}
\end{equation}
The horizontal dashed lines in all four panels represent this same
constant. The finite-range curves show the off-diagonal energy-gap
interference that precedes the plateau. At small occupation they stay
near $C_T$, with spectral excursions most visible for $D=2^8$.
Broader occupation gives a systematic suppression at $L=D$ in these
realizations. Increasing the range to $L=8D$ brings every scan within
$2.1\%$ of the inverse-dimension plateau over its entire occupation
range. The comparison separates the two spectral inputs to
Theorem~\ref{thm:window}: occupied energy weight sets the exact residual,
while finite-range gap interference controls the approach to it.

%% file: eth_four_sector_methods.tex
\subsection{Sparse many-body dynamics and four temporal orders}
\label{app:ethfour}
The local preparations in Eq.~\eqref{eq:localpreps} allow a size comparison
without selecting new energy shells. At bath sizes $N=8,10,12,14,16$, set
\begin{equation}
 X_{A,N}=\frac{Q_2}{2d_N},\qquad X_{B,N}=\frac{Q_1}{d_N},\qquad
 d_N=2^{N+1}.
\end{equation}
Both preparations have the exact tracial energy-dephased state. The full
$H_N$ evolves these differences throughout the waiting interval. Their
observable contrast is local at preparation and remains finite with
increasing volume, unlike a perturbation diluted over a selected basis
vector of the entire bath.
The reference is the infinite-temperature Gibbs state of the interacting
Hamiltonian.

During the short observation window, contract the process on the region
$S,1,\ldots,r$, with $r=4$, retaining its exact reduced initial difference
$X_{j,N}^{(r)}(\tau)$. The restricted Hamiltonian keeps every term inside
this region and removes its coupling to site $r+1$. In particular, its
right endpoint has no artificial $-0.25Z_r$ field. The number-resolved
self-contraction uses the dual cMPS Gram recursion
\begin{align}
 \dot G_0&=Q_r^\dagger G_0+G_0Q_r,\qquad
 G_0(0)=|I\rangle\langle I|,\nonumber\\
 \dot G_n&=Q_r^\dagger G_n+G_nQ_r
 +\sum_\nu J_\nu^\dagger G_{n-1}J_\nu,\qquad G_n(0)=0,
 \label{eq:ethfour-Gram}
\end{align}
where $Q_r$ is the restricted Liouville generator. Thus
$\widetilde e_{j,n}^2=\operatorname{vec}(X_{j,N}^{(r)})^\dagger
G_n(T)\operatorname{vec}(X_{j,N}^{(r)})$ includes all $15^n$ species and
all ordered insertion times, for $n=1,2,3,4$. It is a cMPS
self-contraction, not a selection of discrete measurement strings.

\subsubsection{A uniform short-window locality bound}
The window approximation has an explicit error independent of $N$ and
$\tau$. For any connected interval $I$ containing $S$, the sum of norms
of Hamiltonian terms meeting $I$ is at most $b|I|$, with $b=3.65$.
Starting from a system operator, $p$ nested commutators have support
contained in $S,1,\ldots,p$ and norm at most
$(2b)^pp!$ times the original norm. Intervening system maps enlarge no
support and multiply this bound by at most $R$ each.

Expand all free intervals in nested commutators. If the total order is
$p$, the multinomial sum over its distribution among the intervals has
weight at most $T^p/p!$. Terms through $p=r$ coincide for the full and
restricted Hamiltonians: a contribution sensing the removed bond must
first propagate from $S$ across $r+1$ bonds. Each of the two remaining
tails is bounded by a geometric series. Consequently, for $z=2bT<1$,
\begin{align}
 |C^{(n)}-\widetilde C^{(n)}|
 &\le R^n\|X\|_1\eta_r(T),\nonumber\\
 \eta_r(T)&=\frac{2z^{r+1}}{1-z},\nonumber\\
 |e_n-\widetilde e_n|
 &\le\|X\|_1\eta_r(T)\sqrt{\frac{x^n}{n!}}.
 \label{eq:ethfour-locality}
\end{align}
The same argument applies to a mixed state correlated with the excluded
region, because the final bound uses its trace norm. Only the observation
window is restricted; the waiting dynamics still contains every bath spin.
At $T=0.02$, $r=4$, $\eta_4=1.55359\times10^{-4}$.
Using $\|X_A\|_1\le0.73875$ and $\|X_B\|_1\le0.3$ gives the
sector-norm uncertainties in Table~\ref{tab:ethfour-errors}.

\begin{table}[t]
\caption{Uniform locality bounds $|e_{j,n}-\widetilde e_{j,n}|$ for
the four-sector many-body profiles, at $T=0.02$ and $r=4$. The bounds
hold for every waiting time and every $N\ge r+1$.}
\label{tab:ethfour-errors}
\begin{ruledtabular}
\begin{tabular}{ccc}
Sector $n$ & Preparation $A$ & Preparation $B$\\
1 & $3.144\times10^{-5}$ & $1.277\times10^{-5}$\\
2 & $6.087\times10^{-6}$ & $2.472\times10^{-6}$\\
3 & $9.624\times10^{-7}$ & $3.909\times10^{-7}$\\
4 & $1.318\times10^{-7}$ & $5.352\times10^{-8}$
\end{tabular}
\end{ruledtabular}
\end{table}

\subsubsection{Random-trace contraction and finite-size profiles}
Fifty independent normalized random-phase vectors estimate the reduced
differences at 41 equally spaced waiting times on $[0,8]$ for $N\le14$,
and at 31 times on $[0,6]$ for $N=16$. For each
vector $|r_k\rangle$, propagate both $|r_k\rangle$ and $Q_j|r_k\rangle$
under the full sparse Hamiltonian. The Hermitian partial trace of their
dyad has expectation $X_{j,N}^{(r)}(\tau)$, with the factor $1/2$ included
for $A$. Its exactly known zero trace is imposed on each sample.
Writing these reduced samples as $x_k$, the squared-norm estimator is
\begin{equation}
 \widehat e_n^2=\frac{1}{K(K-1)}
 \sum_{k\ne l}\operatorname{Re}(x_k^\dagger G_nx_l),\qquad K=50.
 \label{eq:ethfour-estimator}
\end{equation}
Independent sample pairs make this estimator unbiased. Delete-one
jackknife errors quantify random-trace uncertainty; the deterministic
locality bound \eqref{eq:ethfour-locality} is applied separately.
Figure~\ref{fig:ethfour} shows the four orders after removing their
factorial weights. The initially visible contrast in $A$ rapidly falls,
while $B$ develops a delayed system response from its initial
system--bath correlation. This ordering change is present throughout
the four computed orders and persists across the sampled bath sizes.

For $N=8$, direct density-matrix evolution also gives the local-window
norms without trace sampling. Combining those values with
Eq.~\eqref{eq:ethfour-locality} and the omitted-sector bound
\begin{equation}
 \sum_{n>4} e_{j,n}^2\le\|X_j\|_1^2
 \left(e^x-\sum_{n=0}^4\frac{x^n}{n!}\right)
\end{equation}
gives full-process intervals
\begin{equation}
\begin{array}{c|cc}
 \tau&E_A(\tau)&E_B(\tau)\\\hline
 0 &[0.006441,0.006507]&[0.000135,0.000168]\\
 2 &[0.000193,0.000278]&[0.004776,0.004804].
\end{array}
 \label{eq:ethfour-full-crossing}
\end{equation}
This is an inversion of process-response ordering for the local
preparation family. The delayed rise of $B$ distinguishes it from a
monotone decay comparison. The equal-population shell example in
Eq.~\eqref{eq:fullintervals} gives a complementary crossing with both
endpoint distances smaller than their initial values.

\subsubsection{Fixed-control probabilities}
For Fig.~\ref{fig:ethoperational}, contract the reduced waiting-time
boundary with the coherent instrument cMPS for
\(h_{\rm c}=0.3Z_S+0.2X_S\) and the six terminal effects in
Eq.~\eqref{eq:ethoperationalpovm}. The control changes the local transfer
generator to
\begin{equation}
 Q_r+\sum_\nu a_\nu J_\nu
 =-i[H_r+h_{\rm c}\otimes I,\,\cdot\,] .
\end{equation}
Here \(a_\nu\) are the coefficients of the system control generator in
the canonical insertion basis. Applying the same contraction to the
tracial reference gives \(p_{*,a,\pm}=1/6\). The full-chain output
distance and its local approximation obey
\begin{equation}
 |D_{\rm TV}^{(j)}-\widetilde D_{\rm TV}^{(j)}|
 \le\tfrac12\|X_j\|_1\eta_{r,\rm c}(T),
 \label{eq:ethcontrolledlocality}
\end{equation}
where \(\eta_{r,\rm c}\) is Eq.~\eqref{eq:ethfour-locality} with
\(b\) replaced by \(b+\|h_{\rm c}\|\). To obtain this bound, sum the
probability differences with their maximizing signs. The resulting
terminal observable has operator norm at most one, so the same
nested-commutator estimate applies. The two TV errors are at most
\(9.35\times10^{-5}\) and \(3.80\times10^{-5}\) for \(A\) and \(B\),
respectively. Exact reduced states are used at \(N=8\); for the larger
sizes the probabilities use the same random-trace samples as the sector
calculation. The initial reduced boundaries are known analytically.

%% file: random_matrix_appendix.tex
\section{Random-matrix preparations, cMPS sectors, and visible frequencies}
\label{app:random}
\subsection{A fixed Hamiltonian and a common stationary process}
For Eq.~\eqref{eq:randomH}, order the independent uniform bath energies and draw $R_{ab}$ uniformly on $[-1,1]$. Set
\begin{align}
 \widetilde B_{ab}&=\frac{R_{ab}+R_{ba}}2
 \mathbf1_{0<|\epsilon_a-\epsilon_b|<0.5},\\
 B&=\widetilde B\big/\sqrt{\Tr(\widetilde B^2)/D}.
\end{align}
This implements the band restriction in Ref.~\cite{DowlingRelaxation2023}, with a dimension-independent root-mean-square coupling. The waiting-time evolution and every insertion use this same Hamiltonian.

For the shell examples, $I$ consists of the $K=12$ interacting eigenstates closest to energy $0.5$. For $O=P_IOP_I$, define $Q_m=(-i\operatorname{ad}_{H_D})^mO$. Its energy-basis entries are
\begin{equation}
 (Q_m)_{ab}=[-i(E_a-E_b)]^mO_{ab}.
 \label{eq:randomfilter}
\end{equation}
The matrix is Hermitian, has zero trace, and has zero restriction to every degenerate energy block for $m\ge1$. Hence $\mathcal D_{H_D}(Q_m)=0$. With
\begin{equation}
 \rho=\omega_I+\frac{0.9s}{K\|Q_m\|}Q_m,\qquad 0\le s\le1,
 \label{eq:randompositivity}
\end{equation}
the spectrum on the shell is bounded below by $(1-0.9s)/K$; the state is zero outside the shell. These statements prove positivity, normalization, and the common reference without imposing a weak-coupling approximation.

For the shell comparisons, draw independent complex Gaussian $K\times K$ matrices $G_r$ and put $O_r=(G_r+G_r^\dagger)/2$ in the energy basis. The $D=48$ pair uses $(O_{14},m_A=1,s_A=0.640150145158043)$ and $(O_{15},m_B=2,s_B=1)$. Both states are then evolved for a common preparation delay $35$ before defining the waiting-time origin in Fig.~\ref{fig:randomselective}. Equivalently, replace $O_r$ by $\mathcal U_{35}(O_r)$ in Eq.~\eqref{eq:randompositivity}; this preserves the energy populations, positivity, and spectral weights. The model and preparation draws use fixed seeds $20261102$ and $20262102$, respectively. Each complex entry is the sum of two independent unit-variance real Gaussian draws, and the matrices are generated in increasing $r$.

\begin{figure*}[t]
 \includegraphics[width=\textwidth]{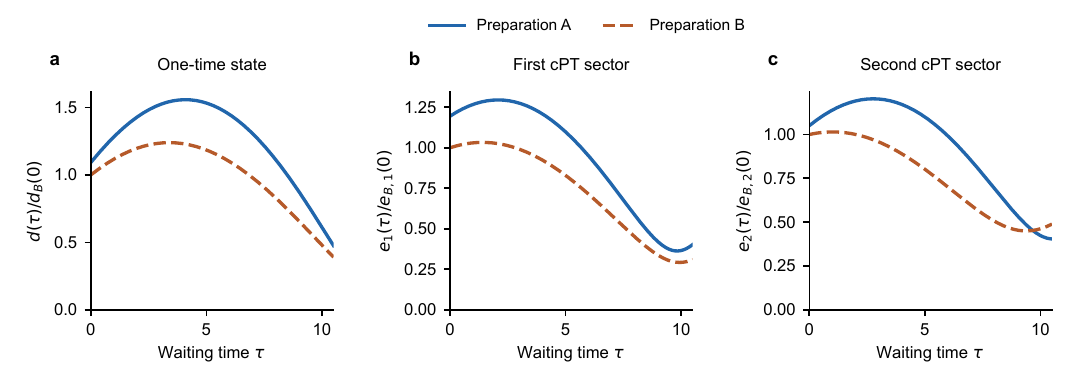}
 \caption{A second sector-selective rate comparison at $D=24$, $K=12$, and $T=4$. Distances are normalized by the corresponding initial distance of $B$. Preparation $A$ contracts more slowly in the state and first-sector distances and more rapidly in the second sector. Only the second sector crosses on the specified interval.}
 \label{fig:random24}
\end{figure*}

At $D=24$, the pair $(O_{20},m_A=2,s_A=0.325620590580897)$ and $(O_{15},m_B=1,s_B=1)$ uses the same construction with no preparation delay. On $[0,10.5]$ it gives
\begin{equation}
 \begin{array}{c|cc}
  &\kappa_A&\kappa_B\\\hline
  \text{state}&0.07910&0.08957\\
  n=1&0.10351&0.11170\\
  n=2&0.09092&0.06817
 \end{array}.
\end{equation}
The second-sector distances change from $(0.07365,0.07015)$ to $(0.02835,0.03429)$. The state and first-sector distances retain their initial ordering throughout this interval (Fig.~\ref{fig:random24}). Thus the selective rate ordering occurs in two fixed environments with different dimensions and preparation pairs.

\subsection{Mixed boundaries and the exact species sums}
Diagonalize a traceless preparation difference as $X=\sum_r w_r\ket{u_r}\bra{u_r}$, allowing signed $w_r$. The cMPS boundary depends linearly on $X$, so these components must be added before forming the overlap. At a first insertion time $t$, write $u_r(t)=U_{\tau+t}u_r$ and
\begin{equation}
 Z_{\mu\nu}(t,s)=\sum_r w_r\Tr_E\!\left[
 U_s\sigma_\mu\ket{u_r(t)}\bra{u_r(t)}\sigma_\nu U_s^\dagger\right].
\end{equation}
Pauli completeness evaluates the two Dyson coefficients of the doubled transfer as
\begin{align}
 e_1^2&=2\int_0^T\!\|\Tr_E\mathcal U_{\tau+t}(X)\|_2^2\dd t,\\
 e_2^2&=\int_0^T\!\dd t\int_0^{T-t}\!\dd s
 \sum_{(\mu,\nu)\ne(0,0)}\!\left[
 \tfrac12\|Z_{\mu\nu}\|_2^2-\tfrac1{16}|\Tr Z_{\mu\nu}|^2\right].
 \label{eq:randommixed}
\end{align}
These are the first two number coefficients of Eq.~\eqref{eq:appsectors}, with all local species summed. In particular, they are independent of a chosen experimental tester.

For a traceless Hermitian qubit operator $Y$, $\|Y\|_1^2=2\|Y\|_2^2$. Substitution in the first line proves Eq.~\eqref{eq:firststateidentity}. The second line retains insertion-generated correlations that the first line integrates out.

\subsection{Continuous-order bounds}
The one-time distance is Lipschitz with
\begin{equation}
 |d_j'(t)|\le\tfrac12\|[H_D,X_j]\|_1=\ell_j.
\end{equation}
Consequently a grid of spacing $h$ certifies $d_A-d_B>0$ on the whole interval whenever its smallest sampled difference exceeds $(\ell_A+\ell_B)h/2$. Equation~\eqref{eq:firststateidentity} similarly gives the Lipschitz bound $2\sqrt T\ell_j$ for $e_{j,1}$.

To bound its integral, let $c_{j,r}=\|\operatorname{ad}_{H_D}^rX_j\|_1$, $r=0,1,2$. Since $d_j^2=\tfrac12\|Y_j\|_2^2$ for $Y_j=\Tr_E\mathcal U_tX_j$,
\begin{equation}
 |(d_j^2)''|\le c_{j,1}^2+c_{j,0}c_{j,2}.
\end{equation}
Composite trapezoidal integration therefore bounds the error in $e_{j,1}^2$ by
\begin{equation}
 \frac{Th^2}{3}(c_{j,1}^2+c_{j,0}c_{j,2}).
\end{equation}
Combining the integral remainder with the Lipschitz extension, at $h=10^{-3}$ the $D=48$ pair has positive lower bounds $7.90\times10^{-4}$ and $1.80\times10^{-3}$ for $d_A-d_B$ and $e_{A,1}-e_{B,1}$ throughout $[0,32.5]$. The $D=24$ pair has corresponding bounds $4.28\times10^{-4}$ and $1.61\times10^{-3}$. These estimates connect the continuous waiting-time ordering directly to the state and sector identities.

\begin{figure*}[t]
 \includegraphics[width=\textwidth]{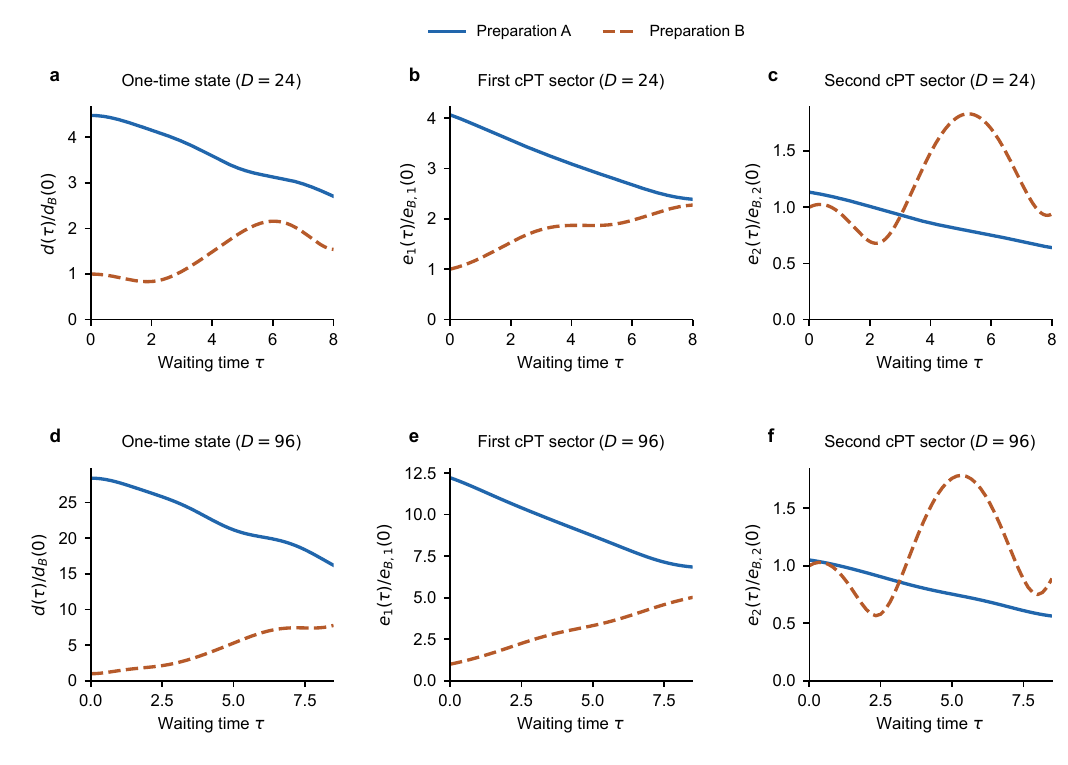}
 \caption{Crossings from local and correlated preparations with $T=4$ and a common infinite-temperature reference. The state and first-sector distances keep their ordering, while the second sector crosses. Each panel is normalized by $B$'s initial distance. The preparation operators have the same form across the two environments.}
 \label{fig:randomcorrelations}
\end{figure*}

\subsection{Exact finite-spectrum residuals and phase-only preparations}
Let $H=\sum_aE_a\ket a\bra a$ have nondegenerate nonzero gaps, and let $X$ have zero energy diagonal. Set $M_{T,n}^{ab}=F_{T,n}(\ket a\bra b)$. Then
\begin{equation}
 F_{T,n}(\mathcal U_\tau X)
 =\sum_{a\ne b}X_{ab}e^{-i(E_a-E_b)\tau}M_{T,n}^{ab}.
\end{equation}
Orthogonality of distinct frequencies under long-time averaging gives
\begin{equation}
 \lim_{L\to\infty}\frac1L\int_0^L e_n(\tau)^2\dd\tau
 =\sum_{a\ne b}|X_{ab}|^2\|M_{T,n}^{ab}\|^2.
 \label{eq:randomresidual}
\end{equation}
The same formula holds in the full Fock norm by the factorial bound. With degenerate gaps, the amplitudes at each common gap are added before squaring, as in Eq.~\eqref{eq:appWiener}.

\begin{proposition}[Phase independence of the averaged residual]
For $\ket{\psi_\phi}=K^{-1/2}\sum_{a\in I}e^{i\phi_a}\ket a$ and $X_\phi=\ket{\psi_\phi}\bra{\psi_\phi}-\omega_I$, the infinite-time mean-square residual of every sector, and of the full cPT, is independent of $\phi$, provided the nonzero shell gaps are nondegenerate.
\end{proposition}
\begin{proof}
The diagonal entries vanish and $|(X_\phi)_{ab}|^2=K^{-2}$ for $a\ne b$ in the shell. Substituting in Eq.~\eqref{eq:randomresidual} gives $K^{-2}\sum_{a\ne b}\|M_{T,n}^{ab}\|^2$, which contains no phase. The full-space statement follows by the same argument before sector projection.
\end{proof}
Finite-time phase-dependent interference and a changed infinite-time residual are therefore separate preparation effects.

For mixed commutator preparations the exact contribution of a visible gap is instead weighted by $|\alpha_j|^2|E_a-E_b|^{2m_j}|(O_j)_{ab}|^2$. A useful finite-band diagnostic is
\begin{equation}
 W_{j,n}(\omega_c)=\sum_{0<|E_a-E_b|\le\omega_c}
 |(X_j)_{ab}|^2\|M_{T,n}^{ab}\|^2.
 \label{eq:slowband}
\end{equation}
It measures the squared amplitudes of the slow visible frequencies, rather than fitting an exponential to an oscillatory curve. The first-sector atom has the closed expression
\begin{equation}
 \|M_{T,1}^{ab}\|^2=2T\|\Tr_E\ket a\bra b\|_2^2,\qquad a\ne b.
\end{equation}
The second-sector atom follows from Eq.~\eqref{eq:randommixed} with boundary $\ket a\bra b$. The fixed-frequency phase cancels from its squared norm, leaving one integral weighted by $T-s$.

The commutator factor $(E_a-E_b)^{m_j}$ suppresses low-frequency weight in every sector. Spectral-edge weights provide the other input to the Fourier asymptotics.

\subsection{Explicit local and correlated preparation operators}
An additional realization replaces Gaussian shell seeds by
\begin{equation}
 O_A=Z_S\otimes I_D,\qquad
 O_B=Y_S\otimes(H_E-\bar\epsilon I_D),
\end{equation}
uses $m_A=m_B=2$, and takes $\omega=I_{2D}/(2D)$. The amplitude factors in Eq.~\eqref{eq:randompositivity} are $(s_A,s_B)=(0.05235645,1)$ at $D=24$ and $(0.03472839,1)$ at $D=96$. Figure~\ref{fig:randomcorrelations} shows second-sector crossings while the state and first sector keep their ordering. At $D=24$, the second-sector interval contractions are $(0.07145,0.00776)$ on $[0,8]$; at $D=96$ they are $(0.07338,0.01392)$ on $[0,8.5]$. These examples distinguish selective crossing visibility from the selective rate ordering in Eq.~\eqref{eq:randomrates}: the first sector also contracts faster for $A$ in these correlated-operator examples.

%% file: semicircle_appendix.tex
\section{All-sector spectral-edge theorem and its autonomous dilation}
\label{app:edgeproof}
Write $\beta_{\ell}=\ell+3/2$ and let $\mu_{\rm sc}$ be the probability measure
in Eq.~\eqref{eq:semicirclemeasure}. All statements in this appendix concern
the $D\to\infty$ limiting process unless a finite $D$ is explicitly indicated.

\subsection{Convergence of the physical processes}
The functions $g_0=1$ and $g_1=(1-b^2/4)_+$ are bounded and continuous.
The preparations in Eq.~\eqref{eq:wddephasing} are positive, normalized,
and have environment marginal $I_D/D$.
In the bath eigenbasis the nonstationary part is
$\epsilon_{\ell} g_{\ell}(b)X_S/2$. Each Pauli species
$R_{\mu\nu}(X)=\sigma_\mu X\sigma_\nu/2$ sends a system matrix unit to
a single matrix unit times a phase and a factor $1/2$.
The two initial units are complementary throughout a given word.
Their waiting-time phases are $\pm2b\tau$, while subsequent free intervals
supply a shift $bs$ with $|s|\le2T$.
Thus every scalar coefficient is a sum of at most two terms of modulus
$\epsilon_{\ell}2^{-n-1}|\phi_{\ell}(\pm2\tau+s)|$, where
\begin{equation}
 \phi_{\ell}(s)=\int g_{\ell}(b)e^{ibs}\,d\mu_{\rm sc}(b).
\end{equation}
Weak convergence of the empirical measures gives coefficient convergence.
For Gaussian Dyson baths the second moments are bounded along an
almost-sure convergent sequence; this makes their characteristic functions
equicontinuous and convergence uniform on compact time intervals.
Boundedness of $g_{\ell}$ and the factorial estimate then give full Fock-space
convergence on every compact waiting-time interval.

\subsection{Upper and lower bounds in every sector}
Direct integration yields Eq.~\eqref{eq:wdphi}. The Bessel expansion gives,
uniformly for $s$ in any fixed bounded interval,
\begin{equation}
 \begin{split}
 \phi_{\ell}(2\tau+s)={}&A_{\ell}(2\tau)^{-\beta_{\ell}}
 \cos\!\left(4\tau+2s-\frac{\pi\beta_{\ell}}{2}\right)\\
 &+O(\tau^{-\beta_{\ell}-1}),
 \end{split}
 \label{eq:edgeBessel}
\end{equation}
where $A_0=1/\sqrt\pi$ and $A_1=3/(2\sqrt\pi)$.
The matrix-unit paths give Eq.~\eqref{eq:edgeupper}; summing its
factorial weights proves $E_{\ell}=O(\tau^{-\beta_{\ell}})$.

For a lower bound at a fixed order $n\ge1$, take $n-1$ insertions
$R_{zz}$ and a final $R_{x0}$. Conjugation by $Z$ changes the sign of
each off-diagonal matrix unit without changing its orientation.
The corresponding scalar coefficient is exactly
\begin{equation}
 C^{(n)}=(-1)^{n-1}\epsilon_{\ell}2^{-n}
 \phi_{\ell}(2\tau+2t_n).
 \label{eq:edgeword}
\end{equation}
Integrating the earlier $n-1$ times therefore gives
\begin{equation}
 e_{\ell,n}^2\ge
 \frac{\epsilon_{\ell}^2}{4^n}
 \int_0^T\frac{t^{n-1}}{(n-1)!}
 |\phi_{\ell}(2\tau+2t)|^2\,dt.
 \label{eq:edgelowerintegral}
\end{equation}
Define
\begin{equation}
 m_{n,T}=\min_{\theta\in[0,2\pi]}
 \int_0^T\frac{t^{n-1}}{(n-1)!}\cos^2(\theta+4t)\,dt.
\end{equation}
Continuity gives a minimum. It is positive: a zero minimum would force a
cosine to vanish almost everywhere on an interval with positive weight.
The uniform remainder in Eq.~\eqref{eq:edgeBessel} is smaller by one
inverse power of $\tau$, so Eq.~\eqref{eq:edgelowerintegral} gives
$e_{\ell,n}=\Omega(\tau^{-\beta_{\ell}})$.
The first-sector lower bound and the summable upper bound prove the
full-norm result.

For the output-retaining coefficients, trace preservation gives
\begin{equation}
 e_{\ell,n}^{\rm scalar}\le\sqrt2\,e_{\ell,n}^{\rm out}.
 \label{eq:edgeoutputlower}
\end{equation}
The same matrix-unit paths, now ending in an operator rather than a trace,
give the upper bound in Eq.~\eqref{eq:edgeupper} also for the output
norm, including $n=0$.
Hence every fixed positive output sector and the complete output norm have
exponent $\beta_{\ell}$. The zero output sector alone has Bessel zeros and
need not have a pointwise two-sided asymptotic bound.

\subsection{Initial ordering in all sectors}
A traced Pauli word is either identically zero or is the sum of the two
complementary matrix-unit paths. Their phases have opposite arguments.
Because $\mu_{\rm sc}$ and $g_{\ell}$ are even, the two Fourier functions agree;
the Pauli prefactors either cancel or add with equal magnitude.
The active set of words is consequently independent of $\ell$.
On an active word the coefficient magnitude is
$\epsilon_{\ell}2^{-n}|\phi_{\ell}(s)|$ at $\tau=0$, with $|s|\le2T$.

The semicircle moments are
$\int b^2\,d\mu_{\rm sc}=1$ and $\int b^4\,d\mu_{\rm sc}=2$.
Using $\cos u\ge1-u^2/2$ gives
\begin{equation}
 \phi_1(s)\ge\int g_1\,d\mu_{\rm sc}
 -\frac{s^2}{2}\int b^2g_1\,d\mu_{\rm sc}
 =\frac34-\frac{s^2}{4}.
\end{equation}
Together with $|\phi_0(s)|\le1$ and
$\epsilon_1/\epsilon_0=4$, this proves coefficientwise domination
by $3-4T^2$. Squaring, integrating, and summing proves
Eq.~\eqref{eq:edgeinitial}. Both sector distances are nonzero by
Eq.~\eqref{eq:edgeword}. For $T=0.4$ the initial ratio is at least $2.36$.

\subsection{A normal-state autonomous realization}
Let $M_b$ be multiplication by $b$ on $L^2(\mu_{\rm sc})$. On
$\mathbb C^2_S\otimes\mathbb C^2_R\otimes L^2(\mu_{\rm sc})$, set
\begin{align}
 H_{\rm dil}&=(Z_S-Z_R)\otimes M_b,\nonumber\\
 |\Omega\rangle&=|\Phi^+\rangle_{SR}\otimes1,\nonumber\\
 |\psi_{\ell}\rangle&=
 \sqrt{I+\epsilon_{\ell}X_S\otimes g_{\ell}(M_b)}\,|\Omega\rangle.
 \label{eq:edgedilation}
\end{align}
The vectors are normalized because
$\langle\Phi^+|X_S|\Phi^+\rangle=0$, and $H_{\rm dil}|\Omega\rangle=0$.
Thus the reference is a stationary normal state of a bounded autonomous
Hamiltonian. All system interventions commute with the $R$ evolution,
which cancels from their Heisenberg products.
At each $b$, the maximally entangled trace identity then gives the
conditional system density $(I+\epsilon_{\ell}g_{\ell}(b)X_S)/2$.
Its integral is precisely the limiting multitime functional.
This establishes the same physical cPT for arbitrary system operations,
including the controlled terminal experiment below.
The construction avoids interpreting a formal continuous diagonal
distribution as a trace-class density operator.

\subsection{Fixed continuous control and exact output probabilities}
\label{app:edgecontrol}
Apply $hY_S$ only during the window.
At fixed $b$, the controlled generator in column vectorization is
\begin{equation}
 Q_b+\mathcal G
 =-i[I\otimes(bZ+hY)-(bZ+hY)^T\otimes I].
 \label{eq:edgecontroltransfer}
\end{equation}
This is the transfer obtained by contracting the process cMPS with the
deterministic coherent control, with terminal boundary
$\langle\!\langle M_{a,\pm}|$.
Its coefficients are $a_{y0}=-2ih$, $a_{0y}=2ih$, so
$K_G^2=e^{8h^2T}$.
The bath integral of the controlled system output has Bloch components
\begin{align}
 \frac{r_{\ell,x}(\tau)}{\epsilon_{\ell}}
 &=\int g_{\ell}(b)c_b\cos(2b\tau)\,d\mu_{\rm sc}\nonumber\\
 &\quad-\int g_{\ell}(b)\frac b{r_b}s_b\sin(2b\tau)\,d\mu_{\rm sc},\nonumber\\
 \frac{r_{\ell,z}(\tau)}{\epsilon_{\ell}}
 &=-\int g_{\ell}(b)\frac h{r_b}s_b\cos(2b\tau)\,d\mu_{\rm sc}\nonumber\\
 &\quad+\int g_{\ell}(b)\frac{hb}{r_b^2}(1-c_b)\sin(2b\tau)\,d\mu_{\rm sc},
 \label{eq:edgeBloch}
\end{align}
where $r_b=\sqrt{b^2+h^2}$,
$c_b=\cos(2r_bT)$, and $s_b=\sin(2r_bT)$.
Parity gives $r_{\ell,y}=0$.
The reference output is $I/2$ and
$p_{\ell}(a,\pm)=(1\pm r_{\ell,a})/6$, hence
$D_{\ell}=(|r_{\ell,x}|+|r_{\ell,z}|)/6$ exactly.

Near the positive edge write $b=2-u$.
The weighted measure is
\begin{equation}
 g_{\ell}(2-u)\,d\mu_{\rm sc}(2-u)
 =\frac1\pi u^{\ell+1/2}[1+O(u)]\,du.
\end{equation}
The negative endpoint has the reflected expansion. Applying the Fourier
endpoint expansion to the smooth coefficient functions in
Eq.~\eqref{eq:edgeBloch} gives Eq.~\eqref{eq:TVellipse}; the interior
contributes a remainder of higher order by integration by parts.
If $\det[u,v]\ne0$, its smallest singular value is positive.
Consequently the leading two-component vector has a positive lower
bound for every phase, and the one-order-smaller remainder preserves it
at all sufficiently large $\tau$. This proves Eq.~\eqref{eq:TVexponents}.

At $\tau=0$, the same moment inequalities give
\begin{align}
 r_{1,x}(0)&\ge
 \epsilon_1\left[\frac34-2T^2
 \left(\frac12+\frac34h^2\right)\right],\nonumber\\
 |r_{0,x}(0)|+|r_{0,z}(0)|
 &\le\epsilon_0(1+2|h|T).
\end{align}
For $T=0.4,h=0.5$ these are $0.424$ and $0.28$.
Also $\cos(0.8\sqrt{4.25})\ne1$.
The initial TV ordering and eventual ratio therefore prove a genuine
crossing for this one fixed control and POVM.

\subsection{Exact continuous contraction used in the figure}
In the double cMPS transfer, Pauli completeness leaves a four-dimensional
invariant subspace. For two bath values $b,b'$ its generator is
\begin{equation}
 A_{b-b'}(z)=
 \operatorname{diag}(0,-2i(b-b'),2i(b-b'),0)
 +z(\mathbf1\mathbf1^T-I_4/4).
\end{equation}
The trace boundary is $(1,0,0,1)$.
Contracting $e^{TA_{b-b'}(z)}$ with the two initial coherences and
integrating $g_{\ell}(b)g_{\ell}(b')\,d\mu_{\rm sc}(b)d\mu_{\rm sc}(b')$
gives the self-overlap; $z=1$ sums every sector.
The measured curves instead use the single process--control transfer
in Eq.~\eqref{eq:edgecontroltransfer} and the fixed POVM boundary.
These are respectively the self-contraction and operational contraction
of the same continuous process.

%% file: mpemba_probability_appendix.tex
\section{Occurrence probabilities for a fixed preparation rule}
\label{app:probability}
\subsection{Compact-time transfer of a strict preparation ordering}
The finite-environment probabilities have a direct relation to the
continuum result. Let $\mu_D$ be the empirical bath spectral measure and
suppose $\mu_D\Rightarrow\mu_{\rm sc}$ almost surely. The preparations,
control, and measurement are fixed functions of the bath coordinate, as
in Eqs.~\eqref{eq:wddephasing} and \eqref{eq:fixedpovm}.

\begin{proposition}[Probability of a compact-window advantage]
\label{prop:occurrence-limit}
For every compact waiting-time interval, the finite-environment full
distances, each fixed sector distance, and the fixed-readout TV distances
converge uniformly almost surely to their semicircle limits. Fix a finite
sector cutoff $K$. If the limiting distances have a strict initial ordering
and a strict reversed ordering on a compact interval $I$, then the
probability that these orderings hold simultaneously for the full norm,
the first $K$ sectors, and the fixed readout tends to one as $D\to\infty$.
For the preparations, $T$, and control used in
Theorems~\ref{thm:edge} and \ref{thm:operational}, such an $I$ exists for
every fixed $K$ and every prescribed finite duration.
\end{proposition}

\begin{proof}
For a sector or the full norm, write the exact doubled cMPS contraction as
\begin{equation}
 d_{j,D}(\tau)^2=
 \iint K_{j,T}(b,b';\tau)\,\dd\mu_D(b)\dd\mu_D(b').
 \label{eq:probability-double-kernel}
\end{equation}
The kernel includes both preparation filters. It is jointly continuous
in $(b,b',\tau)$. Free propagation is unitary, while the insertion
transfer has bounded norm at fixed $T$. Thus the full kernel is bounded
uniformly in bath energies and waiting time by a constant depending on
$T$ and the preparation amplitude. Each fixed-sector kernel has the same
property, with its factorial order factor.

Weak convergence implies convergence of the product measures. To make
the convergence uniform in $\tau$, first restrict both energies to a
compact set carrying all but an arbitrarily small mass of $\mu_D$ and
$\mu_{\rm sc}$. Weak convergence supplies this uniform tightness.
On this set the family of kernels is uniformly continuous in $\tau$.
A finite time net therefore reduces the uniform error to finitely many
weak-convergence errors and a uniformly small continuity error. The
bounded kernel controls the discarded energy tails. Letting the tail
mass and net spacing vanish proves uniform convergence of the squared
distances. The inequality $|\sqrt{x}-\sqrt{y}|\le\sqrt{|x-y|}$ gives
uniform convergence of the distances themselves.

For the fixed readout, each Bloch component is a single integral of a
bounded continuous process--control coefficient against $\mu_D$.
The same tightness and finite-net argument proves uniform convergence
of all three components. Since
$d_{j,D}^{\rm TV}=(|r_x|+|r_y|+|r_z|)/6$, the TV distances converge
uniformly as well. In particular, a finite sample retains $r_y$ even
though the even limiting measure makes it vanish.

There are finitely many distances when $K$ is fixed. Their strict
initial gaps and their continuous, strictly positive reversed gaps on
$I$ have a positive common minimum. Uniform convergence preserves all
these signs for almost every realization at all sufficiently large
$D$. Bounded convergence of the event indicators gives probability
one in the limit. Finally, the initial-order bounds and asymptotic
ratios in Theorems~\ref{thm:edge} and \ref{thm:operational} provide such
an interval $I$ of any fixed finite duration. Its starting time can
depend on $K$.
\end{proof}

The proposition holds the waiting interval and $K$ fixed while taking
$D\to\infty$. It transfers the strict continuum ordering to an
experimental window. Finite-size recurrences determine how far that
window can be extended at a given $D$.

\subsection{Exact finite-size fluctuation competition}
For a finite bath with distinct eigenvalues $b_i$, the diagonal part of
the doubled contraction gives an exact infinite-waiting-time average.
Put
\begin{equation}
 \begin{aligned}
 r_n(0)&=\frac{(15T/4)^n-(-T/4)^n}{2n!},\qquad n\ge1,\\
 r_{\rm full}(0)&=\frac{e^{15T/4}-e^{-T/4}}2.
 \end{aligned}
 \label{eq:filtered-diagonal-kernel}
\end{equation}
For either a sector or the independently summed full norm,
\begin{equation}
 \lim_{L\to\infty}\frac1L\int_0^L d_{j,D}(\tau)^2\dd\tau
 =\frac{\epsilon_j^2r(0)}{2D^2}\sum_{i=1}^D g_j(b_i)^2.
 \label{eq:filtered-finite-plateau}
\end{equation}
Indeed, the exact transfer contains phases
$e^{-2i(b_i-b_{i'})\tau}$; averaging eliminates every $i\ne i'$ term.
At $b=b'$, the reduced insertion matrix
$\mathbf1\mathbf1^{\mathsf T}-I_4/4$ has eigenvalues $15/4$ and $-1/4$.
Its two projections onto the coherence boundary give
Eq.~\eqref{eq:filtered-diagonal-kernel}.

The ratio of the filtered and unfiltered plateaus is independent of
sector and of $T$:
\begin{align}
 \frac{\langle d_{A,D}^2\rangle_\infty}
      {\langle d_{B,D}^2\rangle_\infty}
 &=\frac{16}{D}\sum_i(1-b_i^2/4)_+^2\nonumber\\
 &\longrightarrow 16\int(1-b^2/4)^2\dd\mu_{\rm sc}(b)=10.
 \label{eq:filtered-plateau-ratio}
\end{align}
The final integral is $1-\langle b^2\rangle/2+
\langle b^4\rangle/16=5/8$. Thus the preparation filter suppresses the
continuum tail but leaves a larger finite-size mean-square residual for
the chosen amplitudes. Whenever the finite-$D$ ratio exceeds one, the
filtered distance must return above the unfiltered distance arbitrarily
late: an eventual opposite inequality would contradict
Eq.~\eqref{eq:filtered-plateau-ratio}. The finite-window occurrence
curves measure the interval between these two regimes.

\input{mpemba_probability_table.tex}

\subsection{Finite-environment events and sector coverage}
The random variable is the bath Hamiltonian. The preparation pair is
always $A=(0.8,(1-b^2/4)_+)$ and $B=(0.2,1)$, with the same reference
$I/(2D)$. The positive part preserves positivity for eigenvalues outside
$[-2,2]$. All four dimensions use $500$ independent realizations and
$1201$ waiting times on $[0,60]$. We compute sectors $1$ through $16$
and the full $z=1$ transfer separately. The control and terminal POVM
are identical across realizations.

The sustained event requires the initial raw distance of $A$ to be
larger, followed by $\overline d_A<\overline d_B$ over consecutive grid
points spanning at least two units. The root-mean-square window has
$\Delta=2$. Its admissible starts lie in $[0,58]$, so its entire
integration interval remains inside the sampled trajectory. The
integration uses composite Simpson quadrature on a spacing of $0.0125$,
with further refinement at a classification boundary. The
reported durations are grid-resolved run lengths. A common-interval
event intersects the two inequalities before computing its duration.
This differs from requiring each inequality to persist somewhere in
the trajectory.

Table~\ref{tab:occurrenceGUE} separates instantaneous crossings from
sustained preparation advantages. Every realization has an instantaneous
full-norm crossing, including $D=2^8$, where most advantages disappear
under the root-mean-square comparison. At $D=2^{10}$, requiring an
additional $5\%$ distance reduction changes the full-process count from
$353$ to $279$ and the TV count from $224$ to $156$. Both larger
dimensions retain $500/500$ events under that margin. Their sustained
occurrence estimates have $95\%$ Wilson intervals $[0.9924,1]$.

The longest full-process advantage has median and interquartile range
$2.65$ with $[1.90,3.40]$ at $D=2^{10}$, $9.73$ with $[8.30,11.06]$
at $D=2^{12}$, and $22.15$ with $[20.00,24.05]$ at $D=2^{14}$.
The spectral filter suppresses the continuum edge contribution, while
the discrete spectrum produces residual fluctuations. Increasing $D$
extends the interval in which the filtered contribution stays below
the unfiltered one. This explains the growth of both occurrence
probability and advantage duration in Fig.~\ref{fig:mpembaprobability}.

%% file: mpemba_probability_table.tex
\begin{table*}[!t]
\caption{Occurrence counts out of $500$ fixed GUE realizations per dimension.
All sustained events use Eq.~\eqref{eq:occurrence-envelope} and a duration of at least two units.
The last column requires all $16$ resolved sectors to have an advantage on one common interval.}
\label{tab:occurrenceGUE}
\begin{ruledtabular}
\setlength{\tabcolsep}{5pt}
\renewcommand{\arraystretch}{1.15}
\begin{tabular}{crrrrrr}
$D$ & \shortstack{Instantaneous\\full} & \shortstack{Sustained\\full} & \shortstack{Sustained\\TV} & \shortstack{Common\\full/TV} & \shortstack{Each sector\\$n=1,\ldots,16$} & \shortstack{Common interval\\$n=1,\ldots,16$} \\
\hline
$2^{8}$ & 500 & 3 & 2 & 1 & 3 & 2 \\
$2^{10}$ & 500 & 353 & 224 & 196 & 331 & 278 \\
$2^{12}$ & 500 & 500 & 500 & 500 & 500 & 500 \\
$2^{14}$ & 500 & 500 & 500 & 500 & 500 & 500 \\
\end{tabular}
\end{ruledtabular}
\end{table*}

%% file: dyson_sector_theory.tex
\section{Joint Dyson symmetries and the sector contraction}
\label{app:dyson}
Let $L=2D$, with even $D$, and let $G=(Z+Z^\dagger)/2$, where the entries
of $Z$ are independent centered complex Gaussians. Before subtracting the
trace and fixing $\Tr H^2/L=1$, our draws are
\begin{equation}
 \begin{aligned}
 H_1&=\operatorname{Re}G,\quad H_2=G,\quad
 H_4=\tfrac12(G+JG^*J^\dagger),\\
 J&=I_2\otimes\begin{pmatrix}0&I_{D/2}\\-I_{D/2}&0\end{pmatrix}.
 \end{aligned}
 \label{eq:jointdysondraw}
\end{equation}
The symmetries are imposed on the joint Hamiltonian. In its Pauli
decomposition, the GOE constraint makes $B_x,B_z$ real symmetric and $B_y$
imaginary antisymmetric. In the GSE case, $B_x,B_z$ are even and $B_y$ is odd
under $J_EK$. These correlated constraints preserve the joint symmetry while
allowing noncommuting couplings. The Frobenius normalization fixes the time
unit and preserves the antiunitary constraints.

\subsection{Exact number-resolved transfer}
Use row vectorization and define
\begin{equation}
 \begin{aligned}
 Q&=-i(H\otimes I-I\otimes H^{\mathsf T}),\\
 R_{\mu\nu}&=\tfrac12(\sigma_\mu\otimes I_D)
       \otimes(\sigma_\nu\otimes I_D)^{\mathsf T},
 \end{aligned}
 \label{eq:dysonQR}
\end{equation}
where $(\mu,\nu)\ne(0,0)$. Let $|\ell\rangle=\operatorname{vec}(I_L)$.
The scalar process sector is the ordered-simplex amplitude
\begin{equation}
 c_{\bm\nu}(\bm s;x)=\langle\ell|
 R_{\nu_n}e^{Q(s_n-s_{n-1})}\cdots
 R_{\nu_1}e^{Qs_1}|x\rangle.
\end{equation}
Introduce its Gram matrix $G_n(T)$ by integrating
$c_{\bm\nu}^*c_{\bm\nu}$ and summing all $15^n$ words. A translation of
all insertion times, followed by separating the first insertion, gives
\begin{align}
 \dot G_0&=Q^\dagger G_0+G_0Q,\quad G_0(0)=|\ell\rangle\langle\ell|,\\
 \dot G_n&=Q^\dagger G_n+G_nQ+\mathcal B(G_{n-1}),\quad G_n(0)=0,\\
 \mathcal B(K)&=\sum_{(\mu,\nu)\ne(0,0)}R_{\mu\nu}^\dagger K R_{\mu\nu}.
 \label{eq:dysonGram}
\end{align}
This is the coefficient of order $n$ in the doubled cMPS transfer exponential.
If the local Liouville indices are grouped together, Pauli completeness yields
\begin{equation}
 \mathcal B(K)=I_4\otimes\Tr_{\mathrm{local}\,4}K-\tfrac14K.
 \label{eq:paulicompletenessGram}
\end{equation}
For $x_j(\tau)=\operatorname{vec}[U_\tau(\rho_j-\omega)U_\tau^\dagger]$,
the measured quantity is exactly
\begin{equation}
 e_{j,n}(\tau)^2=x_j(\tau)^\dagger G_n(T)x_j(\tau).
 \label{eq:dysonsectorquad}
\end{equation}
No insertion times or operator words are sampled. The three sectors are
computed as three coefficients of the same continuous transfer.

\subsection{Symmetry selection of the stationary block}
Write $\mathcal D_H(X)=\sum_aP_aXP_a$, retaining complete degenerate energy
blocks. For a simple real-symmetric $H$, any Hermitian operator with
$X^*=-X$ has $\langle a|X|a\rangle=0$ in a real eigenbasis. Consequently
\begin{equation}
 \mathcal D_H(Y_S\otimes I_E)=0\qquad\text{(GOE, simple spectrum)}.
 \label{eq:GOEoddblock}
\end{equation}
For a GSE Hamiltonian with isolated Kramers doublets, an antiunitary-even
Hermitian operator compresses to a scalar in each doublet; an antiunitary-odd
one compresses to a traceless Hermitian $2\times2$ matrix. To prove this,
choose a Kramers basis in the doublet, in which $\Theta=i\sigma_yK$.
Conjugation by $\Theta$ fixes $I$ and reverses the three Pauli matrices.
The even and odd statements follow by expanding the compressed operator.
Here $Z_S\otimes I_E$ is even and $Y_S\otimes I_E$ is odd. The latter can
therefore retain a nonzero stationary block in GSE, unlike
Eq.~\eqref{eq:GOEoddblock}. Its visibility in sector $n$ is determined by
$F_{T,n}\mathcal D_H(X)$.

This selection rule concerns the time-averaged amplitude. Its mean-square
distance also retains all nonzero visible frequencies. For finite $L$, a
nonzero visible finite sum of frequencies cannot converge to zero at every
late time: the time average of its squared norm is the positive sum of the
equal-gap weights. This gives a precise separation between stationary-block
selection, interval acceleration, and asymptotic equilibration.

\subsection{Marked gap measures and ensemble averages}
For one realization write $X_g=\sum_{E_a-E_b=g}P_aXP_b$ and
$v_{n,g}=F_{T,n}(X_g)$. Equal gaps are combined at this stage. Then
\begin{align}
 F_{T,n}(\mathcal U_\tau X)&=\sum_g e^{-ig\tau}v_{n,g},\\
 e_n(\tau)^2&=\sum_{g,h}e^{-i(g-h)\tau}
                   \langle v_{n,h},v_{n,g}\rangle.
 \label{eq:dysonfreqsquare}
\end{align}
The inner product is conjugate linear in its first argument. Define the
complex marked measure
\begin{equation}
 \nu_{n,H}=\sum_{g,h}\langle v_{n,h},v_{n,g}\rangle\,
                   \delta_g\otimes\delta_h.
 \label{eq:dysonmarked}
\end{equation}
Its Fourier transform along $(\tau,-\tau)$ equals the squared sector norm.
A waiting-time average on $[0,A]$ replaces the exponential by
$e^{-i(g-h)A/2}\operatorname{sinc}[(g-h)A/2]$; an infinite average retains
$g=h$. In an ensemble, both the random gaps and their marks are averaged:
\begin{equation}
 \mathbb E_\beta\overline{e_n^2}
 =\mathbb E_\beta\left[\sum_{g\in\mathcal G(H)}\|v_{n,g}\|^2\right].
\end{equation}
The two gaps contain four energy indices. Their marked distribution generally
requires more information than the ordinary two-level correlation function.
For a preparation filter $p(-i\operatorname{ad}_H)$, the mark acquires the
factor $p(-ig)\overline{p(-ih)}$.

\begin{proposition}[Stability under marked-measure perturbations]
Suppose two models have finite averaged marked measures
$\bar\nu_n,\bar\nu'_n$ and
$\|\bar\nu_n-\bar\nu'_n\|_{\mathrm{TV}}\le\eta_n$, where the total variation
is the norm of a complex measure. Then, at every waiting time,
\begin{equation}
 |\mathbb E e_n(\tau)^2-\mathbb E e'_n(\tau)^2|\le\eta_n.
\end{equation}
If $\sum_n\eta_n<\infty$, the corresponding full-Fock squared norms differ
by at most $\sum_n\eta_n$.
\end{proposition}
\begin{proof}
The Fourier kernel in Eq.~\eqref{eq:dysonfreqsquare} has modulus one. The
defining inequality for total variation gives the sector bound; absolute
summability permits the sector sum. For two preparations, an ordering margin
larger than the sum of their two errors is consequently preserved.
\end{proof}
Ordinary weak convergence of unfolded levels supplies neither this marked
measure bound nor the uniform summability needed for the full process.

\subsection{What a symmetry label can classify}
\begin{proposition}[No compulsory reversal for the fixed pair]
For even $D\ge4$, the pair in Eq.~\eqref{eq:dysonfixedpair}, a fixed finite waiting
interval, and sectors $n=1,2,3$, each of the three normalized Dyson
ensembles has a positive-probability set with no distance reversal.
\end{proposition}
\begin{proof}
Choose a normalized traceless $H_0=I_S\otimes H_E$ in the desired symmetry
space. For GSE, choose $H_E$ with Kramers-paired diagonal entries. The
environment is a spectator. Unitary covariance of the complete local
superoperator basis makes the sector norm of $Y_S$ equal to that of $Z_S$.
Thus $e_{A,n}(\tau)/e_{B,n}(\tau)=0.54/0.50=1.08$ for every waiting time
and each active sector. For a finite number of sectors their strict squared
distance margins have a positive minimum. The transfer exponential and the
waiting propagator depend continuously on $H$, uniformly on a compact time
interval. These margins persist on an open neighborhood of $H_0$ in the
normalized symmetry space. Gaussian draws followed by trace subtraction and
norm fixing assign positive probability to every such neighborhood.
\end{proof}
This statement concerns fixed finite dimension and time; typicality in a
large-dimension limit is a separate ensemble question.

The stationary-block rule above is a direct symmetry statement. A Mpemba
classification additionally depends on the preparation and the visible maps.
Indeed, take any traceless Hermitian $X$ and positive preparations
$\rho_A=\omega+aX$, $\rho_B=\omega+bX$, with $a>b>0$. Linearity gives
\begin{equation}
 e_{A,n}(\tau)=\frac ab e_{B,n}(\tau)
\end{equation}
for every active sector and every $\tau$, in every Dyson class. There is no
crossing. Thus a Dyson label alone cannot imply a strong or selective effect.

A useful conditional classification follows when a model has an ordered
relaxation expansion
\begin{equation}
 \begin{gathered}
 F_{T,n}(\mathcal U_\tau X)
 =\sum_{j=1}^r e^{-\lambda_j\tau}V_{n,j}(X)
       +o(e^{-\lambda_r\tau}),\\
 0<\lambda_1<\cdots<\lambda_r.
 \end{gathered}
\end{equation}
The sector rate is the first $\lambda_j$ with $V_{n,j}(X)\ne0$.
Cancellation of the common slowest mode in every active sector is precisely
$X\in\bigcap_n\ker V_{n,1}$; cancellation on a specified subset uses the
intersection over that subset. Nonzero leading coefficients in the competing
preparation then give the corresponding strict rate advantage. A full-norm
rate further requires a summable uniform remainder bound across $n$.
The expansion is a dynamical hypothesis, separate from Dyson symmetry.

\subsection{Spectral form factors and process-visible weights}
For a unitarily invariant ensemble, conjugation covariance makes
$\Phi_t(X)=\mathbb E[U_tXU_t^\dagger]$ scalar on the traceless subspace.
The trace of $\Phi_t$ as a superoperator is $\mathbb E|\Tr U_t|^2$; its
identity eigenvalue is one. The remaining $L^2-1$ equal eigenvalues yield
Eq.~\eqref{eq:formfactortwirl}. In particular,
\begin{equation}
 \begin{aligned}
 \mathbb E\, C^{(1)}_\nu(s;X,\tau)
   &=\alpha_L(\tau+s)\Tr[J_\nu(X)],\\
 \alpha_L(t)&=\frac{\mathbb E|\Tr U_t|^2-1}{L^2-1}.
 \end{aligned}
\end{equation}
For a traceless Hermitian $X$ and the qubit Pauli convention used here,
\begin{equation}
 \|\mathbb E F_{T,1}(\mathcal U_\tau X)\|^2
  =2\|\Tr_E X\|_2^2\int_0^T|\alpha_L(\tau+s)|^2\dd s.
\end{equation}
This identity uses a fixed $X$ independent of the Hamiltonian draw and
concerns the squared mean amplitude. Equation~\eqref{eq:quenchedvariance}
adds the fluctuation term for the mean-square process distance.

Even complete spectral information does not determine a cPT. For two qubits,
the real Hamiltonians $H_0=aZ_S+bZ_E$ and
$H_1=\operatorname{CNOT}H_0\operatorname{CNOT}^\dagger=aZ_S+bZ_SZ_E$
are isospectral. Starting from $|+\rangle\langle+|\otimes I_E/2$, their
system off-diagonal matrix elements are respectively
$e^{-2ia\tau}/2$ and $e^{-2ia\tau}\cos(2b\tau)/2$.
The partition and eigenvectors change the process although every eigenvalue
statistic agrees.

%% file: auxiliary_results.tex
\input{random_matrix_results.tex}

\subsection{Symmetry classes, quenched processes, and spectral averaging}
Theorem~\ref{thm:edge} applies to all three Dyson bath classes because their
empirical one-point measures have the same semicircle limit.
Their antiunitary symmetries and gap correlations instead influence
finite-dimensional fluctuations and recurrences.
The noncommuting comparison tests how joint spectral and eigenvector
structure enters the sector-visible kernels.

For a fixed-realization process deviation $Y_H$, the ensemble mean-square
distance separates into the mean process and its sample fluctuations:
\begin{equation}
 \mathbb E_H\|Y_H\|^2=\|\mathbb E_HY_H\|^2
 +\mathbb E_H\|Y_H-\mathbb E_HY_H\|^2.
 \label{eq:quenchedvariance}
\end{equation}
The second term is the sample-to-sample cPT fluctuation. The
crossings concern fixed Hamiltonians, while the
GOE/GUE/GSE comparison measures variability across realizations. The
almost-sure semicircle limit gives the process of Theorem~\ref{thm:edge}.

For a noncommuting random bath we instead consider
\begin{equation}
 H_{\rm nc}^{(\beta)}=H_S\otimes I+I\otimes H_E
 +\lambda\sum_{\alpha=x,y,z}\sigma_\alpha\otimes B_\alpha^{(\beta)}.
 \label{eq:noncommutingwd}
\end{equation}
The Dyson constraint is imposed on the full $H_{\rm nc}^{(\beta)}$.
Assigning the bath coefficients independently to a Dyson ensemble does not
in general preserve the joint antiunitary symmetry. Our normalized Gaussian
draws obey $H=H^*$ for GOE and $H=JH^*J^\dagger$ for GSE, where
$J=I_S\otimes J_E$ and $J_EJ_E^*=-I_E$; GUE has no imposed antiunitary.
Their unique Pauli decomposition gives Eq.~\eqref{eq:noncommutingwd}.
The joint eigenvectors and the system--environment partition then enter each
visible kernel. The fixed-frequency residual has the exact ensemble decomposition
\begin{equation}
 \mathbb E_\beta\,\overline{e_n^2}
 =\mathbb E_\beta\!\left[
 \sum_{\omega\in\mathcal G(H)}
 \bigl\|M_{n,\omega}^{(\beta)}(X_\omega)\bigr\|^2\right],
 \label{eq:wdkernelaverage}
\end{equation}
with equal-gap amplitudes combined before squaring. Thus a symmetry-dependent
strong or selective effect requires a difference in the joint visible kernels
$M_{n,\omega}^{(\beta)}$, not merely a difference in the one-point density.
The banded example has a real joint Hamiltonian and a structured bath;
the Gaussian comparison below isolates the joint Dyson constraint.

\input{dyson_sector_results.tex}
\input{gse_operational_appendix.tex}

For a unitarily invariant Hamiltonian and a fixed traceless operator $X$
independent of the Hamiltonian draw,
Haar covariance gives the exact identity
\begin{equation}
 \mathbb E[U_tXU_t^\dagger]
 =\frac{\mathbb E|\Tr U_t|^2-1}{L^2-1}X,\qquad L=2D.
 \label{eq:formfactortwirl}
\end{equation}
Two-level correlations determine the ensemble-mean one-insertion amplitude
through the spectral form factor. The mean-square multitime distance also
involves marked gap pairs, generally four energy indices and eigenvector
overlaps. Appendix~\ref{app:dyson} derives this distinction and the stability
bound for the marked measures.
\input{dyson_probability_results.tex}

\subsection{ETH many-body realization}
Consider an open mixed-field Ising environment and a locally coupled probe,
\begin{align}
 H_{E,N}={}&\sum_{j=1}^{N-1}Z_jZ_{j+1}+0.3\sum_{j=1}^NZ_j
 +1.1\sum_{j=1}^NX_j\nonumber\\
 &+0.25Z_1-0.25Z_N,\\
 H_N={}&0.7625Z_S+H_{E,N}+0.15X_SX_1.
 \label{eq:ethH}
\end{align}
The noncommuting fields give a nonintegrable many-body environment. Its thermal interpretation follows the eigenstate thermalization framework~\cite{Deutsch1991,Srednicki1994,Rigol2008Thermal,Dalessio2016}. Experiments on isolated quantum systems establish the physical relevance of local relaxation and entanglement growth~\cite{Trotzky2012,Kaufman2016}. The process is contracted from the joint Liouville cMPS; the bath remains a dynamical quantum memory throughout the observation window.

For a fixed energy shell, we prepare equal-population phase states in the interacting energy eigenbasis. Their diagonal energy distribution and local energy density are identical, while their off-diagonal phase patterns change the process-visible spectral amplitudes. The stationary reference is the uniform microcanonical state in the shell. Under microcanonical--canonical equivalence, its local thermodynamic limit is a Gibbs reference. The relevant environmental correlator is
\begin{equation}
 C_E(t)=\langle B(t)B(0)\rangle_\beta-\langle B\rangle_\beta^2,
\end{equation}
with spectral density $S_E(\omega)=\int e^{i\omega t}C_E(t)\dd t$. ETH controls the smooth matrix-element envelope, while conservation laws and finite-size gaps determine the low-frequency tail. Thermal and nonthermal stationary ensembles are distinguished by their conserved quantities~\cite{Rigol2007,Mori2018}; here the common stationary process retains the full interacting Hamiltonian.

\subsubsection{A finite-time ordering crossing of the full cPT norm}
For $N=8$ and a shell of $K=32$ eigenstates, the common reference is $\sigma_N=K^{-1}\sum_{a\in I}|E_a\rangle\langle E_a|$. Each phase preparation has the same energy populations, so its trace-norm difference from $\sigma_N$ is $D=2(1-1/K)$. With $T=0.02$ and $x=15T/4$, the omitted sectors obey
\begin{equation}
 \sum_{n>2}e_{i,n}^2\le D^2(e^x-1-x-x^2/2)
 =2.68971\times10^{-4}.
\end{equation}
Combining this tail with the first two cMPS self-contractions gives the intervals
\begin{equation}
\begin{array}{c|cc}
 \tau & \|\Delta\Upsilon_A\|&\|\Delta\Upsilon_B\|\\\hline
 0 &[0.08087,\,0.08252]&[0.06560,\,0.06763]\\
 42 &[0.00621,\,0.01755]&[0.03551,\,0.03913]
\end{array}
\label{eq:fullintervals}
\end{equation}
for the phase preparations $A=163$ and $B=46$. The disjoint intervals reverse
their ordering, and both distances decrease between the endpoints. The tail
covers every omitted sector, so the conclusion concerns the full process norm.
Appendix~\ref{app:eth} gives the contraction and quadrature bounds.

\begin{figure}[tp]
\includegraphics[width=\linewidth]{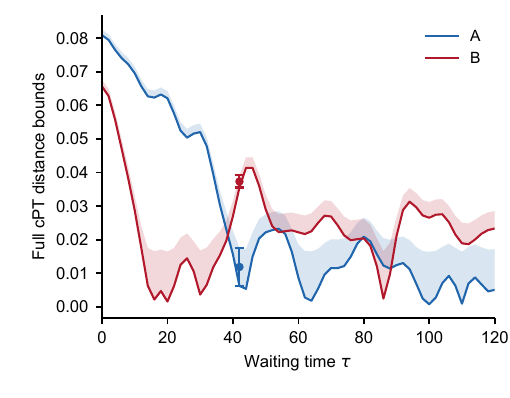}
\caption{Full-process finite-time ordering crossing in the $N=8$ environment. The shaded intervals bound the complete cPT norm using the first two sector contractions and the factorial tail. At $\tau=0$, $A$ is farther from the common stationary process; at $\tau=42$, the intervals show that it is closer.}
\label{fig:n8}
\end{figure}

\subsubsection{Larger environment and sector selectivity}
\begin{figure*}[tp]
\includegraphics[width=\textwidth]{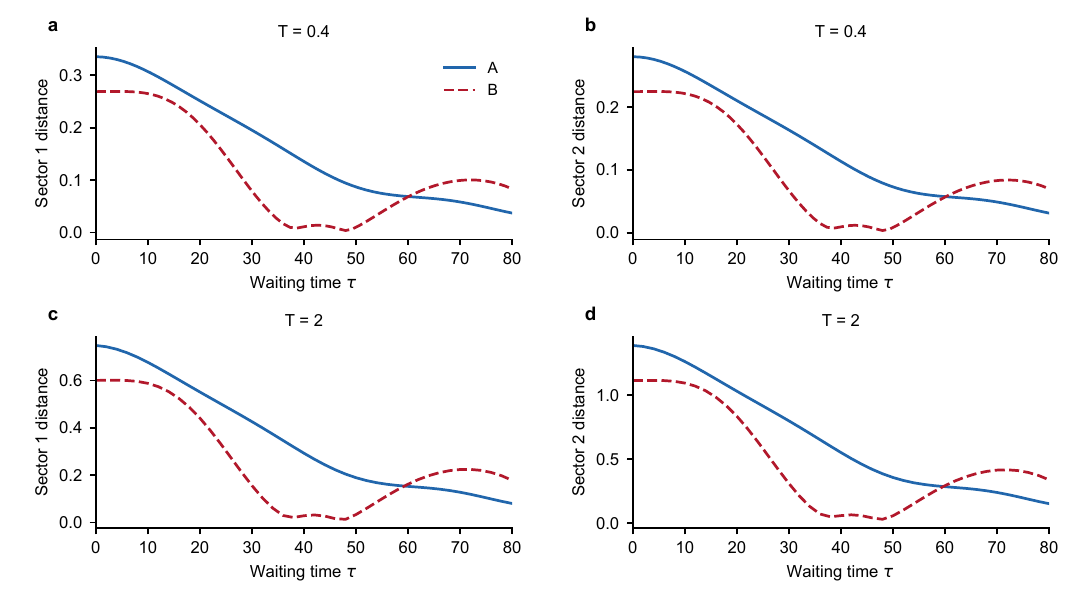}
\caption{The first and second cPT sectors in an $N=10$ environment, evaluated from the joint cMPS contraction. The same two phase preparations reverse their initial ordering in both sectors and both windows. These are fixed-sector distances; the full-norm tail bound is used separately in Fig.~\ref{fig:n8}.}
\label{fig:n10}
\end{figure*}

The same cMPS contraction gives the first two sectors for $N=10$. For the
pair labelled $\texttt{shell\_phase\_163}$ and $\texttt{shell\_phase\_46}$,
the initial ordering reverses at a finite waiting time in both windows, with
late-window RMS ratios near $0.56$ for $T=0.4$ and near $0.54$ for $T=2$.
This extends the many-body process-distance crossing beyond the two-spin
exchange model.
Figure~\ref{fig:n10} displays these two-sector comparisons.

The phase ensemble makes the preparation dependence explicit. Fifty equal-population shell states have the same diagonal energy distribution and the same stationary reference, yet their cPT curves separate because their off-diagonal phases enter the process-visible spectral amplitudes. The ensemble resolves preparation information at fixed energy populations (Fig.~\ref{fig:ensemble}, Appendix~\ref{app:eth}).

\subsubsection{Local preparations across bath sizes and temporal orders}
The preparation filter can also be implemented locally. The operators
in Eq.~\eqref{eq:localpreps} have the same energy-dephased state $I/d_N$
at every $N$. Their positivity and initial process ordering are
independent of system size. Figure~\ref{fig:ethfour} extends this family
to $N=8,10,12,14,16$ bath spins and $n=1,2,3,4$ temporal orders, at
$T=0.02$. The waiting dynamics contains the full many-body bath.
A short-window cMPS contraction supplies the four norms, with the
uniform locality error in Eq.~\eqref{eq:ethfour-locality}.
Appendix~\ref{app:ethfour} derives this contraction and its error bound.

The common energy populations make the origin of the ordering change
explicit: the preparations change temporal response at fixed stationary
reference. The delayed response of the initially hidden correlation in
$B$ persists as the bath grows. For $N=8$, the factorial tail also turns
the four-sector result into the full-process intervals in
Eq.~\eqref{eq:ethfour-full-crossing}.

\subsection{Finite-time ordering crossing and sector-selective acceleration}
Using the finite-interval contraction in Eq.~\eqref{eq:intervalrate},
preparations 33 and 20 in the $N=8$ chain at $T=2$ have $(\kappa_{A,1},\kappa_{B,1})=(0.49818,0.09661)$ and $(\kappa_{A,2},\kappa_{B,2})=(0.38418,0.01234)$ on $[0,2]$. Both sectors reverse their initial ordering. For preparations 1 and 38, the first-sector contractions are $(0.36101,0.21130)$, whereas the second-sector contractions are $(0.28171,0.31296)$. Only the first sector reverses and accelerates. These pairs exhibit simultaneous and selective process acceleration, respectively.

The waiting-time comparison keeps the dynamics fixed. It retains the
non-Markovian response inside the observation window and gives a direct
continuous-time process interpretation of a Mpemba effect.

\input{eth_four_sector_figure.tex}

\subsection{Exact two-qubit process mechanisms}
Two solvable system--memory models separate state relaxation, process
relaxation, and correlated preparation. Their joint semigroups have standard
continuous reservoir dilations and repeated-interaction limits~\cite{Hudson1984,Attal2006}. All process norms use the cMPS self-contraction of Appendix~\ref{app:transfer}.

\subsubsection{A slow correlation hidden from every unperturbed marginal}
Let $\mathsf S$ be swap, $P_s=(I-\mathsf S)/2$, $P_t=I-P_s$, and $\rho_*=I_4/4$. Define $U_t=e^{-igt\mathsf S}$ and
\begin{align}
 \mathcal T(X)&=\Tr(P_sX)P_s+\tfrac13\Tr(P_tX)P_t,\\
 \mathcal E_t(X)&=u_tU_tXU_t^\dagger+v_t\mathcal T(X)
 +w_t\Tr(X)\rho_*,\label{eq:twoqubitmap}
\end{align}
where $u_t=e^{-(\gamma+\epsilon)t}$, $v_t=e^{-\epsilon t}-u_t$, and $w_t=1-e^{-\epsilon t}$. These nonnegative weights and commuting projections give a CPTP semigroup. Compare
\begin{equation}
 \rho_A=\tfrac12(I+X)\otimes\eta,\quad
 \rho_B=\tfrac12(I+bZ)\otimes\eta,\quad
 \eta=\tfrac12(I+mZ).
\end{equation}
Writing $c=\cos(gt)$ and $s=\sin(gt)$, the system trace distances from $I/2$ are
\begin{align}
 d_A(t)&=\tfrac12u_t\sqrt{c^4+m^2c^2s^2+m^2s^4},\\
 d_B(t)&=\tfrac12u_t(bc^2+ms^2).\label{eq:twoqubitdist}
\end{align}
For $b=0.2$ and $m=0.6$, the squared difference is nonnegative at every time,
so the system distances never reverse.

The joint process retains a slow correlation invisible in those distances. With $R=XX+YY+ZZ$, $\mathcal T(\rho_A)=\rho_*$ but $\mathcal T(\rho_B)-\rho_*=bmR/12$. The latter component decays at $\epsilon$, while $A$ has rate $\gamma+\epsilon$. A system rotation $V_z=(I-iZ_S)/\sqrt2$ and a delayed $Z_S$ readout detect it:
\begin{equation}
 \Tr[Z_S\mathcal E_d(V_z(R/4)V_z^\dagger)]
 =e^{-(\gamma+\epsilon)d}\sin(2gd).
 \label{eq:hidden}
\end{equation}
For $d=0.35$, $g=1$, $\gamma=0.8$, and $\epsilon=0.05$, the response is
$0.47844$ and uses operations on $S$ alone.

Figure~\ref{fig:twoqubit}(b) uses a finite-width tester: a recorded branch selects an $X_S$ readout at $u\in[0,0.02]$ with probability $0.8$, or the rotation at $u$ followed by a $Z_S$ readout after $s\in[0.35,0.37]$ with probability $0.2$. Both times are uniformly distributed and then discarded. For $m=0.6$ and $b=0.2$, its output TV distances are $(0.40201,0.01825)$ at $\tau=0$ and $(3.99\times10^{-9},7.14\times10^{-4})$ at $\tau=20$. This is a process-readout reversal in the absence of a state-distance reversal. Appendix~\ref{app:twoqubit} derives the tester and its asymptotic slow coefficient.

\subsubsection{Constrained separable preparations cannot remove the slow mode}
A second benchmark gives an exact constrained optimization. Let
\begin{equation}
 \sigma=\frac{I-mZZ}{4},\quad p_*=\frac{1+m}{4},\quad
 \sigma_t=\frac{\sigma-p_*P_s}{1-p_*},
\end{equation}
and $\mathcal P(X)=\Tr(P_sX)P_s+\Tr(P_tX)\sigma_t$. The fixed generator
\begin{equation}
 \mathcal L(X)=-i[gZZ,X]+\gamma(\mathcal P-I)(X)
 +\epsilon[\Tr(X)\sigma-X]
 \label{eq:thermalbenchmark}
\end{equation}
has Gibbs state $\sigma$ at $\beta g=\operatorname{atanh}m$ and admits energy-eigenoperator jumps with detailed balance, in the thermal semigroup setting~\cite{Davies1974,AlickiLendi2007}.

Set $m=1/50$, $v=23/25$, and impose $\rho_S=\rho_M=I/2$ and $\Tr(ZZ\rho)=-v$. The necessary and sufficient two-qubit positive-partial-transpose criterion~\cite{Peres1996,PPT1996} gives the sharp constraint
\begin{equation}
 \inf_{\rho\in\mathcal C_{\rm sep}}\Tr(P_s\rho)=\frac v2,
 \qquad \frac v2-p_* =\frac{41}{200}>0.
 \label{eq:sepbound}
\end{equation}
Consider the entangled preparation
\begin{equation}
 \rho_e=\frac{I+\frac{v-m}{2}(XX+YY)-vZZ}{4}.
 \label{eq:entprep}
\end{equation}
It is Bell diagonal and positive for the stated parameters; its partial
transpose is not positive, as shown explicitly in Appendix~\ref{app:separable}.
This preparation has the prescribed marginals and energy and satisfies
$\mathcal P(\rho_e)=\sigma$. Its process distance has decay exponent
$\gamma+\epsilon$, while every constrained separable state retains a
system-visible mode with exponent $\epsilon$.
Figure~\ref{fig:twoqubit}(c) and Appendix~\ref{app:separable} give the
resulting uniform eventual advantage and the sharp separable optimization.

%% file: random_matrix_results.tex
\subsection{Sector-selective relaxation in a random-matrix environment}
A qubit coupled to a banded random environment isolates preparation-dependent temporal response without spatial transport. We take the Hamiltonian of Appendix G in Ref.~\cite{DowlingRelaxation2023},
\begin{align}
 H_D&=H_S\otimes I_D+I_2\otimes H_E+\lambda X_S\otimes B,\\
 H_S&=0.25Z_S+0.10X_S,\qquad
 H_E=\sum_{a=1}^D\epsilon_a\ket a\bra a.
 \label{eq:randomH}
\end{align}
The bath energies are sampled uniformly on $[0,1]$. A real symmetric coupling matrix has zero diagonal and vanishes whenever $|\epsilon_a-\epsilon_b|\ge0.5$. We fix $\Tr(B^2)/D=1$ and $\lambda=0.12$, so changes of environment dimension preserve the coupling scale. Each curve uses one fixed autonomous Hamiltonian. Random-matrix environments also provide a setting for non-Markovian reduced dynamics beyond a semigroup description~\cite{Znidaric2011}.

Let $P_I$ project onto $K$ eigenstates of the interacting $H_D$ and set $\omega_I=P_I/K$. Our preparations are
\begin{equation}
 \rho_j=\omega_I+\alpha_jQ_j,\qquad
 Q_j=(-i\operatorname{ad}_{H_D})^{m_j}O_j,\quad m_j\ge1,
 \label{eq:randompreps}
\end{equation}
where $O_j=P_IO_jP_I$ is Hermitian and
$|\alpha_j|\|Q_j\|\le0.9/K$. They are positive on the common support and obey $\mathcal D_{H_D}(\rho_j)=\omega_I$. Thus a speed comparison changes the preparation while keeping the stationary process and Hamiltonian fixed.

For a qubit the first scalar cPT sector has a useful exact interpretation. Write $d_j(\tau)=\tfrac12\|\Tr_E\mathcal U_\tau(\rho_j-\omega_I)\|_1$. Pauli completeness gives
\begin{equation}
 e_{j,1}(\tau)^2=4\int_0^T d_j(\tau+t)^2\dd t.
 \label{eq:firststateidentity}
\end{equation}
This identity connects the contraction to the state-distance integral in Eq.~\eqref{eq:selective-window}.

Figure~\ref{fig:randomselective} shows this separation at $D=48$, $K=12$, and $T=4$. At the preparation time, $A$ is farther in the state distance and both cPT sectors. Throughout $0\le\tau\le32.5$, the state and first-sector distances retain this ordering. The second-sector distances instead change from
\begin{align}
 (e_{A,2}(0),e_{B,2}(0))&=(0.08852,0.07848),\\
 (e_{A,2}(32.5),e_{B,2}(32.5))&=(0.01582,0.02717).
\end{align}
Both second-sector endpoint distances are smaller than their initial values.
The finite-interval comparison uses Eq.~\eqref{eq:intervalrate} and its
analogue for $d_j$. The contractions on $[0,32.5]$ are
\begin{equation}
 \begin{array}{c|cc}
  &\kappa_A&\kappa_B\\\hline
  \text{state}&0.05102&0.06869\\
  n=1&0.05360&0.06202\\
  n=2&0.05299&0.03264
 \end{array}.
 \label{eq:randomrates}
\end{equation}
Preparation $A$ has a speed advantage in the second cPT sector, while $B$ contracts faster in the state and first sector. Appendix~\ref{app:random} supplies the preparation operators, the continuous-order bounds, and a second realization at $D=24$.

The spectral distinction between preparation classes is exact. Under nondegenerate nonzero energy gaps, equal-population pure shell states have the same infinite-time mean-square residual in each sector, independently of their phases. Their finite-time differences arise from interference between visible frequencies. Mixed preparations in Eq.~\eqref{eq:randompreps} also modify the spectral weights through the factor $(E_a-E_b)^{m_j}$. This connects the finite-interval examples to the full-process filtering theorem and identifies the separate spectral input required for a long-time rate law.

%% file: dyson_sector_results.tex
We compare a fixed laboratory pair,
\begin{equation}
 \begin{aligned}
 \rho_A&=\frac{I+0.54Y_S\otimes I_D}{2D},\\
 \rho_B&=\frac{I+0.50Z_S\otimes I_D}{2D},\qquad
 \omega=\frac{I}{2D}.
 \end{aligned}
 \label{eq:dysonfixedpair}
\end{equation}
Both states are positive for every realization and Dyson class. The reference
is the same stationary preparation, independent of $H$. We take $D=8$ and
$T=1.2$. The exact number-resolved
double cMPS transfer computes $n=1,2,3$ together; its dual Gram recursion is
given in Eq.~\eqref{eq:dysonGram}.

One GSE realization resolves a higher-sector advantage with the opposite
first-sector ordering. Its initial ratios are $(1.07318,1.07306,1.07471)$,
and its ratios at $\tau=12$ are $(1.10624,0.91397,0.84810)$.
The interval contractions are
\begin{equation}
 \begin{array}{c|rr}
 &\kappa_A&\kappa_B\\\hline
 n=1&0.152886&0.155414\\
 n=2&0.147279&0.133907\\
 n=3&0.150488&0.130754
 \end{array}.
 \label{eq:GSEselectiverates}
\end{equation}
Preparation $A$ starts farther in all three sectors. It contracts more slowly
in the first sector and more rapidly in the second and third, where its
distance crosses that of $B$. A GOE realization has the same higher-sector
pattern, with rate differences $(-0.011975,0.019410,0.023956)$.
These examples isolate a multitime speed advantage that the first-sector rate
does not detect.

\begin{figure*}[t]
\includegraphics[width=\textwidth]{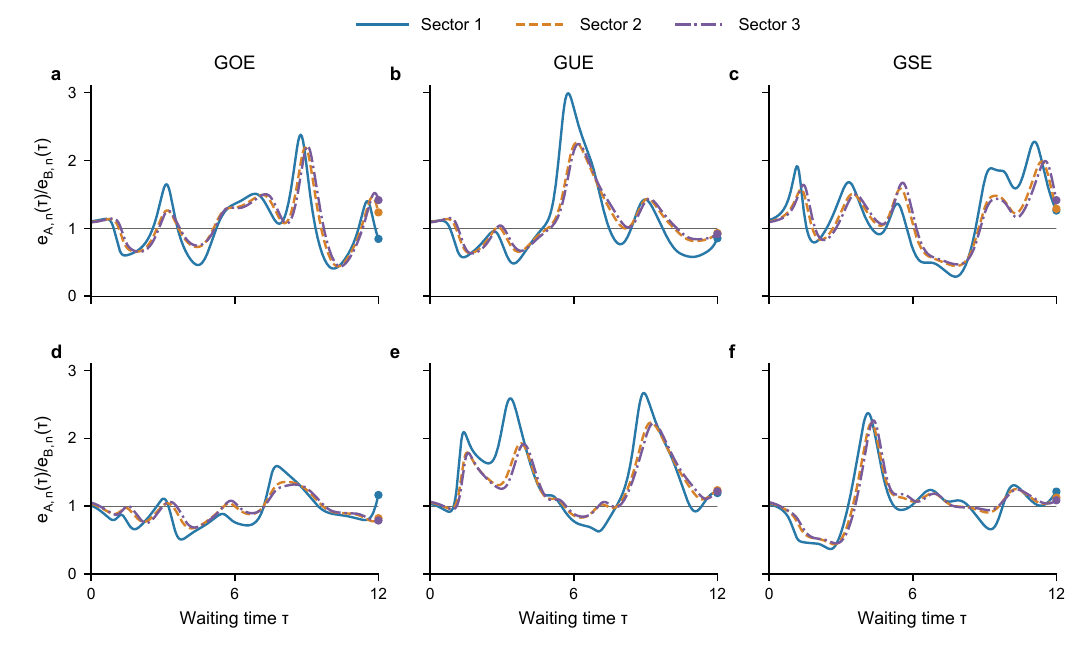}
\caption{Sector coverage of acceleration in joint Dyson ensembles. Each panel
shows the three cMPS sector-distance ratios for the pair in
Eq.~\eqref{eq:dysonfixedpair}, at $D=8$ and $T=1.2$.
Columns are GOE, GUE, and GSE. The rows use common Gaussian draws before
the symmetry projection.
The unit line marks equal distances. In (a), $A$ contracts faster in sector
one and slower in sectors two and three; in (b), all three contract faster
and cross. In (d), $A$ contracts slower in sector one and faster in sectors
two and three, which cross. These are comparisons on $[0,12]$. The interval
rate uses the logarithmic
endpoint contraction, so intermediate oscillations are retained.}
\label{fig:dysonsectors}
\end{figure*}

The distribution of patterns is also a process property. After using nine
realizations per class to choose the common amplitude ratio, we keep that
pair fixed for $41$ independent realizations per class. Requiring $A$ to
start farther in all three sectors and both preparations to have positive
interval contractions gives
\begin{center}
\begin{tabular}{c|ccc}
 & GOE & GUE & GSE\\\hline
 All three cross and accelerate & $13/41$ & $18/41$ & $11/41$\\
 Selective rate advantage & $8/41$ & $3/41$ & $6/41$\\
 Selective rate and a crossing & $4/41$ & $3/41$ & $3/41$
\end{tabular}
\end{center}
In the last two rows the faster sectors form a nonempty proper subset of
$\{1,2,3\}$. A partial set of crossings with faster contraction in all three
is counted separately from a selective rate advantage. The $95\%$ Wilson
intervals for simultaneous three-sector crossing are respectively
$[0.196,0.470]$, $[0.299,0.590]$, and $[0.157,0.419]$.
The patterns coexist within the Dyson classes; their symmetry labels alone
do not determine a Mpemba classification. The classification is fixed by
the preparation-weighted visible kernels. Figure~\ref{fig:dysonsectors}
compares two fixed realizations using the same laboratory pair.
At $D=12$, ten additional fixed realizations per class yield simultaneous
three-sector crossing in $6$, $6$, and $5$ cases, respectively. One GSE
realization again has a slower first-sector contraction and faster second-
and third-sector contractions, with a third-sector crossing.

%% file: gse_operational_appendix.tex
\subsubsection{A fixed-control readout in the noncommuting GSE bath}
\label{app:gseoperational}
The sector-selective \(D=12\) GSE realization also gives an output
crossing under one fixed local control. We retain its Hamiltonian and the
preparations in Eq.~\eqref{eq:dysonfixedpair}, with the stationary reference
\(\omega=I/(2D)\). During \(T=1.2\), apply \(h_{\rm c}=0.3X_S\) and
measure \(M_\pm=(I\pm X_S)/2\). Define
\begin{align}
 U_{\rm c}(T)&=\exp[-i(H+0.3X_S\otimes I_E)T],\nonumber\\
 X_j(\tau)&=\mathcal U_\tau(\rho_j-\omega).
 \label{eq:gseoperationalX}
\end{align}
The process--control cMPS contraction with a terminal effect gives
\begin{align}
 p_{j,\pm}(\tau)
 &=\Tr\!\left[(M_\pm\otimes I_E)
 U_{\rm c}(T)\mathcal U_\tau(\rho_j)U_{\rm c}^\dagger(T)\right],
 \nonumber\\
 p_{*,\pm}
 &=\Tr\!\left[(M_\pm\otimes I_E)
 U_{\rm c}(T)\omega U_{\rm c}^\dagger(T)\right]=\tfrac12 .
 \label{eq:gseoperationalprobabilities}
\end{align}
Thus both outcome probabilities enter the distance to the equally
controlled reference:
\begin{align}
 D_{\rm TV}^{(j)}(\tau)
 &=\frac12\sum_{s=\pm}|p_{j,s}(\tau)-p_{*,s}|\nonumber\\
 &=\left|\Tr\!\left[(M_+\otimes I_E)
 U_{\rm c}(T)X_j(\tau)U_{\rm c}^\dagger(T)\right]\right|.
 \label{eq:gseoperationaltv}
\end{align}
The second equality follows because the two probability differences sum
to \(\Tr X_j=0\).

\begin{figure}[t]
\includegraphics[width=\columnwidth]{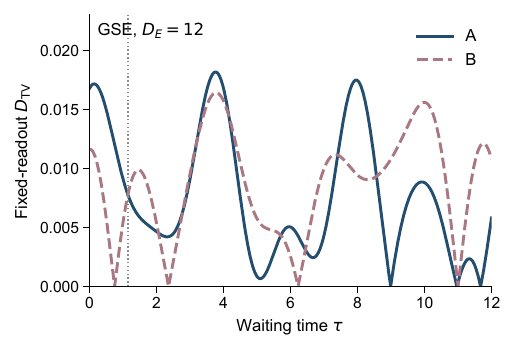}
\caption{Fixed-readout TV distances in a noncommuting GSE environment with
\(D=12\). The control \(0.3X_S\) acts for \(T=1.2\), followed by the
binary \(X_S\) measurement. Both preparations are compared with the same
controlled stationary reference. The dotted line marks their first
crossing. Subsequent oscillations resolve the finite-time character of
the measured ordering.}
\label{fig:gseoperational}
\end{figure}

Initially the TV distances are \((0.01678,0.01162)\) for \((A,B)\).
They first cross near \(\tau=1.15\); at \(\tau=12\), they are
\((0.00581,0.01081)\) [Fig.~\ref{fig:gseoperational}]. Both endpoint
distances are smaller than their initial values. In the same realization,
the sector ratios \(e_{A,n}/e_{B,n}\) at \(\tau=12\) are
\((1.186,1.034,0.996)\) for \(n=1,2,3\): only the third has reversed its
endpoint ordering, while the interval contraction is faster for \(A\) in
the second and third sectors. A single coherent control mixes temporal
orders, making a preparation advantage visible in its output even when
the lower-sector distance orderings persist at the endpoint.

%% file: dyson_probability_results.tex
\subsection{Four-sector occurrence in independent joint environments}
The fixed laboratory pair in Eq.~\eqref{eq:dysonfixedpair} also permits
an occurrence comparison without adapting the preparation to a
realization. We use $500$ fresh realizations per class at each $D=8,12$,
keep $T=1.2$, and compute $n=1,\ldots,4$ by the exact doubled transfer.
The rate is the logarithmic endpoint contraction on $[0,12]$.
All-four eligibility requires $A$ to start farther and both preparations
to contract in every resolved sector. A selective rate event then
requires a nonempty proper subset of faster sectors.

Table~\ref{tab:occurrenceDyson} gives counts with the unconditional
denominator $500$. At $D=8$, simultaneous four-sector crossing and
acceleration occur in $199$, $163$, and $62$ realizations for GOE, GUE,
and GSE. Selective rates with at least one crossing occur in $44$, $53$,
and $26$ realizations. At $D=12$, these selective counts are $48$, $59$,
and $42$. In $40$, $44$, and $41$ realizations, respectively, a higher
sector contracts faster while the first does not. Four-sector
decomposition therefore resolves preparation orderings that a
first-sector rate cannot identify.

\input{dyson_probability_table.tex}

\begin{figure*}[t]
\includegraphics[width=\textwidth]{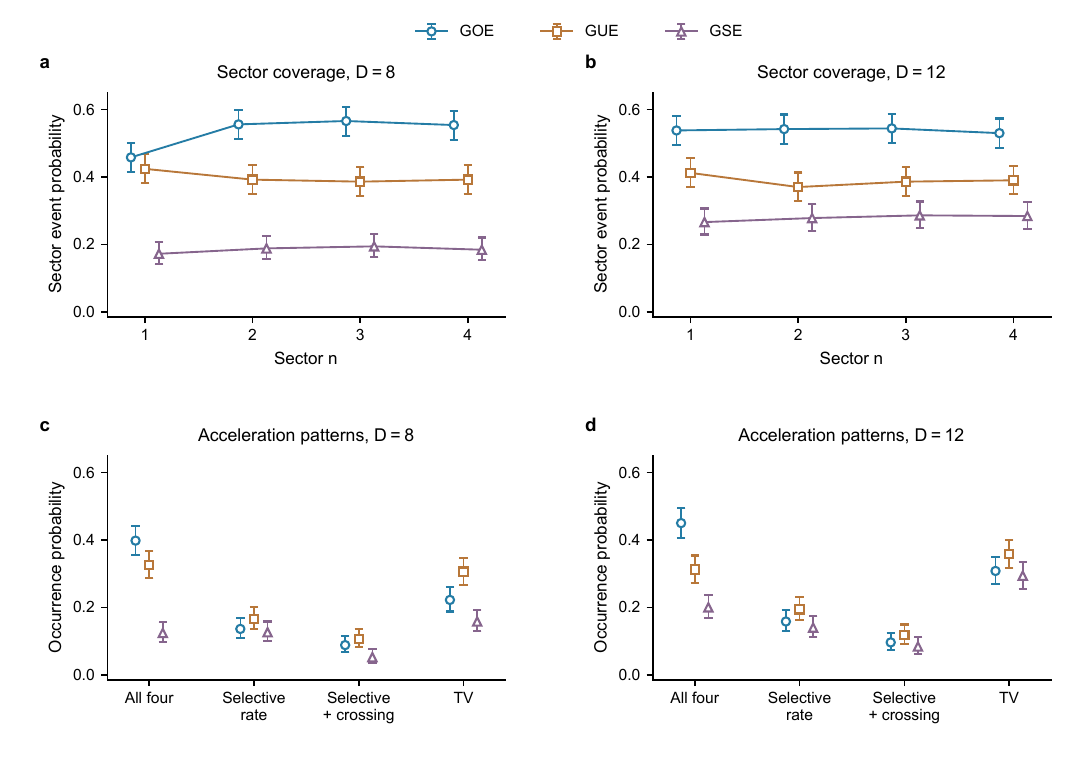}
\caption{Occurrence of finite-window acceleration across joint Dyson
classes for a fixed laboratory preparation pair. Each class and each
environment dimension use $500$ independent realizations. The observation
window is $T=1.2$ and the waiting interval is $[0,12]$.
(a,b) Per-sector probabilities of an initially farther preparation
crossing and contracting faster, with both contraction rates positive.
(c,d) Probabilities of simultaneous four-sector events, selective rate
advantages, selective rates accompanied by a crossing, and fixed-control
TV crossings with faster positive contraction. Selective rate events
require all-four eligibility; TV events use their own initial ordering.
Error bars are $95\%$ Wilson intervals with denominator $500$.
The shared and selective classifications concern the four resolved
sectors and the stated finite interval.}
\label{fig:dysonprobability}
\end{figure*}

Figure~\ref{fig:dysonprobability}(a,b) resolves the crossing probability
by sector. Panels (c,d) compare the shared and selective rate patterns
with the actual fixed-readout event. The latter uses a continuous
$0.5Y_S$ control during the window and the six-effect POVM in
Eq.~\eqref{eq:fixedpovm}. Its initial-order and positive-contraction
requirements are evaluated in TV itself. All three symmetry classes
support both shared and selective finite-window advantages; the
relative probabilities depend on the class and preparation pair.
The stationary-block selection rules in Appendix~\ref{app:dyson}
explain why the same laboratory operator can retain different
components under these joint symmetries.

Every within-class probability uses independent Hamiltonians. Across
classes, the same Gaussian draw is projected onto the three symmetries
before normalization. Class differences therefore retain this pairing
when their uncertainty is assessed. The earlier single realizations
and three-sector comparison in Fig.~\ref{fig:dysonsectors} remain
separate from this fresh four-sector sample.

%% file: dyson_probability_table.tex
\begin{table*}[!t]
\caption{Four-sector finite-interval events for $500$ fresh joint-Hamiltonian realizations
per class and environment dimension. All-four eligibility requires the initial ordering and
positive contraction of both preparations in every resolved sector. Selective events additionally
require a nonempty proper subset of faster sectors. Raw rates concern $[0,12]$.}
\label{tab:occurrenceDyson}
\begin{ruledtabular}
\setlength{\tabcolsep}{5pt}
\renewcommand{\arraystretch}{1.15}
\begin{tabular}{ccrrrrrr}
$D$ & Class & Eligible & \shortstack{All four\\cross} & \shortstack{Selective\\rate} & \shortstack{Selective rate\\+ crossing} & \shortstack{Higher rate\\not first} & \shortstack{TV\\crossing} \\
\hline
8 & GOE & 425 & 199 & 68 & 44 & 42 & 111 \\
8 & GUE & 500 & 163 & 83 & 53 & 34 & 153 \\
8 & GSE & 490 & 62 & 63 & 26 & 35 & 79 \\
12 & GOE & 484 & 225 & 79 & 48 & 40 & 154 \\
12 & GUE & 499 & 156 & 97 & 59 & 44 & 179 \\
12 & GSE & 500 & 100 & 70 & 42 & 41 & 147 \\
\end{tabular}
\end{ruledtabular}
\end{table*}

%% file: eth_four_sector_figure.tex
\begin{figure*}[tp]
\includegraphics[width=\textwidth]{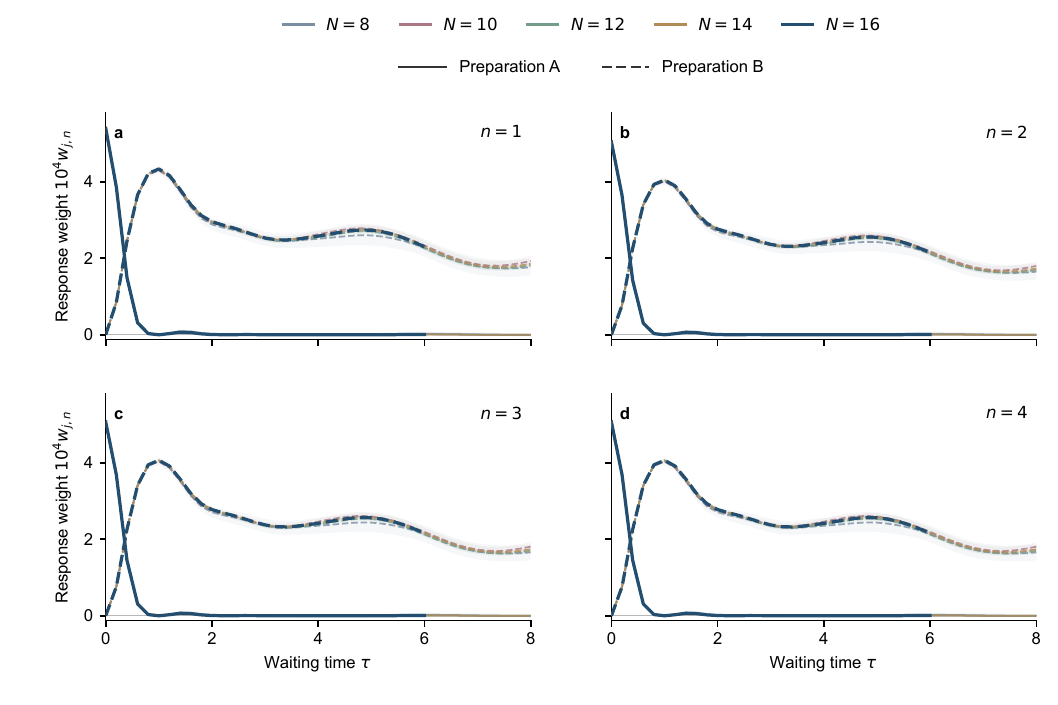}
\caption{Temporal response of local preparations in the mixed-field
Ising environment. The bath contains $N=8,10,12,14,16$ spins, in
addition to the probe qubit. Panels show $n=1,2,3,4$ cPT sectors after
removing their factorial weight, $w_{j,n}=n!e_{j,n}^2/(15T/4)^n$, with
$T=0.02$. Solid and dashed curves denote $A$ and $B$; colors denote
bath size. There are 41 waiting times on $[0,8]$ for $N\le14$ and
31 on $[0,6]$ for $N=16$, each using 50 independent random-phase trace
vectors. Ribbons show two jackknife standard errors
of the squared contraction. The delayed response
of $B$ and the rapidly reduced response of $A$ appear in all four
orders. These are finite-time profiles of the preparation filter.}
\label{fig:ethfour}
\end{figure*}

%% file: references.bib
@CONTROL{apsrev42Control,
  title="0"
}

@article{FigueroaRomero2020,
  author={Figueroa-Romero, Pedro and Modi, Kavan and Pollock, Felix A.},
  title={Equilibration on average in quantum processes with finite temporal resolution},
  journal={Physical Review E},
  volume={102},
  pages={032144},
  year={2020},
  doi={10.1103/physreve.102.032144}
}

@article{DowlingRelaxation2023,
  author={Dowling, Neil and Figueroa-Romero, Pedro and Pollock, Felix A. and Strasberg, Philipp and Modi, Kavan},
  title={Relaxation of Multitime Statistics in Quantum Systems},
  journal={Quantum},
  volume={7},
  pages={1027},
  year={2023},
  doi={10.22331/q-2023-06-01-1027}
}

@article{DowlingEquilibration2023,
  author={Dowling, Neil and Figueroa-Romero, Pedro and Pollock, Felix A. and Strasberg, Philipp and Modi, Kavan},
  title={Equilibration of multitime quantum processes in finite time intervals},
  journal={SciPost Physics Core},
  volume={6},
  pages={043},
  year={2023},
  doi={10.21468/scipostphyscore.6.2.043}
}

@article{Wassner2026,
  author={Wassner, Clara and Fuksa, Jon{\'a}{\v{s}} and Eisert, Jens and White, Gregory A. L.},
  title={An operational continuum limit of quantum combs},
  journal={arXiv preprint arXiv:2601.16974}, year={2026}, eprint={2601.16974}, archivePrefix={arXiv}
}

@article{Fuksa2026,
  author={Fuksa, Jon{\'a}{\v{s}} and Wassner, Clara and Eisert, Jens and White, Gregory A. L.},
  title={Dilation theorem for continuum quantum stochastic processes},
  journal={arXiv preprint arXiv:2609.35403}, year={2026}, eprint={2609.35403}, archivePrefix={arXiv}
}

@article{Chiribella2009,
  author={Chiribella, Giulio and D'Ariano, Giacomo Mauro and Perinotti, Paolo},
  title={Theoretical framework for quantum networks},
  journal={Physical Review A},
  volume={80},
  pages={022339},
  year={2009},
  doi={10.1103/physreva.80.022339}
}

@article{Chiribella2008,
  author={Chiribella, G. and D'Ariano, G. M. and Perinotti, P.},
  title={Quantum Circuit Architecture},
  journal={Physical Review Letters},
  volume={101},
  pages={060401},
  year={2008},
  doi={10.1103/physrevlett.101.060401}
}

@article{Pollock2018,
  author={Pollock, Felix A. and Rodr{\'{i}}guez-Rosario, C{\'{e}}sar and Frauenheim, Thomas and Paternostro, Mauro and Modi, Kavan},
  title={Non-Markovian quantum processes: Complete framework and efficient characterization},
  journal={Physical Review A},
  volume={97},
  pages={012127},
  year={2018},
  doi={10.1103/physreva.97.012127}
}

@article{MilzModi2021,
  author={Milz, Simon and Modi, Kavan},
  title={Quantum Stochastic Processes and Quantum non-Markovian Phenomena},
  journal={PRX Quantum},
  volume={2},
  pages={030201},
  year={2021},
  doi={10.1103/prxquantum.2.030201}
}

@article{Pollock2018Markov,
  author={Pollock, Felix A. and Rodr{\'{i}}guez-Rosario, C{\'{e}}sar and Frauenheim, Thomas and Paternostro, Mauro and Modi, Kavan},
  title={Operational Markov Condition for Quantum Processes},
  journal={Physical Review Letters},
  volume={120},
  pages={040405},
  year={2018},
  doi={10.1103/physrevlett.120.040405}
}

@article{White2020,
  author={White, G. A. L. and Hill, C. D. and Pollock, F. A. and Hollenberg, L. C. L. and Modi, K.},
  title={Demonstration of non-Markovian process characterisation and control on a quantum processor},
  journal={Nature Communications},
  volume={11},
  pages={6301},
  year={2020},
  doi={10.1038/s41467-020-20113-3}
}

@article{Verstraete2010,
  author={Verstraete, F. and Cirac, J. I.},
  title={Continuous Matrix Product States for Quantum Fields},
  journal={Physical Review Letters},
  volume={104},
  pages={190405},
  year={2010},
  doi={10.1103/physrevlett.104.190405}
}

@article{Haegeman2013,
  author={Haegeman, Jutho and Cirac, J. Ignacio and Osborne, Tobias J. and Verstraete, Frank},
  title={Calculus of continuous matrix product states},
  journal={Physical Review B},
  volume={88},
  pages={085118},
  year={2013},
  doi={10.1103/physrevb.88.085118}
}

@article{Hudson1984,
  author={Hudson, R. L. and Parthasarathy, K. R.},
  title={Quantum Ito's formula and stochastic evolutions},
  journal={Communications in Mathematical Physics},
  volume={93},
  pages={301},
  year={1984},
  doi={10.1007/bf01258530}
}

@article{Attal2006,
  author={Attal, St{\'{e}}phane and Pautrat, Yan},
  title={From Repeated to Continuous Quantum Interactions},
  journal={Annales Henri Poincar{\'{e}}},
  volume={7},
  pages={59},
  year={2006},
  doi={10.1007/s00023-005-0242-8}
}

@article{Gough2009,
  author={Gough, J. and James, M.R.},
  title={The Series Product and Its Application to Quantum Feedforward and Feedback Networks},
  journal={IEEE Transactions on Automatic Control},
  volume={54},
  pages={2530},
  year={2009},
  doi={10.1109/tac.2009.2031205}
}

@book{WisemanMilburn2010,
  author={Wiseman, Howard M. and Milburn, Gerard J.},
  title={Quantum Measurement and Control},
  publisher={Cambridge University Press},
  year={2010},
  doi={10.1017/cbo9780511813948}
}

@article{JacobsSteck2006,
  author={Jacobs, Kurt and Steck, Daniel A.},
  title={A straightforward introduction to continuous quantum measurement},
  journal={Contemporary Physics},
  volume={47},
  pages={279},
  year={2006},
  doi={10.1080/00107510601101934}
}

@article{Dalibard1992,
  author={Dalibard, Jean and Castin, Yvan and M{\o}lmer, Klaus},
  title={Wave-function approach to dissipative processes in quantum optics},
  journal={Physical Review Letters},
  volume={68},
  pages={580},
  year={1992},
  doi={10.1103/physrevlett.68.580}
}

@article{Breuer2009,
  author={Breuer, Heinz-Peter and Laine, Elsi-Mari and Piilo, Jyrki},
  title={Measure for the Degree of Non-Markovian Behavior of Quantum Processes in Open Systems},
  journal={Physical Review Letters},
  volume={103},
  pages={210401},
  year={2009},
  doi={10.1103/physrevlett.103.210401}
}

@article{Rivas2014,
  author={Rivas, {\'{A}}ngel and Huelga, Susana F and Plenio, Martin B},
  title={Quantum non-Markovianity: characterization, quantification and detection},
  journal={Reports on Progress in Physics},
  volume={77},
  pages={094001},
  year={2014},
  doi={10.1088/0034-4885/77/9/094001}
}

@article{Davies1974,
  author={Davies, E. B.},
  title={Markovian master equations},
  journal={Communications in Mathematical Physics},
  volume={39},
  pages={91},
  year={1974},
  doi={10.1007/bf01608389}
}

@book{AlickiLendi2007,
  author={Alicki, Robert and Lendi, Karl},
  title={Quantum Dynamical Semigroups and Applications},
  publisher={Springer Berlin Heidelberg},
  year={2007},
  doi={10.1007/3-540-70861-8}
}

@article{Deutsch1991,
  author={Deutsch, J. M.},
  title={Quantum statistical mechanics in a closed system},
  journal={Physical Review A},
  volume={43},
  pages={2046},
  year={1991},
  doi={10.1103/physreva.43.2046}
}

@article{Srednicki1994,
  author={Srednicki, Mark},
  title={Chaos and quantum thermalization},
  journal={Physical Review E},
  volume={50},
  pages={888},
  year={1994},
  doi={10.1103/physreve.50.888}
}

@article{Rigol2007,
  author={Rigol, Marcos and Dunjko, Vanja and Yurovsky, Vladimir and Olshanii, Maxim},
  title={Relaxation in a Completely Integrable Many-Body Quantum System: An Ab Initio Study of the Dynamics of the Highly Excited States of 1D Lattice Hard-Core Bosons},
  journal={Physical Review Letters},
  volume={98},
  pages={050405},
  year={2007},
  doi={10.1103/physrevlett.98.050405}
}

@article{Rigol2008Thermal,
  author={Rigol, Marcos and Dunjko, Vanja and Olshanii, Maxim},
  title={Thermalization and its mechanism for generic isolated quantum systems},
  journal={Nature},
  volume={452},
  pages={854},
  year={2008},
  doi={10.1038/nature06838}
}

@article{Dalessio2016,
  author={D'Alessio, Luca and Kafri, Yariv and Polkovnikov, Anatoli and Rigol, Marcos},
  title={From quantum chaos and eigenstate thermalization to statistical mechanics and thermodynamics},
  journal={Advances in Physics},
  volume={65},
  pages={239},
  year={2016},
  doi={10.1080/00018732.2016.1198134}
}

@article{Mori2018,
  author={Mori, Takashi and Ikeda, Tatsuhiko N and Kaminishi, Eriko and Ueda, Masahito},
  title={Thermalization and prethermalization in isolated quantum systems: a theoretical overview},
  journal={Journal of Physics B: Atomic, Molecular and Optical Physics},
  volume={51},
  pages={112001},
  year={2018},
  doi={10.1088/1361-6455/aabcdf}
}

@article{GogolinEisert2016,
  author={Gogolin, Christian and Eisert, Jens},
  title={Equilibration, thermalisation, and the emergence of statistical mechanics in closed quantum systems},
  journal={Reports on Progress in Physics},
  volume={79},
  pages={056001},
  year={2016},
  doi={10.1088/0034-4885/79/5/056001}
}

@article{Linden2009,
  author={Linden, Noah and Popescu, Sandu and Short, Anthony J. and Winter, Andreas},
  title={Quantum mechanical evolution towards thermal equilibrium},
  journal={Physical Review E},
  volume={79},
  pages={061103},
  year={2009},
  doi={10.1103/physreve.79.061103}
}

@article{Reimann2008,
  author={Reimann, Peter},
  title={Foundation of Statistical Mechanics under Experimentally Realistic Conditions},
  journal={Physical Review Letters},
  volume={101},
  pages={190403},
  year={2008},
  doi={10.1103/physrevlett.101.190403}
}

@article{ShortFarrelly2012,
  author={Short, Anthony J and Farrelly, Terence C},
  title={Quantum equilibration in finite time},
  journal={New Journal of Physics},
  volume={14},
  pages={013063},
  year={2012},
  doi={10.1088/1367-2630/14/1/013063}
}

@article{LiebRobinson1972,
  author={Lieb, Elliott H. and Robinson, Derek W.},
  title={The finite group velocity of quantum spin systems},
  journal={Communications in Mathematical Physics},
  volume={28},
  pages={251},
  year={1972},
  doi={10.1007/bf01645779}
}

@article{Bravyi2006,
  author={Bravyi, S. and Hastings, M. B. and Verstraete, F.},
  title={Lieb-Robinson Bounds and the Generation of Correlations and Topological Quantum Order},
  journal={Physical Review Letters},
  volume={97},
  pages={050401},
  year={2006},
  doi={10.1103/physrevlett.97.050401}
}

@article{Nachtergaele2006,
  author={Nachtergaele, Bruno and Sims, Robert},
  title={Lieb-Robinson Bounds and the Exponential Clustering Theorem},
  journal={Communications in Mathematical Physics},
  volume={265},
  pages={119},
  year={2006},
  doi={10.1007/s00220-006-1556-1}
}

@article{Mpemba1969,
  author={Mpemba, E B and Osborne, D G},
  title={Cool?},
  journal={Physics Education},
  volume={4},
  pages={312},
  year={1969},
  doi={10.1088/0031-9120/4/3/312}
}

@article{Klich2019,
  author={Klich, Israel and Raz, Oren and Hirschberg, Ori and Vucelja, Marija},
  title={Mpemba Index and Anomalous Relaxation},
  journal={Physical Review X},
  volume={9},
  pages={021060},
  year={2019},
  doi={10.1103/physrevx.9.021060}
}

@article{LuRaz2017,
  author={Lu, Zhiyue and Raz, Oren},
  title={Nonequilibrium thermodynamics of the Markovian Mpemba effect and its inverse},
  journal={Proceedings of the National Academy of Sciences},
  volume={114},
  pages={5083},
  year={2017},
  doi={10.1073/pnas.1701264114}
}

@article{Carollo2021,
  author={Carollo, Federico and Lasanta, Antonio and Lesanovsky, Igor},
  title={Exponentially Accelerated Approach to Stationarity in Markovian Open Quantum Systems through the Mpemba Effect},
  journal={Physical Review Letters},
  volume={127},
  pages={060401},
  year={2021},
  doi={10.1103/physrevlett.127.060401}
}

@article{Nava2024,
  author={Nava, Andrea and Egger, Reinhold},
  title={Mpemba Effects in Open Nonequilibrium Quantum Systems},
  journal={Physical Review Letters},
  volume={133},
  pages={136302},
  year={2024},
  doi={10.1103/physrevlett.133.136302}
}

@article{Strachan2025,
  author={Strachan, David J. and Purkayastha, Archak and Clark, Stephen R.},
  title={Non-Markovian Quantum Mpemba Effect},
  journal={Physical Review Letters},
  volume={134},
  pages={220403},
  year={2025},
  doi={10.1103/physrevlett.134.220403}
}

@article{Bhore2025,
  author={Bhore, Tanmay and Su, Lei and Martin, Ivar and Clerk, Aashish A. and Papi{\'{c}}, Zlatko},
  title={Quantum Mpemba effect without global symmetries},
  journal={Physical Review B},
  volume={112},
  pages={L121109},
  year={2025},
  doi={10.1103/1td3-2vwf}
}

@article{AresReview2025,
  author={Ares, Filiberto and Calabrese, Pasquale and Murciano, Sara},
  title={The quantum Mpemba effects},
  journal={Nature Reviews Physics},
  volume={7},
  pages={451--460},
  year={2025},
  doi={10.1038/s42254-025-00838-0}
}

@article{Rylands2024,
  author={Rylands, Colin and Klobas, Katja and Ares, Filiberto and Calabrese, Pasquale and Murciano, Sara and Bertini, Bruno},
  title={Microscopic Origin of the Quantum Mpemba Effect in Integrable Systems},
  journal={Physical Review Letters},
  volume={133},
  pages={010401},
  year={2024},
  doi={10.1103/PhysRevLett.133.010401}
}

@article{Moroder2024,
  author={Moroder, Michael and Culhane, Ois{\'i}n and Zawadzki, Krzysztof and Goold, John},
  title={Thermodynamics of the Quantum Mpemba Effect},
  journal={Physical Review Letters},
  volume={133},
  pages={140404},
  year={2024},
  doi={10.1103/PhysRevLett.133.140404}
}

@article{Joshi2024,
  author={Joshi, Lata Kh. and Franke, Johannes and Rath, Aniket and Ares, Filiberto and Murciano, Sara and Kranzl, Florian and Blatt, Rainer and Zoller, Peter and Vermersch, Beno{\^\i}t and Calabrese, Pasquale and Roos, Christian F. and Joshi, Manoj K.},
  title={Observing the Quantum Mpemba Effect in Quantum Simulations},
  journal={Physical Review Letters},
  volume={133},
  pages={010402},
  year={2024},
  doi={10.1103/PhysRevLett.133.010402}
}

@article{AharonyShapira2024,
  author={Aharony Shapira, Shahaf and Shapira, Yotam and Markov, Jovan and Teza, Gianluca and Akerman, Nitzan and Raz, Oren and Ozeri, Roee},
  title={Inverse Mpemba Effect Demonstrated on a Single Trapped Ion Qubit},
  journal={Physical Review Letters},
  volume={133},
  pages={010403},
  year={2024},
  doi={10.1103/PhysRevLett.133.010403}
}

@article{ZhangStrong2025,
  author={Zhang, Jie and Xia, Gang and Wu, Chun-Wang and Chen, Ting and Zhang, Qian and Xie, Yi and Su, Wen-Bo and Wu, Wei and Qiu, Cheng-Wei and Chen, Ping-Xing and Li, Weibin and Jing, Hui and Zhou, Yan-Li},
  title={Observation of quantum strong Mpemba effect},
  journal={Nature Communications},
  volume={16},
  pages={301},
  year={2025},
  doi={10.1038/s41467-024-54303-0}
}

@article{LonghiEnt2025,
  author={Longhi, Stefano},
  title={Quantum Mpemba effect from initial system-reservoir entanglement},
  journal={APL Quantum},
  volume={2},
  pages={026133},
  year={2025},
  doi={10.1063/5.0266143}
}

@article{Kaufman2016,
  author={Kaufman, Adam M. and Tai, M. Eric and Lukin, Alexander and Rispoli, Matthew and Schittko, Robert and Preiss, Philipp M. and Greiner, Markus},
  title={Quantum thermalization through entanglement in an isolated many-body system},
  journal={Science},
  volume={353},
  pages={794},
  year={2016},
  doi={10.1126/science.aaf6725}
}

@article{Trotzky2012,
  author={Trotzky, S. and Chen, Y-A. and Flesch, A. and McCulloch, I. P. and Schollw{\"{o}}ck, U. and Eisert, J. and Bloch, I.},
  title={Probing the relaxation towards equilibrium in an isolated strongly correlated one-dimensional Bose gas},
  journal={Nature Physics},
  volume={8},
  pages={325},
  year={2012},
  doi={10.1038/nphys2232}
}

@article{Wiener1930,
  author={Wiener, Norbert},
  title={Generalized harmonic analysis},
  journal={Acta Mathematica},
  volume={55},
  pages={117},
  year={1930},
  doi={10.1007/bf02546511}
}

@article{Peres1996,
  author={Peres, Asher},
  title={Separability Criterion for Density Matrices},
  journal={Physical Review Letters},
  volume={77},
  pages={1413},
  year={1996},
  doi={10.1103/physrevlett.77.1413}
}

@article{PPT1996,
  author={Horodecki, Micha{\l} and Horodecki, Pawe{\l} and Horodecki, Ryszard},
  title={Separability of mixed states: necessary and sufficient conditions},
  journal={Physics Letters A},
  volume={223},
  pages={1},
  year={1996},
  doi={10.1016/s0375-9601(96)00706-2}
}

@article{Znidaric2011,
  author={{\v{Z}}nidari{\v{c}}, Marko and Pineda, Carlos and Garc{\'{i}}a-Mata, Ignacio},
  title={Non-Markovian Behavior of Small and Large Complex Quantum Systems},
  journal={Physical Review Letters},
  volume={107},
  pages={080404},
  year={2011},
  doi={10.1103/physrevlett.107.080404}
}

@article{Breuer2016,
  author={Breuer, Heinz-Peter and Laine, Elsi-Mari and Piilo, Jyrki and Vacchini, Bassano},
  title={Colloquium : Non-Markovian dynamics in open quantum systems},
  journal={Reviews of Modern Physics},
  volume={88},
  pages={021002},
  year={2016},
  doi={10.1103/revmodphys.88.021002}
}

@article{Wigner1958,
  author={Wigner, Eugene P.},
  title={On the Distribution of the Roots of Certain Symmetric Matrices},
  journal={Annals of Mathematics},
  volume={67},
  pages={325--327},
  year={1958},
  doi={10.2307/1970008}
}

@article{Dyson1962,
  author={Dyson, Freeman J.},
  title={The Threefold Way. Algebraic Structure of Symmetry Groups and Ensembles in Quantum Mechanics},
  journal={Journal of Mathematical Physics},
  volume={3},
  pages={1199--1215},
  year={1962},
  doi={10.1063/1.1703863}
}

@article{DumitriuEdelman2002,
  author={Dumitriu, Ioana and Edelman, Alan},
  title={Matrix models for beta ensembles},
  journal={Journal of Mathematical Physics},
  volume={43},
  pages={5830--5847},
  year={2002},
  doi={10.1063/1.1507823}
}
